\documentclass[english, 12pt]{article}
\usepackage[T1]{fontenc}
\usepackage[utf8]{inputenc}
\usepackage[top=1.25in, bottom=1.25in, left=1.25in, right=1.25in]{geometry}
\usepackage[pdftex]{graphicx}
\usepackage{verbatim}
\usepackage[width=0.8\textwidth]{caption}
\usepackage{amsmath,amssymb,amsopn,dsfont,mathrsfs,bm}
\usepackage[english]{babel}
\usepackage{lmodern}
\usepackage{flafter, float,placeins,longtable,array}
\usepackage[authoryear]{natbib}
\usepackage{euscript}
\usepackage[table]{xcolor}
\usepackage{comment}
\usepackage{color}
\usepackage{lscape,subfig,threeparttable}
\usepackage{adjustbox}

\usepackage{comment}
\usepackage{listings}
\usepackage{color} 
\definecolor{mygreen}{RGB}{28,172,0} 
\definecolor{mylilas}{RGB}{170,55,241}

\usepackage{siunitx}

\usepackage{pdflscape}
\usepackage{afterpage}

\usepackage{multirow}
\usepackage{enumitem}
\usepackage{booktabs}
\usepackage{stmaryrd}
\usepackage[hidelinks]{hyperref}
\definecolor{navyblue}{rgb}{0.0, 0.0, 0.5}
\hypersetup{breaklinks=true,colorlinks=true, citecolor=navyblue, linkcolor=black, urlcolor=black}
\makeatletter

\newtheorem{thm}{Theorem}[section]
\newtheorem{cor}[thm]{Corollary}
\newtheorem{lem}[thm]{Lemma}
\newtheorem{prop}[thm]{Proposition}
\newtheorem{hyp}{Assumption}

\numberwithin{equation}{section}

\newcommand{\ConvNor}[1]{\stackrel{d}{\to} \mathcal{N}\left(0,#1\right)}

\newcommand{\norm}[1]{\big\Vert#1\big\Vert}
\newcommand{\abs}[1]{\left\vert#1\right\vert}
\newcommand{\set}[1]{\left\{#1\right\}}

\newcommand{\N}{\mathbb N}
\newcommand{\R}{\mathbb R}

\newcommand{\E}{\mathbb{E}}

\newcommand{\plim}{\mathrm{plim}}
\newcommand{\eps}{\varepsilon}

\newcommand{\ind}[1]{\mathbf{1}\{ #1 \}}

\DeclareMathOperator*{\argmax}{arg\,max}
\DeclareMathOperator*{\argmin}{arg\,min}

\renewcommand{\section}{\@startsection{section}{2}{0mm}{-1.5\baselineskip}{1\baselineskip}{\normalfont\Large\bfseries}}
\renewcommand{\subsection}{\@startsection{subsection}{2}{0mm}{-1.2\baselineskip}{1\baselineskip}{\normalfont\normalsize\bfseries}}
\renewcommand{\subsubsection}{\@startsection{subsubsection}{3}{0mm}{-0.8\baselineskip}{0.4\baselineskip}{\normalfont\normalsize\itshape}}

\title{A Simple and Computationally Trivial Estimator for Grouped Fixed Effects Models\footnote{\linespread{1}\selectfont An earlier version of this paper based on the third chapter of my Ph.D.~dissertation at CREST was circulated under the title ``Make the Difference!~Computationally Trivial Estimators for Grouped Fixed Effects Models''. I thank Jad Beyhum, Stéphane Bonhomme, Xavier D'Haultf\oe uille, Bruno Furtado, Elena Manresa, Pauline Rossi, Ao Wang, Martin Weidner, Andrei Zeleneev, two anonymous referees, an associate editor, the co-editor Aureo De Paula, and seminar participants at CREST, LMU Munich, UChicago, 2024  Causal Inference and Machine Learning Workshop in Groningen, Oxford EET 2023, and Bristol ESG 2022 for helpful comments. This research is supported by the French National Research Agency grants ANR-17-CE26-0015-041, ANR-18-EURE-0005, ANR-11-LABX-0047, ANR-17-EURE-0001, and the European Research Council grant ERC-2018-CoG-819086-PANEDA.}}
\author{Martin Mugnier\thanks{Paris School of Economics, \href{mailto:martin.mugnier@psemail.eu}{martin.mugnier@psemail.eu}}}
\date{}

\begin{document}
\maketitle

\thispagestyle{empty}
	
\begin{abstract}
\linespread{1}\selectfont
\noindent This paper introduces a new fixed effects estimator for linear panel data models with clustered time patterns of unobserved heterogeneity. The method avoids non-convex and combinatorial optimization by combining a preliminary consistent estimator of the slope coefficient, an agglomerative pairwise-differencing clustering of cross-sectional units, and a pooled ordinary least squares regression. Asymptotic guarantees are established in a framework where $T$ can grow at any power of $N$, as both $N$ and $T$ approach infinity. Unlike most existing approaches, the proposed estimator is computationally straightforward and does not require a known upper bound on the number of groups. As existing approaches, this method leads to a consistent estimation of well-separated groups and an estimator of common parameters asymptotically equivalent to the infeasible regression controlling for the true groups. An application revisits the statistical association between income and democracy. 
	
\end{abstract}
\thispagestyle{empty}

\newpage
\pagenumbering{arabic}

\section{Introduction} 
Suppose a sample of panel data $\{(y_{it},x_{it}):1\leq i\leq N, 1\leq t\leq T\}$ is observed and consider a linear regression model with grouped fixed effects:
\begin{equation}\label{eq:model}
\left\{
\begin{array}{ll}
	y_{it}=x_{it}'\beta +\alpha_{g_it}+v_{it} \\
	\E[v_{it}]=\E[\alpha_{g_it}v_{it}]=\E[x_{itk}v_{it}]=0 
	\end{array} \right., \quad i=1,\ldots, N, \;t=1,\ldots,T, \; k=1,\ldots,K,
\end{equation}
where $i$ denotes cross-sectional units, $t$ denotes time periods, $y_{it} \in \R$ is a dependent variable, and $x_{it}\in\R^K$ is a vector of explanatory covariates uncorrelated with the zero-mean random variable $v_{it}\in\R$ but possibly arbitrarily correlated with the unobserved group membership variable  $g_i\in\{1,\ldots,G\}$ and the group-time effect $\alpha_{g_it}\in\R$. 

This paper focuses on the estimation of and inference on the unknown slope parameter $\beta\in\R^K$ and the group-time effects $(\alpha_{1t},\ldots,\alpha_{Gt})'\in\R^{G}$, as well as the consistent estimation of the group memberships $g_i\in\set{1,\ldots,G}$ and the number of groups $G$, within an asymptotic framework in which $N/T^\nu\to0$ for some constant $\nu>0$, as $N$ and $T$ diverge while $K$ and $G$ remain fixed. 

Since they are special cases of interactive fixed effects with factor loadings confined to a finite set,\footnote{Note that $\alpha_{g_it}=\lambda_i'f_t$ for any $\lambda_i'\equiv\left(c_1\mathbf{1}\{g_i=1\},\ldots,c_G\mathbf{1}\{g_i=G\}\right)$, $f_t'\equiv\left(\alpha_{1t}/c_1,\ldots,\alpha_{Gt}/c_G\right)$, and $c\equiv(c_1,\ldots,c_G)'\in(\R\backslash\{0\})^{G}$. Thus $\lambda_1/c,\ldots, \lambda_N/c$ lie in the finite set of vertices of the unit simplex of $\R^G$. Reciprocally, if $\tilde f_t\in\R^r$ for some $r\in\N\backslash\{0\}$ and $\tilde \lambda_i\in\Lambda\subset\R^r$ with $\abs{\Lambda}=\tilde G$, then there exist $(\tilde g_1,\ldots,\tilde g_N)'\in\{1,\ldots, \tilde G\}^N$ and $(\tilde\alpha_{11},\ldots,\tilde\alpha_{\tilde GT})'\in\R^{\tilde GT}$ such that  $\tilde \lambda_i'\tilde f_t=\tilde\alpha_{\tilde g_it}$. \label{footnote:factor_analytic}}  grouped fixed effects (GFE hereafter) provide a parsimonious yet flexible device to accommodate cross-sectional correlations and a few unrestricted trends of unobserved heterogeneity.  Since their introduction in economics \citep{HahnMoon2010, BM2015}, GFE have gained considerable interest in both methodological and applied work \citep[e.g.,][]{SSP2016, cheng2019clustering, BLM2019,GU2019, ChetverikovManresa2021, BonhommeLamadonManresa2022, mugnier_JMP,Janys2024}.  

Treating GFE as interactive fixed effects, however, leads to two main issues. First, parametric-rate inference for the slope parameter is generally not available when $T$ grows very slowly with $N$ \citep[e.g.,][]{Bai2009, moon2019nuclear, BeyhumGautier2023}. Second, although \cite{Higgins2022} shows that parametric-rate inference remains possible under some circumstances, his proposed method, like most other interactive fixed effects methods, solves a high-dimensional non-convex least-squares problem typically subject to local minima. Both issues can lead to poor inferences, especially in microeconometric datasets where $T$ may be much smaller than $N$. 

Addressing the combinatorial nature of GFE, on the other hand,  introduces two main difficulties. First, GFE estimators, defined as pseudo joint maximum likelihood estimators optimizing over all possible partitions of cross-sectional units in $G$ groups, encounter a challenging non-convex and combinatorial optimization problem. If P$\neq$NP, exact solutions in polynomial time are unattainable for most real-world datasets of interest.\footnote{GFE estimators are instances of minimum sum-of-squares clustering, which is NP-hard unless both $G$ and $T$ are fixed \citep[see, e.g.,][]{InabaKatohImai1994, Aloise2009NPhardnessOE,MAHAJAN201213}.} Second, existing GFE estimators typically require the number of groups, or an upper bound for it, to be known to the analyst. While BIC-type criteria are known to achieve consistent model selection when $N$ and $T$ grow at the same rate, to the best of my knowledge, no such result is available in the asymptotic regimes considered here.

This paper provides a three-step estimation procedure free of these limitations. The first step leverages the low-rank factor structure of the linear GFE model to offer a computationally simple preliminary estimator of the slope coefficient. Though some emphasis is put on smooth and convex regularized nuclear-norm estimation, the interactive fixed effects literature provides several candidates further discussed. The second step examines residual correlations within triads of units to construct distances between units that, for each pair of units, are shown to be asymptotically zero if and only if both units belong to the same group, whenever group-time effects are well separated. An agglomerative clustering based on distance thresholding, in turn, is a natural choice. The third step computes an ordinary least squares regression (OLS) that controls for interactions of time and estimated group dummies. 

While the novel procedure combines disparate ideas from the existing panel data literature, the combination itself, along with its appealing large-sample statistical properties, appears to be new. Specifically, consistent model selection -- i.e., consistent estimation of the number of groups -- is achieved jointly with the estimation of other parameters by merging similar units into groups at a rate governed by time-dependence conditions. An advantage over traditional BIC-type selection methods is that it does not require estimating an arbitrary range of possibly misspecified models, determined by an upper bound $G_{\max}\geq G$, and applying some selection rule. The latter is often computationally cumbersome for GFE estimators. Since $G$ is unknown, however, regularization is necessary, and there is no free lunch. A computationally simple, data-driven procedure for determining the minimal tolerated distance to merge units is proven to be theoretically valid and demonstrates reasonable performance in finite samples. 

Unlike standard GFE estimators, the proposed method is both computationally ``trivial'' and exact, as the combinatorial optimization problem of GFE -- often solved using heuristics that provide local solutions -- is replaced by a sequential agglomerative clustering procedure with polynomial-time complexity. Fast implementations of agglomerative clustering procedures are already available in standard software used by economists (e.g., \textsf{R} or MATLAB). In addition, the preliminary estimation step can be performed via convex optimization, for which computationally efficient algorithms also exist.\footnote{A full implementation in MATLAB is  available  at \href{https://github.com/martinmugnier/TPWD-Estimators}{https://github.com/martinmugnier/TPWD-Estimators}.} 

The main theoretical contribution of this paper is to provide standard regularity conditions under which  the first two steps yield consistent estimation of well-separated groups. As in similar methods, this leads to an estimator of the slope parameter and the group-fixed effects asymptotically equivalent to the infeasible regression controlling for the true group memberships. Compared to existing approaches, the proposed method is computationally simple, endogenously determines the number of groups using a theoretically valid data-driven selection rule, and relies on minimal assumptions about the covariates -- requiring only sufficient conditional variation through a restricted eigenvalue condition. This generalization rules out cross-sectional correlation in idiosyncratic errors, as the new estimator uses triad-specific comparisons rather than cross-sectional averages to recover the group structure.

While the main text focuses on a simple linear model with a homogeneous slope to convey the core proof ideas, the versatility of the new approach is illustrated through several extensions with a growing number of groups, heterogeneous slope coefficients (in time or across units), and multiplicative network models (e.g., gravity trade equations), all discussed in Appendix Section~\ref{sec:extensions}. This paper does not address inference on group memberships. \cite{Dzemski2024}  develop pointwise valid inference methods in a model with group-specific slopes, but ruling out time-varying group-specific effects, provided a preliminary estimator of the slopes is available. If this approach could be adapted to accommodate a homogeneous slope and time-varying effects, the estimator proposed in the present paper could plausibly be used.

The finite-sample performance of the new estimator is compared to that of state-of-the-art alternatives across various Monte Carlo experiments calibrated to the empirical application: potentially misspecified GFE, spectral, and post-spectral estimators with different pre-specified group numbers as proposed in \cite{BM2015} and \cite{ChetverikovManresa2021}, nuclear-norm regularized (NNR) and nuclear-norm (NN) estimators from \cite{moon2019nuclear}, and the interactive fixed effects (IFE) estimator studied in \cite{Bai2009}, initialized with either the NNR estimator or random draws. Without covariates, the new method outperforms all estimators except the well-specified GFE, closely matching its performance for moderate values of \(T\). It achieves remarkably homogeneous clustering in terms of Precision, Recall, and Rand Index, which are defined later. With a scalar covariate, the new estimator significantly improves bias, root mean square error, and coverage compared to all alternatives, except the well-specified GFE, GFE with a well-specified BIC criterion (\(G_{\max} \geq G\)), and the well-specified IFE in terms of bias. For \(T \geq 20\), its performance approaches that of the infeasible pooled OLS regression, with confidence intervals based on a consistent asymptotic variance estimator aligning closely with the nominal 95\% level.

The usefulness of the method is demonstrated by revisiting \cite{BM2015}’s analysis of the statistical relationship between income and democracy in a large panel of countries from 1970 to 2000. This association may be confounded by critical junctures in history that led to similar unobserved development paths for some countries and different ones for others. While the authors find statistically significant approximations of the GFE estimates for the effect of lagged log-income per capita on a democracy index ranging from 0.061 to 0.089, depending on the pre-specified number of groups, and suggest fewer than 10 groups, the proposed method identifies 4 groups, an effect of 0.07, and a larger cumulative income effect of 0.258 (vs. from 0.104 to 0.151). The preliminary regularized nuclear-norm estimator delivers point estimates of 0.016 and 0.078, respectively. 

\paragraph{Related Literature.}

This paper contributes to the extensive literature on estimating panel data models with interactive fixed effects \citep[e.g.,][]{Pesaran2006, Bai2009, MoonWeidner2015, MoonWeidner2017, BonhommeLamadonManresa2022, armstrong2022robust, BeyhumGautier2023}. While convex nuclear-norm regularized estimators \citep[e.g.,][]{moon2019nuclear, chernozhukov2019inference} are computationally simpler than non-convex least squares estimators \citep[e.g.,][]{Bai2009}, they may converge at slower-than-parametric rates. The first contribution of this paper is to leverage off-the-shelves bounds on the rate of convergence for such estimators, as derived in this literature, to construct computationally simple estimators of the special case of grouped fixed effects models that converge at the parametric rate. \cite{ChetverikovManresa2021} independently employed similar ideas, though they impose some factor structure on the covariates and a known upper bound on the number of groups in order to apply spectral clustering techniques. Differently, the method proposed in this paper relies on a plausibly weaker restricted eigenvalue condition and achieves model selection simultaneously. Using the nuclear-norm regularized estimator in the first step, it requires two regularization hyperparameters, which serve as data-driven substitutes for a known upper bound $G_{\max}\geq G$ and the BIC (or AIC) model selection criteria often used to select the number of factors in interactive fixed effects models \citep[e.g.,][]{Bai2003}. These hyperparameters, however, depend on fundamental sampling properties of the data, such as cross-sectional and time dependence, rather than prior information about the maximal dimension of the model.\footnote{Alternatively, if a tuning-free nuclear-norm estimator is used in the first step, such as \cite{moon2019nuclear}'s nuclear norm estimator, the method requires only one hyperparameter for the clustering step. While the available assumptions ensuring a sufficiently fast rate of convergence for the nuclear-norm estimator are relatively high-level compared to those for the regularized nuclear-norm estimator, both methods exhibit nearly identical finite-sample properties in the Monte Carlo simulations reported in Section~\ref{sec:mc_sim}.}

This paper also contributes to the rapidly growing literature on estimating grouped fixed effects models. \cite{BM2015}'s GFE estimator, an extension of $k$-means clustering to handle covariates, solves an NP-hard optimization problem. Algorithms that provide fast solutions may fail to converge to the true estimate values defined as a global minimum, and the same drawback applies to extensions and other non-convex estimators \citep[e.g.,][]{SSP2016, Ando_Bai_2022, mugnier_JMP,Lumsdaine2023}. In contrast, the inferential theory developed in this paper is valid for a computationally simple estimator that replaces a known upper bound on the number of groups with the willingness to merge units based on estimated pairwise distances.  Since inference is conducted on a true population parameter, these results contrast with those of \cite{Pollard_1981, Pollard_1982}, who provide asymptotic theory for the solution to the population $k$-means sum-of-squares problem in the cross-sectional case, which corresponds to a pseudo-true value. Similarly, \cite{Lewis_et_al2022} introduced a fuzzy clustering procedure that performs well in simulations, but whose large sample properties, however, have not been derived in the large \(N,T\) framework and thus apply to a pseudo-true value. Its implementation requires the number of groups to be specified by the analyst. 

The closest approach is that of \cite{ChetverikovManresa2021}'s spectral and post-spectral estimators. While computationally straightforward, these estimators require prior knowledge about the number of groups or a consistent estimator (not provided by the authors) and the theoretical validity of the spectral (resp.~post-spectral) estimator crucially rests upon a factor (resp.~grouped) structure for the covariates. Such assumptions bring the model closer to random or correlated random effects in the spirit of \cite{Pesaran2006} and could be restrictive in practice. Differently, the method proposed in this paper does not impose a specific model for the covariates or a known upper bound on the number of groups. Recently, \cite{Mehrabani2023} adapts ``sum-of-norm'' convex clustering \citep[e.g.,][]{Hocking2011, Tan2015} to a linear panel data model with latent group structure but time-constant group effects. Extending this approach to accommodate time-varying effects seems challenging. Albeit close in spirit, the proposed procedure differs from the binary segmentation algorithm developed in \cite{KeLiZhang2016} and \cite{WANG2021272}. Another approach involves applying spectral clustering to some dissimilarity matrix  \citep[see, e.g.,][]{Ng_2002, vonluxburg2007tutorial, ChetverikovManresa2021, brownlee_Lugosi2022, LuGuVolgushev_2022}. One  limitation is to introduce additional complexity and tuning parameters, as  \(L\) eigenvectors of the dissimilarity matrix need to be computed and clustered (and \(L\) must be chosen), typically by approximating an NP-hard $k$-means solution, which this paper aims to avoid in order to ensure valid inference. 

Finally, this paper establishes a connection between the mature statistical and operation research literature on clustering problems and the recent grouped fixed-effects literature. Although agglomerative or hierarchical clustering methods are well established in the former, their adaptation to the latter is relatively new \citep[a few examples include][]{Vogt2016,Chen2019, MammenWilke2022}. The advantage lies both in computational efficiency and valid inference, breaking the high-dimensionality of $k$-means by considering agglomerative approaches and leveraging the discrete structure of the econometric model. Hence, this paper can be seen as the first application and analysis of an agglomerative clustering method in the econometric panel data model~\eqref{eq:model}.\footnote{Since the first arXiv version of this paper, \cite{FreemanWeidner2023} also propose a hierarchical clustering algorithm but do not explicitly verify that it meets the approximation conditions for their large sample theory to be valid. Differently from $k$-means \citep[e.g.,][]{BonhommeLamadonManresa2022,GrafLuschgy2002}, it remains  unclear when these conditions are met.}  The pairwise distance used in the clustering step has already been employed in the mathematical statistics and econometric literature to study topological properties of the graphon \citep[e.g.,][]{Lovasz2012,ZhangLevinaZhu2017, Zeleneev2020, Auerbach2022}. More generally, dyad, triad, or tetrad comparisons have proven useful in a variety of other econometric contexts \citep[see, e.g.,][]{HONORE1994241,Graham2017, Charbonneau2017, jochmans_2017}.   

\medskip 
The rest of the paper is organized as follows. Section~\ref{sec:a_simple_two_step_estimator} outlines the three-step estimation procedure. Section~\ref{sec:large_sample_properties} discusses the large-sample properties, including uniform consistency for the grouping structure and asymptotic normality at parametric rates. Section~\ref{sec:mc_sim} presents the results of a small-scale Monte Carlo simulation. Section~\ref{sec:emp_app} reports the results of the empirical application. Section~\ref{sec:conc}  concludes. All proofs are provided in the Appendix. Additional material can be found in the Supplementary Material \citep{mugnier2022SM}, with section numbers denoted as S.1, etc.

\paragraph{Notation.} Let $\ind{\cdot}$ denote the indicator function. For any set $S\subset \R^p$ (for any $p\geq 1$), let $S^*\equiv S\backslash\{0\}$ and $\vert S\vert $ denote the cardinality of $S$. For any set $I\subset\N$, all $k\in\N^*$ such that $k\leq\vert I\vert$, let $\mathscr P_k(I)$ denote the set of subsets of $I$ with cardinality $k$. The operators $\overset{p}{\to}$, $\overset{d}{\to}$, and $\plim$ denote convergence in probability, convergence in distribution, and probability limit, respectively. All vectors are column vectors. $\Vert\cdot\Vert$ denotes the Euclidean norm. For an $m\times n$ real matrix $A$ of rank $r$, let $A'$ denote its transpose, $\Vert A\Vert_{\textrm{F}}\equiv[{\rm Tr}(A'A)]^{1/2}$ its Frobenius norm, $\Vert A\Vert_\infty\equiv\sigma_1(A)$  its spectral norm,  $\Vert A\Vert_{\max}\equiv\max_{i=1,\ldots,m;j=1,\ldots,n}\abs{a_{ij}}$ its max norm, and $\Vert A \Vert_1\equiv\sum_{i=1}^r\sigma_i(A)$ its nuclear norm, where $\sigma_1(A)\geq\cdots\geq\sigma_r(A)>0$ denote the positive singular values of $A$. 

\section{Three-step estimation}\label{sec:a_simple_two_step_estimator}

In this section, I introduce the three-step triad pairwise-differencing (TPWD) estimator $\widehat\theta$ for the parameter 
$\theta\equiv(G, g_1,\ldots, g_N,\alpha_{11},\ldots,\alpha_{GT}, \beta')'\in\Theta$
of Model~\eqref{eq:model}, where 
$$\Theta\equiv\bigcup_{g\in\N^*}\Theta_{g} \quad \text{with} \quad \Theta_{g}\equiv\set{g}\times\set{1,\ldots,g}^N\times\mathcal A^{gT}\times\mathcal B,$$
 for some subsets $\mathcal B\subset\R^K$ and $\mathcal A\subset\R$. The dependence of $\Theta$ on $N,T$ is omitted. Sections~\ref{sec:prelim_con_est}--\ref{sec:clstr_algo} describe the three key components of  $\widehat \theta$:  a preliminary consistent estimator of $\beta$, a measure of pairwise distances between units, and an agglomerative clustering algorithm. Section~\ref{sec:tpwd_est} formally defines $\widehat \theta$.

\subsection{Preliminary consistent estimation of the slope coefficient}\label{sec:prelim_con_est}

The first component of the three-step TPWD estimator is a preliminary consistent estimator $\widehat \beta^1$ of $\beta$:
\begin{equation}\label{eq:preliminary_est}
	\norm{\widehat\beta^1-\beta}=o_p(1) \quad \text{ as } \min(N,T)\to\infty.
\end{equation}
Intuitively, since covariates can be arbitrarily correlated with the unobserved grouped fixed effects, the clustering problem is simpler in the approximate ``pure'' grouped fixed effects model: 
$$y_{it}-x_{it}' \widehat\beta^1= \alpha_{g_it}+v_{it}+o_p(1), \quad i=1,\ldots,N, \; t=1,\ldots,T,$$
as  $\min(N,T)\to\infty$.
 
The interactive fixed effects literature offers several computationally simple estimators that verify~\eqref{eq:preliminary_est} under various identifying assumptions \citep[e.g.,][]{moon2019nuclear, chernozhukov2019inference, beyhum2019squareroot, BeyhumGautier2023}. The large-sample properties of \( \widehat{\theta} \), established in Section~\ref{sec:large_sample_properties}, do not depend on the convergence rate of the preliminary estimator, which is typically slower than \( \sqrt{NT} \), as long as a lower bound on this rate is known and used to calibrate the clustering tuning parameter. Such bounds are available for all the references mentioned above, provided certain regularity conditions hold.

As an illustrative example, consider below the nuclear-norm regularized estimator. Under a known upper bound on the number of groups and a factor structure on the covariates, another example is the correlated random effects estimators proposed by \cite{ChetverikovManresa2021}. Simulation results reported in Section~\ref{sec:mc_sim_full}, however, suggest that while \cite{ChetverikovManresa2021}'s  spectral norm estimator may outperform the nuclear-norm regularized estimator under a well-specified covariate model, the iterated version of the triad pairwise-differencing estimator based on the nuclear-norm regularized estimator may outperform \cite{ChetverikovManresa2021}'s spectral and post-spectral estimators for small values of $T$.\footnote{For large values of \(T\), which is not the typical setting for which GFE models were developed, both estimators exhibit equivalent performance (although spectral and post-spectral estimators require the number of groups as input); in this case, however, many other estimators compete.}

\paragraph{Nuclear-norm regularization.} 

Let $Y\equiv(y_{it})_{i=1,\ldots,N;t=1,\ldots,T}\in\R^{N\times T}$ and $X_k\equiv(x_{it,k})_{i=1,\ldots,N;t=1,\ldots,T} \in \R^{N\times T}$ for all $k\in\set{1,\ldots,K}$. For all $v=(v_1,\ldots,v_K)'\in\R^K$, let  $v\cdot X\equiv\sum_{k=1}^KX_kv_k$. For any value of the regularization parameter $\psi_{NT}\in(0,\infty)$, let $Q_{\psi_{NT}}$ denote the nuclear-norm regularized concentrated objective function: 
\begin{equation*}\label{eq:crit_nuc_norm_reg}
	Q_{\psi_{NT}}(\beta)\equiv\min_{\Gamma\in\R^{N\times T}}\set{\frac{1}{2NT}\Vert Y-\beta\cdot X - \Gamma \Vert^2_{\textrm{F}}+\frac{\psi_{NT}}{\sqrt{NT}}\Vert \Gamma\Vert_1},
\end{equation*}
for all $\beta\in\R^K$. A nuclear-norm regularized (NNR) estimator is defined as a solution to the following minimization problem:
\begin{equation}\label{eq:beta_first_step}
	\widehat\beta^1(\psi_{NT})\in\argmin_{\beta\in \R^K} Q_{\psi_{NT}}(\beta).
\end{equation}
Under regularity conditions,  $\widehat\beta^1(\psi_{NT})$ is unique with probability approaching one \cite[see, e.g.,][]{moon2019nuclear}. 

Instead of setting $\psi_{NT}=0$ and optimizing over all $G^{\rm guess}$ ($G^{\rm guess}<<N$) unobserved grouped trend patterns
$$\Gamma\in\set{\Lambda F': \Lambda\in \set{0,1}^{N\times G^{\rm guess}},F\in\R^{T\times G^{\rm guess}},\sum_{j=1}^{G^{\rm guess}}\Lambda_{ij}=1\quad  i=1,\ldots,N},$$
which leads to the  NP-hard problem solved by \cite{BM2015}'s GFE estimator, the NNR estimator penalizes the nuclear norm (the sum of singular values) of an unrestricted matrix $\Gamma\in\R^{N\times T}$ of individual trends. This convexifies the rank-constrained problem solved by \cite{Bai2009}'s interactive fixed effects (IFE) estimator -- which sets $\psi_{NT}=0$ and optimizes over $\Gamma\in\{\Lambda F': \Lambda\in\R^{N\times G^{\rm guess}}, F\in\R^{T\times G^{\rm guess}} \}$ -- thus solving the ``local minima'' problem.\footnote{The nuclear norm $\Vert \Gamma\Vert_1$ is the convex envelope of rank$(\Gamma)$ over the set of matrices with spectral norm at most one.} 

The tuning parameter $\psi_{NT}$ performs model regularization without the need to set both $G^{\rm guess}\geq G$ and a model selection rule. As proved in Section~\ref{sec:nuc_norm_cond}, under regularity assumptions and a rate condition on $\psi_{NT}$, the interactive fixed effects structure of grouped fixed effects is sufficient for $\widehat\beta^1(\psi_{NT})$ to achieve a convergence rate arbitrarily close to $\sqrt{\min(N,T)}$. In the Monte Carlo experiments, the theoretically justified choice $\psi_{NT}\equiv\log(\log(T))/\sqrt{16\min(N,T)}$ is employed.

\paragraph{Computation.} Minimization problem~\eqref{eq:beta_first_step}  is convex and can be efficiently solved using modern optimization techniques. Algorithm 1 below follows a straightforward iterative strategy that alternates between the proximal gradient algorithm proposed by \cite{Mazumder2010} and a projection step. For any real matrix $A$ of rank $r$ and any scalar $\tau\in[0,\infty)$, let $\mathbf S_\tau(A)\equiv UD_\tau V'$, where $D_\tau\equiv{\rm diag}[(\sigma_1(A)-\tau)_+,\ldots,(\sigma_r(A)-\tau)_+]$, $UDV'$ is the compact singular value decomposition of $A$, $D\equiv{\rm diag}(\sigma_1(A),\ldots, \sigma_r(A))$, and $t_+\equiv \max(t,0)$. For all $\beta\in\R^K$ and $\Gamma\in\R^{N\times T}$, let
\[
 L_{\psi_{NT}}(\beta,\Gamma)\equiv\frac{1}{2NT}\Vert{Y-\beta\cdot X - \Gamma}\Vert^2_{\textrm{F}}+\frac{\psi_{NT}}{\sqrt{NT}}\Vert \Gamma\Vert_1.
\]
\begin{table}[H]
	\label{tab:algo1}
	\centering	
	\begin{adjustbox}{max width={0.99\linewidth},center}
		\begin{threeparttable}[h]
			\begin{tabular}{p{15cm}}
				\toprule
				{\bf Algorithm 1} {\scshape nuclear-norm regularized Estimator} \\
				\midrule
				{\bf Input:} $Y$, $X$, $\psi_{NT}>0$, $\eps>0$. \\
				{\bf Output:} $\widehat \beta^1(\psi_{NT},\eps)$.
				\begin{enumerate}
					\item {\bf Initialize} $\beta^{\rm old}=0\in\R^K$ and $\Gamma^{\rm old}=0\in\R^{N\times T}$.
					\item {\bf Repeat}
					\begin{itemize}
						\item $\Gamma^{\rm new} \leftarrow \mathbf S_{\sqrt{NT}\psi_{NT}}(Y-\beta^{\rm old}\cdot X)$.
						\item $\beta^{\rm new} \leftarrow \argmin_{b\in\R^K}\Vert{Y-b\cdot X-\Gamma^{\rm new}}\Vert^2_{\textrm{F}}$. 
						\item {\bf If} $L_{\psi_{NT}}(\beta^{\rm old},\Gamma^{\rm old})-L_{\psi_{NT}}(\beta^{\rm new},\Gamma^{\rm new})<\eps$ {\bf exit}.
						\item $\beta^{\rm old} \leftarrow \beta^{\rm new}$.
						\item $\Gamma^{\rm old} \leftarrow \Gamma^{\rm new}$.
					\end{itemize}
					\item {\bf Return} $\widehat \beta^1(\psi_{NT},\eps) \leftarrow \beta^{\rm new}$.
				\end{enumerate}
				\\
				\bottomrule
			\end{tabular}
		\end{threeparttable}
	\end{adjustbox}
\end{table}
Algorithm 1 alternates between soft-thresholding the singular values of the residual matrix $Y-\beta^{\rm old}\cdot X$ to obtain $\Gamma^{\rm new}$, and performing a pooled OLS regression of the residual matrix $Y-\Gamma^{\rm new}$ on $X$ to obtain $\beta^{\rm new}$, iterating until convergence. Commonly employed statistical software includes fast routines for both tasks (e.g., \textsf{R} or MATLAB function \texttt{svd}). By convexity and due to the uniqueness of the minimum of the objective function $L_{\psi_{NT}}$, if Algorithm 1 converges, it will converge to $\widehat\beta^1(\psi_{NT})$. In addition, the optimization error can be made arbitrarily small by selecting a sufficiently small value for $\eps$, which justifies abstracting from optimization errors hereafter.  

Since Algorithm 1 may be slow in practice, an alternative approach, based on Lemma 1 from \cite{moon2019nuclear}, uses a built-in optimization tool to minimize the closed-form expression of the convex concentrated objective function
	\[
	Q_{\psi_{NT}}(\beta)=\sum_{r=1}^{\min(N,T)}q_{\psi_{NT}}\left(\sigma_r\left(\frac{Y-\beta\cdot X}{\sqrt{NT}}\right)\right),
	\]
	where, for any $s\in[0,\infty)$,
	\[
		q_{\psi_{NT}}(s)=\left\{
		\begin{array}{ll}
			\frac{1}{2}s^2, &\text{ for } s<\psi_{NT}, \\
			\psi_{NT} s-\frac{\psi_{NT}^2}{2}, &\text{ for } s\geq \psi_{NT}.
		\end{array}\right.
	\]
For example, an implementation using MATLAB's \texttt{fminsearch} is significantly faster and provides estimates that are nearly identical to those obtained from Algorithm 1 with $\eps=10^{-9}$.

\subsection{Pairwise distance between cross-sectional units}\label{sec:pairwise_dist}

The second component of the TPWD estimator is a measure of distance between any two units $i$ and $j$, informative of whether $i$ and $j$ belong to the same group. This can be constructed by exploiting the linear structure of the model and the preliminary consistent estimator $\widehat \beta^1$.  Let $\widehat v_{it}\equiv y_{it}-x_{it}'\widehat\beta^1$ denote the first-step residual. The empirical distance between $i$ and $j$, denoted $\widehat d_{\infty,1}^2(i,j)$, is given by
\begin{equation}\label{eq:distance_ing}
	\widehat d_{\infty,1}^2(i,j)\equiv\max_{k \in \set{1,\ldots,N}\backslash\set{i,j}}\abs{\frac{1}{T}\sum_{t=1}^T(\widehat v_{it}-\widehat v_{jt})\widehat v_{kt}}.
\end{equation}
Let $\widehat D\equiv (\widehat d_{\infty,1}^2(i,j))_{(i,j)\in\{1,\ldots,N\}^2}$ denote the symmetric dissimilarity matrix that collects all pairwise distances. 

The distance $\widehat d_{\infty,1}^2$ is borrowed from the statistical literature on graphon estimation \citep[e.g.,][]{Lovasz2012, zhang2017estimating, Zeleneev2020, Auerbach2022}. While alternative distances could be of interest, e.g.,
\begin{equation*}
	\widehat d_{\infty,2}^2(i,j)\equiv\frac{1}{N}\sum_{k=1}^N\abs{\frac{1}{T}\sum_{t=1}^T(\widehat v_{it}-\widehat v_{jt})\widehat v_{kt}} \quad  \text{or} \quad \widehat d_{2}^2(i,j)\equiv \frac{1}{T}\sum_{t=1}^T(\widehat v_{it}-\widehat v_{jt})^2,
\end{equation*}
the distance $\widehat d_{\infty,1}^2$ performs well in simulations and is convenient to prove large sample properties. For example, following the present proof techniques, using $\widehat d_{\infty,2}^2$ would rule out asymptotically negligible groups in order to ensure consistency of the subsequent clustering step, while $\widehat d_2^2$ could be contaminated by  heteroskedastic error variances which would prevent consistency. Similarly, the distance could be based on a self-normalized sum to account for heteroskedastic variances across pairs, which might provide some finite-sample improvements. This is not required for deriving asymptotic results under heteroskedastic errors and will not be pursued further here.

Why is the empirical distance $\widehat d^2_{\infty,1}(i,j)$ informative about whether $g_i=g_j$? The intuition is as follows.\footnote{See also p.~14 in \cite{Zeleneev2020} in a network setting.} Since $\widehat\beta^1\overset{p}{\to}\beta$ as $\min(N,T)\to\infty$, it holds with probability approaching one that $\widehat v_{it}\approx \alpha_{g_it}+v_{it}$. Under weak time dependence, tails, and cross-sectional independence restrictions on the error terms,  it then holds ``uniformly'' over $i$, $j$, and $k\in \set{1,\ldots,N}\backslash\set{i,j}$ that
\begin{align*}
		\frac{1}{T}\sum_{t=1}^T(\widehat v_{it}-\widehat v_{jt})\widehat v_{kt} \approx 	\frac{1}{T}\sum_{t=1}^T(\alpha_{g_it}-\alpha_{g_jt})\alpha_{g_kt}.
\end{align*}
If $g_i=g_j$, then $\alpha_{g_it}-\alpha_{g_jt}=0$ and 
\begin{align*}
	\widehat d_{\infty,1}^2(i,j)&=\max_{k \in \set{1,\ldots,N}\backslash\set{i,j}}\abs{\frac{1}{T}\sum_{t=1}^T(\widehat v_{it}-\widehat v_{jt})\widehat v_{kt}} 
	\approx \max_{k \in \set{1,\ldots,N}\backslash\set{i,j}}\abs{\frac{1}{T}\sum_{t=1}^T\underbrace{(\alpha_{g_it}-\alpha_{g_jt})}_{=0}\alpha_{g_kt}}=0.
\end{align*}
Reciprocally, if
\begin{equation}\label{eq:reciprocal}
	\widehat d_{\infty,1}^2(i,j)=\max_{k \in \set{1,\ldots,N}\backslash\set{i,j}}\abs{\frac{1}{T}\sum_{t=1}^T(\widehat v_{it}-\widehat v_{jt})\widehat v_{kt}}\approx 0,
\end{equation}
then necessarily $g_i=g_j$. To see this, suppose that $g_i\neq g_j$. Then, under the weak condition that each group has at least two units asymptotically, there exist $k^*,l^*\in\set{1,\ldots,N}\backslash\set{i,j}$ such that $g_{k^*}=g_i$ and $g_{l^*}=g_j$. Equation~\eqref{eq:reciprocal} implies
\begin{align}\label{eq:reciprocal2}
		\frac{1}{T}\sum_{t=1}^T(\widehat v_{it}-\widehat v_{jt})\widehat v_{k^*t}\approx\frac{1}{T}\sum_{t=1}^T(\alpha_{g_it}-\alpha_{g_jt})\alpha_{g_{k^*}t} & \approx 0, \\
	\frac{1}{T}\sum_{t=1}^T(\widehat v_{it}-\widehat v_{jt})\widehat v_{l^*t}\approx\frac{1}{T}\sum_{t=1}^T(\alpha_{g_it}-\alpha_{g_jt})\alpha_{g_{l^*}t} & \approx 0.\label{eq:reciprocal3}
\end{align}
Differencing \eqref{eq:reciprocal2}--\eqref{eq:reciprocal3} and using that $\alpha_{g_{k^*t}}=\alpha_{g_it}$ and $\alpha_{g_{l^*t}}=\alpha_{g_jt}$ yields
\begin{equation*}
	\frac{1}{T}\sum_{t=1}^T(\alpha_{g_it}-\alpha_{g_jt})^2\approx 0,
\end{equation*}
a contradiction if groups are ``well separated'', i.e., if for all $(g,\widetilde g)\in\set{1,\ldots,G}^2$ such that $g\neq \widetilde g$, there exists a constant $c_{g,\widetilde g}>0$ such that
$$\frac{1}{T}\sum_{t=1}^T(\alpha_{gt}-\alpha_{\widetilde gt})^2 \geq c_{g,\widetilde g}>0.$$
Section~\ref{sec:large_sample_properties} formalizes this equivalence result by establishing uniform asymptotic control over the remainders in the stochastic approximations.

\subsection{Agglomerative clustering based on thresholding distances}\label{sec:clstr_algo}

Given a preliminary consistent estimator $\widehat\beta^1$ and a dissimilarity matrix $\widehat D$ based on $\widehat\beta^1$, the last component of the TPWD estimator is an agglomerative clustering algorithm that builds clusters from the dissimilarity matrix. 

The main methodological contribution of this paper is to frame the clustering task as the equivalent high-dimensional multiple-testing problem associated with the null hypothesis $H_{0,i,j}:g_i=g_j$ for all $N(N-1)/2$ distinct ordered pairs of units.\footnote{Similar ideas have recently been developed by \cite{Dzemski2024} for different purposes.} If the researcher knew the adjacency matrix of the graph spanned by group memberships,  $W\equiv (\ind{g_i=g_j})_{(i,j)\in\{1,\ldots,N\}^2}$, which is block diagonal up to some permutation of its rows and columns, then they would know the groups and vice-versa. Although \( W \) is not directly observable in practice, it can be consistently estimated based on the results of pairwise tests.

Specifically, Section~\ref{sec:large_sample_properties} and Lemma~\ref{lem:sup_norm_mat_form} provide sufficient conditions under which any sequence of matrices $\widehat  W(c_{NT})\equiv \ind{\widehat D\leq c_{NT}}$ with $c_{NT}\to0^+$ converges in max norm to $W$: as $N$ and $T$ tend jointly to infinity,
$$\norm{\widehat W(c_{NT})-W}_{\max}=\max_{(i,j)\in\{1,\ldots,N\}^2}\abs{\widehat W_{ij}(c_{NT})-W_{ij}}=o_p(1).$$
The result follows from establishing that both the Type I and Type II errors of the test $\phi_{NT}(i,j)=1-\widehat W_{ij}(c_{NT})$ of $H_{0,i,j}$ tend to zero uniformly across $(i,j)$ (or, equivalently, that both its power and confidence level tend to 1). It motivates using thresholding tests based on the entries of $\widehat D$ and basic agglomerative merging rules that involve a number of operations independent of $G$ and bounded by a polynomial in $N$. In fact, it is even more natural to apply agglomerative clustering algorithms based on merging units whose weigthed pairwise distances fall below some threshold. Such an agglomerative step addresses the practical problem of aggregation arising from the fact that, in finite samples, $\widehat W(c_{NT})$ is generally not block diagonal even after an arbitrary number of permutations of its rows and columns.

This idea lies at the heart of Hierarchical Agglomerative Clustering (HAC) algorithms \citep[see, e.g., Chapter 14.3.12 in ][]{hastie2009elements}, which rely on various ``linkage'' functions to measure distances between disjoint clusters $A$ and $B$, where $A,B\subset\{1,\ldots,N\}$, based on the dissimilarity matrix $\widehat D$, and various rules to stop the sequential agglomerative merging process, i.e., to cut the induced dendrogram from $N$ singleton clusters, each containing one unit, to a unique cluster of $N$ units, or vice versa for divisive algorithms. Table~\ref{tab:linkage_functions} summarizes popular linkage functions.
\begin{table}[H]
\centering
\caption{Some popular linkage functions}
\label{tab:linkage_functions}
		\begin{threeparttable}[h]
\begin{tabular}{lc}
\toprule
Linkage function & Formula \\
\midrule
	 \texttt{average\_linkage}$(A,B;\widehat D)$ & $\frac{1}{\abs{A}\abs{B}}\sum_{i\in A}\sum_{j\in B}\widehat d_{\infty,1}^2(i,j)$ \\
	\texttt{complete\_linkage}$(A,B;\widehat D)$ & $\max_{i\in A,j\in B}\widehat d_{\infty,1}^2(i,j)$ \\
 	 \texttt{single\_linkage}$(A,B;\widehat D)$ & $\min_{i\in A,j\in B}\widehat d_{\infty,1}^2(i,j)$\\
	 \bottomrule
	\end{tabular}
				\begin{tablenotes}
				\footnotesize \item {\em Notes:} $A$ and $B$ denote disjoint subsets of $\{1,\ldots,N\}$. $\widehat D\equiv \widehat d_{\infty,1}^2(i,j)\in[0,\infty)^{N\times N}$ denotes the dissimilarity matrix.
			\end{tablenotes}
	\end{threeparttable}
\end{table}
The main theoretical contribution of this paper is to be the first to highlight that such HAC-type algorithms can be successfully applied to GFE models and to provide a distance matrix and cut-off rule which, together with sufficient regularity conditions, yield clustering consistency as $\min(N,T)\to\infty$. 

Algorithm 2 below starts with $N$ singleton groups and iteratively merges the closest groups based on their linkage value, continuing until only one group remains or the smallest  linkage value exceeds a thresholding parameter $c_{NT}\in[0,\infty)$. This process guarantees a final partition into $\widehat G\in\{1,\ldots, N\}$ non-empty groups. In case of ties, the pair $(j^\star,l^\star)$ with the smallest $j^\star$ and smallest $l^\star$ is selected for merging.\footnote{Asymptotically, or if observed variables are continuously distributed, ties do not occur.}
\begin{table}[H]
	\label{tab:algo2}
	\centering	
	\begin{adjustbox}{max width={0.99\linewidth},center}
		\begin{threeparttable}[h]
			\begin{tabular}{p{16cm}}
				\toprule
				{\bf Algorithm 2} {\scshape Hierarchical Clustering Algorithm (HAC)} \\
				\midrule
				{\bf Input:} $\widehat D$, $c_{NT}\geq 0$, \texttt{linkage} function.\\
				{\bf Output:} $\widehat G,\widehat g_1,\ldots,\widehat g_N$.
				\begin{enumerate}
					\item {\bf Initialize}
					$$(G^{\rm old}, g_1^{\rm old}, g_2^{\rm old}, \ldots, g_N^{\rm old}, \mathcal C_1^{\rm old},\mathcal C_2^{\rm old},\ldots,\mathcal C_N^{\rm old}) \leftarrow (N,1,2,\ldots,N,\set{1},\set{2},\ldots,\set{N}).$$
					\item {\bf Repeat:}
					\begin{itemize}
						\item $(j^\star,l^\star) \leftarrow \argmin_{(j,l)\in\set{1,\ldots, G^{\rm old}}^2: j< l}$\texttt{linkage}$\left(\mathcal C_j^{\rm old},\mathcal C_l^{\rm old};\widehat D\right)$. 
						\item {\bf If} \texttt{linkage}$\left(\mathcal C_{j^\star}^{\rm old},\mathcal C_{l^\star}^{\rm old};\widehat D\right)>c_{NT}$ {\bf exit}.
						\item  $G^{\rm new}\leftarrow G^{\rm old}-1$.
						\item $\mathcal C_j^{\rm new}\leftarrow \mathcal C_j^{\rm old}$ for $j$ such that $1\leq j<j^\star$ or $j^\star<j<l^\star$.
						\item $\mathcal C_{j^\star}^{\rm new}\leftarrow \mathcal C_{j^\star}^{\rm old}\cup \mathcal C_{l^\star}^{\rm old}$.
						\item $\mathcal C_j^{\rm new}\leftarrow \mathcal C_{j+1}^{\rm old}$ for $j$ such that  $l^\star\leq j\leq G^{\rm new}$.
						\item  $(\mathcal C_1^{\rm old}, \ldots, \mathcal C_{G^{\rm new}}^{\rm old}, G^{\rm old}) \leftarrow (\mathcal C_1^{\rm new}, \ldots, \mathcal C_{G^{\rm new}}^{\rm new}, G^{\rm new})$.
					\end{itemize}
					\item  {\bf Return}
					$$(\widehat G,\widehat g_1,\ldots,\widehat g_N) \leftarrow \left(G^{\rm old}, \sum_{j=1}^{G^{\rm old}}j\ind{1\in\mathcal C_j^{\rm old}}, \sum_{j=1}^{G^{\rm old}}j\ind{2\in\mathcal C_j^{\rm old}}, \ldots, \sum_{j=1}^{G^{\rm old}}j\ind{N\in\mathcal C_j^{\rm old}}\right).$$	
					\end{enumerate}
				\\
				\bottomrule
			\end{tabular}
		\end{threeparttable}
	\end{adjustbox}
\end{table}
Algorithm 2 can be efficiently implemented using built-in functions available in popular software (e.g., \texttt{hclust} and \texttt{cutree} in \textsf{R}, or \texttt{cluster} and \texttt{linkage} in MATLAB).

\subsection{A three-step triad pairwise-differencing estimator} \label{sec:tpwd_est}
The TPWD estimator  $\widehat\theta\in\Theta$ of $\theta\in\Theta$ is obtained as follows. Fix $(\psi_{NT},c_{NT})\in[0,\infty)^2$, choose a \texttt{linkage} function (e.g, from Table~\ref{tab:linkage_functions}), and perform the following steps.
\begin{enumerate}
	\item {\scshape Preliminary Slope Estimation}: Compute $\widehat{\beta}^1(\psi_{NT})$ using Algorithm 1 from Section~\ref{sec:prelim_con_est}. 
	\item {\scshape Agglomerative Clustering}:
	\begin{itemize}
		\item[(a)] Compute $\widehat D$ as described by Equation~\eqref{eq:distance_ing} in Section~\ref{sec:pairwise_dist}.
		\item[(b)] Compute $\set{\widehat g_1,\ldots,\widehat g_N}$ and $\widehat G=\abs{\set{\widehat g_1,\ldots,\widehat g_N}}$ using Algorithm 2 from Section~\ref{sec:clstr_algo}.
	\end{itemize}
	\item {\scshape Projection Step:} Compute
	$$\left(\widehat\beta',\widehat \alpha_{11},\ldots,\widehat \alpha_{\widehat GT}\right) \in \argmax_{\left(\beta', \alpha_{11},\ldots,\alpha_{\widehat GT}\right)\in\mathcal B\times\mathcal A^{\widehat GT}}\sum_{i=1}^N\sum_{t=1}^T\left(y_{it}-x_{it}'\beta-\alpha_{\widehat g_it}\right)^2.$$
\end{enumerate} 
If $\mathcal B=\R^K$ and $\mathcal A=\R$, the projection step is a pooled OLS regression of $y_{it}$ on $x_{it}$ and the interactions of estimated groups and time dummies.

\medskip
\noindent{\scshape Remark 1 (Regularization Path):}  Consider a finite sample of fixed dimensions $N$ and $T$. If all random variables except group memberships are continuous, then as $c_{NT}\to0$, $\widehat G\to N$ and each group contains a single unit in the limit. Conversely, as $c_{NT}\to +\infty$, $\widehat G\to1$ and a single group contains all units in the limit. Given the low computational cost of the method, a regularization path can be obtained by varying $c_{NT}$ between these extremes. Notably, the clustering step can be vectorized, significantly reducing the computational burden compared to iterative loop-based approaches.

\medskip
\noindent{\scshape Remark 2 (Choice of Tuning Parameters):} Section \ref{sec:consistency_clustering} offers theoretical guidance on selecting $\psi_{NT}$ and $c_{NT}$. Section~\ref{sec:data_driven_selec_rule} proposes a simple data-driven selection rule that performs well across various Monte Carlo simulations. To further enhance the finite-sample performance, the first step can be re-run with $\widehat v_{it}=y_{it}-x_{it}'\widehat\beta$ (instead of using $\widehat v_{it}= y_{it}-x_{it}'\widehat\beta^1$) to obtain new group assignments $\widehat g_1,\ldots,\widehat g_N$. The second and third steps are then repeated, and the process iterated until convergence. The asymptotic results presented in the following section hold for all subsequent iterations, and Monte Carlo simulations further suggest that the iterative procedure achieves notable improvements in precision. Simulations also suggest that removing the tuning parameter $\psi_{NT}$ by using the NN estimator as the preliminary estimator, instead of the NNR estimator,
\[
\widehat\beta^1 \in\argmin_{\beta\in\R^K}\Vert Y-\beta\cdot X\Vert_1,
\]
leads to nearly identical results.

\medskip
\noindent{\scshape Remark 3 (Computation):} The full estimation procedure requires $O(N^3T)$ operations, which is a substantial improvement relative to the NP-hard $k$-means problem underlying \cite{BM2015}'s GFE estimator. Whether the computational cost of a consistent clustering algorithm could be further reduced, for example to $O(N^2T)$, remains to the best of my knowledge an open question. While a current limitation of the method is rather the memory required to store triad differences of residuals (looping over $N(N-1)/2$ rather than considering a full vectorization  when $N$ is large),  it seems hardly possible to cluster units without some measure of distance between them.  When unobserved heterogeneity is assumed to be time-constant, the computation cost can be reduced from $O(N^3T)$ to $O(N^2T)$, and the preliminary estimator can be replaced with any standard fixed effects differencing estimator \citep[e.g.][]{arellano_bond_1991,wooldridge2010econometric}; see Section S.2 of the Supplementary Material  for further discussion.

\section{Large-sample properties}\label{sec:large_sample_properties} 

In this section, I establish the consistency and asymptotic normality at the parametric rate of the TPWD estimator introduced in Section~\ref{sec:a_simple_two_step_estimator}, under conditions similar to those in the existing literature. Consider the data generating process:
\begin{equation}\label{eq:model_true_dgp}
	y_{it} = x_{it}'\beta^0+\alpha_{g_i^0t}^0+v_{it},  \quad i=1,\ldots,N, \; t=1,\ldots,T,
\end{equation}
where $g_i^0\in\set{1,\ldots,G^0}$ denotes group membership, and the $0$ superscripts refer to true parameter values. The asymptotic framework is such that $N$ and $T$ diverge jointly to infinity, denoted by $\min(N,T)\to\infty$. The number of groups $G^0$ is fixed relative to $(N,T)$ but unknown. The discussion on the case of an increasing sequence $G^0=G^0_{NT}$ is relegated to Section S.1 of the Supplementary Material.

\subsection{Consistency of estimated group memberships}\label{sec:consistency_clustering}
Consider the following assumptions.
\begin{hyp}\label{as:prelim_rate}
		$\norm{\widehat \beta^1-\beta^0} =O_p(r_{NT})$ for some deterministic sequence $r_{NT}\to0$ as $\min(N,T)\to\infty$.
\end{hyp}
\begin{hyp}\label{as:linkage}
	For all $(N,T)$, all subsets $A,B\subset\set{1,\ldots, N}$, almost surely
	$$\min_{i\in A,j\in B}\widehat d_{\infty,1}^2(i,j)\leq \texttt{linkage}(A,B,\widehat D)\leq\max_{i\in A,j\in B}\widehat d_{\infty,1}^2(i,j).$$
\end{hyp}
\begin{hyp}\label{as:tuning_param} 
There exist constants $(C,\nu,\kappa)'\in(0,\infty)^2\times(0,1/2)$ such that
\begin{enumerate}[label=(\alph*)]
	\item\label{as:tuning_param_a} $NT^{-\nu}\to0$ as $\min(N,T)\to\infty$.
	\item\label{as:tuning_param_b} $c_{NT}\overset{p}{\to}0$, $c_{NT}r_{NT}^{-1}\overset{p}{\to}\infty$, and $\Pr(c_{NT}T^\kappa \geq C)\to 1$  as $\min(N,T)\to\infty$.
	\end{enumerate} 
\end{hyp}
\begin{hyp}\label{as:sup_norm_cons} ~
	There exist constants $a,b,c,d_1,d_2,M>0$  and a sequence $\tau(t)\leq e^{-at^{d_1}}$ such that:
	\begin{enumerate}[label=(\alph*)]
		\item \label{as:sup_norm_cons_a} $\mathcal A$ is a compact subset of $\R$.
		\item \label{as:sup_norm_cons_b} For all $(i,t) \in\set{1,\ldots,N}\times\set{1,\ldots,T}$: $\Pr(\abs{v_{it}}>m)\leq e^{1-(m/b)^{d_2}}$ for all $m>0$.
		\item \label{as:sup_norm_cons_c} For all $(g,\widetilde g)\in\set{1,\ldots,G^0}^2$ such that $g\neq \widetilde g$: 
		$$\plim_{\min(N,T)\to\infty}T^{-1}\sum_{t=1}^T(\alpha_{gt}^0-\alpha_{\widetilde gt}^0)^2=c_{g,\widetilde g}\geq c.$$
		\item \label{as:sup_norm_cons_d} For all $(i,j,k,g,\widetilde g)\in\mathscr P_3(\set{1,\ldots,N})\times\set{1,\ldots,G^0}^2$ such that $g\neq\widetilde g$, $\set{v_{it}}_t$, $\set{(v_{it}-v_{jt})v_{kt}}_t$, $\set{\alpha_{gt}^0-\alpha_{\widetilde gt}^0}_t$,  $\set{(v_{it}-v_{jt})\alpha_{gt}^0}_t$, and $\set{(\alpha_{gt}^0-\alpha_{\widetilde gt}^0)v_{it}}_t$ are strongly mixing processes with mixing coefficients $\tau(t)$. Moreover, $\E[v_{it}v_{jt}]=0$.
		\item \label{as:sup_norm_cons_e} $\lim_{\min(N,T)\to\infty}\Pr(\min_{g\in\set{1,\ldots,G^0}}\;\sum_{i=1}^N\mathbf{1}\{g_i^0=g\}\geq 2)=1$.
		\item \label{as:sup_norm_cons_f} As $N$ and $T$ tend to infinity:
		$$\sup_{i \in \set{1,\ldots,N}}\; \Pr\left(\frac{1}{T}\sum_{t=1}^T\lVert x_{it}\rVert\geq M\right)=o(T^{-\delta}) \text{ for all } \delta>0.$$
	\end{enumerate}
\end{hyp}
Assumption~\ref{as:prelim_rate} requires $\widehat \beta^1$ to be consistent for $\beta^0$ at a rate bounded below by $r_{NT}^{-1}$. This rate can be slow. Examples of computationally simple estimators satisfying this condition under low-level conditions are provided in Sections~\ref{sec:prelim_con_est} and Appendix Section~\ref{sec:nuc_norm_cond}. Assumption~\ref{as:linkage} ensures that both Type I and Type II classification errors are minimized during the clustering step. It is satisfied for the average, complete, and single linkage functions displayed in Table~\ref{tab:linkage_functions}. While the specific choice of a linkage function is asymptotically innocuous, it may matter in finite samples (see, e.g., Table~\ref{tab:mc_pureGFE_complinkage}).  Assumption~\ref{as:tuning_param} allows $T$ to grow considerably more slowly than $N$ (if $\nu\gg1$). It requires the sequence of clustering tuning parameters to vanish, but not too quickly, at a rate of convergence bounded below by $r_{NT}^{-1}$ and strictly slower than $T^{1/2}$. Probability limits are used because this tuning parameter is allowed to be data-driven and, as such, can be random. Assumptions~\ref{as:sup_norm_cons}\ref{as:sup_norm_cons_a}--\ref{as:sup_norm_cons_b} and \ref{as:sup_norm_cons}\ref{as:sup_norm_cons_d} collect standard moment, tail, and decaying temporal dependence conditions. These assumptions do not impose homoskedasticity but only require uniform bounds on the unconditional variances. Assumption~\ref{as:sup_norm_cons}\ref{as:sup_norm_cons_c} requires groups to be well-separated. Although it is strictly weaker than the ``strong factors'' assumption commonly found in the interactive fixed effects literature, it may break down in very large panels, e.g., in the presence of converging trends.\footnote{Consider the interactive fixed effect version of the model described in Footnote~\ref{footnote:factor_analytic}, where $\lambda_i$ belongs to the set of vertices of $\R^{G^0}$ and $f_t$ collects the group-specific effects at time $t$. Common strong factor assumptions impose $\frac1N\sum_{i=1}^N\lambda_i\lambda_i'\overset{p}{\to}\Sigma_\Lambda>0$ and $\frac1T\sum_{t=1}^Tf_tf_t'\overset{p}{\to}\Sigma_F>0$. Letting $e_g$ denote the $G^0$-vector with 1 in the $g$th coordinate and 0 everywhere else, and $e_{g,\tilde g}:=e_g-e_{\tilde g}$, the continuous mapping theorem implies
\begin{align*}
	\frac{1}{T}\sum_{t=1}^T(\alpha_{gt}^0-\alpha_{\tilde gt}^0)^2&=e_{g,\tilde g}'\left(\frac1T\sum_{t=1}^Tf_tf_t'\right)e_{g,\tilde g}\overset{p}{\to}e_{g,\tilde g}'\Sigma_Fe_{g,\tilde g}>0.
\end{align*}
Hence, groups are well separated under strong factor assumptions. However, if the time trend of a given group is a homothety of another, the strong factor assumption fails but the two groups will be well-separated as long as the homothety is not the identity. Similarly, 
\[
\frac{1}{N}\sum_{i=1}^N\mathbf 1\{g_i^0=g\}=e_g'\left(\frac1N\sum_{i=1}^N\lambda_i\lambda_i'\right)e_g\overset{p}{\to}e_g'\Sigma_\Lambda e_g>0. 
\]
Hence, Assumption~\ref{as:sup_norm_cons}\ref{as:sup_norm_cons_e} is weaker than the strong factor assumption.} Assumption~\ref{as:sup_norm_cons}\ref{as:sup_norm_cons_d} rules out cross-sectional correlation in the error term. Assumption~\ref{as:sup_norm_cons}\ref{as:sup_norm_cons_e} allows for asymptotically negligible groups, but requires that each group contains at least two members with probability approaching one. Assumption~\ref{as:sup_norm_cons}\ref{as:sup_norm_cons_f} is \cite{BM2015}'s Assumption 2(e). It holds if covariates have bounded support or satisfy dependence and tail conditions similar to those of $v_{it}$. All results below are understood up to group relabelling.  
\begin{prop}[Sup-Norm Classification Consistency]
	\label{prop:sup_norm_cons}
	Let Assumptions~\ref{as:prelim_rate}--\ref{as:sup_norm_cons} hold. Then, as $N$ and $T$ tend to infinity,
	\begin{equation}\label{eq:consistency_group_member}
		\max_{i \in \set{1,\ldots,N}}\abs{\widehat g_i-g_i^0}\overset{p}{\to}0
	\end{equation}
	and
	\begin{equation}\label{eq:consistency_grp_number}
		\widehat G-G^0\overset{p}{\to}0.
	\end{equation}
\end{prop}

\subsection{Asymptotic distribution}
The next assumption is useful to establish the asymptotic distribution of $\widehat\beta$ and $\widehat{\alpha}_{gt}$. 
\begin{hyp}\label{as:AN} ~
	\begin{enumerate}[label=(\alph*)]
		\item \label{as:AN_a} For all $g \in \set{1,\ldots,G^0}$: $\plim_{\min(N,T)\to \infty}\frac{1}{N}\sum_{i=1}^N\mathbf{1}\{g_i^0=g\} = \pi_g>0$. 
		\item \label{as:AN_b}  For all $(i,j,t)\in\set{1,\ldots,N}^2\times\set{1,\ldots,T}$: $\E[x_{jt}v_{it}] = 0$.
		\item \label{as:AN_c} There exist positive definite matrices $\Sigma_\beta$ and $\Omega_\beta$ such that
		\begin{align*}
			\Sigma_\beta&  = \underset{\min(N,T)\to \infty}{\plim}\frac{1}{NT}\sum_{i=1}^N\sum_{t=1}^T(x_{it} - \overline{x}_{g_i^0t})(x_{it} - \overline{x}_{g_i^0t})', \\
			\Omega_\beta & = \underset{\min(N,T)\to \infty}{\plim}\frac{1}{NT}\sum_{i=1}^N\sum_{j=1}^N\sum_{t=1}^T\sum_{s=1}^T\E\left[v_{it}v_{js}(x_{it} - \overline{x}_{g_i^0t})(x_{js} - \overline{x}_{g_j^0s})'\right],
		\end{align*}
		where $\overline{x}_{gt}\equiv\left(\sum_{i=1}^N\mathbf{1}\{g_i^0=g\}\right)^{-1}\sum_{i=1}^N\mathbf{1}\{g_i^0=g\}x_{it}$.
		\item \label{as:AN_d} As $N$ and $T$ tend to infinity: $\frac{1}{\sqrt{NT}}\sum_{i=1}^N\sum_{t=1}^T(x_{it} - \overline{x}_{g_i^0t})v_{it} \overset{d}{\to}\mathcal N(0,\Omega_\beta)$.
		\item \label{as:AN_e} For all $(g,t) \in \set{1,\ldots,G^0}\times\set{1,\ldots,T}$: $$\lim_{\min(N,T)\to \infty}\frac{1}{N}\sum_{i=1}^N\sum_{j=1}^N\E\left[\mathbf{1}\{g_i^0=g\}\mathbf{1}\{g_j^0=g\}v_{it}v_{jt}\right] = \omega_{gt}>0.$$
		\item \label{as:AN_f}  For all $(g,t) \in \set{1,\ldots,G^0}\times\set{1,\ldots,T}$: as $N$ and $T$ tend to infinity, $$\frac{1}{\sqrt N}\sum_{i=1}^N\mathbf{1}\{g_i^0=g\}v_{it}\overset{d}{\to}\mathcal N(0, \omega_{gt}).$$
	\end{enumerate}
\end{hyp}
Assumption~\ref{as:AN} ensures that the infeasible least squares estimator has a standard asymptotic distribution. Assumption~\ref{as:AN}(b) is satisfied if the $x_{it}$ are strictly exogenous or predetermined and observations are independent across units. As a special case, lagged outcomes may thus be included in $x_{it}$. The assumption does not allow for spatial lags such as $y_{i-1t}$.
\begin{cor}[Asymptotic Distribution]
	\label{cor:AN}
	Let Assumptions~\ref{as:prelim_rate}--\ref{as:AN} hold. Then, as $N$ and $T$ tend to infinity, 
	\begin{equation}\label{eq:AN_theta_cov}
		\sqrt{NT}(\widehat \beta-\beta^0)\overset{d}{\to}\mathcal N\left(0,\Sigma_\beta^{-1}\Omega_\beta \Sigma_\beta^{-1}\right)
	\end{equation}
	and, for all $t$,
	\begin{equation}\label{eq:AN_group_fe}
		\sqrt{N}(\widehat\alpha_{gt}-\alpha_{gt}^0) \overset{d}{\to}\mathcal N\left(0, \frac{\omega_{gt}}{\pi_g^2}\right), \quad g = 1,\ldots,G^0,
	\end{equation}
	where $\Sigma_\beta, \Omega_\beta, \omega_{gt}$, and $\pi_g$ are defined in Assumption~\ref{as:AN}.
\end{cor}
Consistent plug-in estimates of the asymptotic variances can easily be constructed \citep[see, e.g.,][Supplementary Material]{BM2015}.

\subsection{Choice of tuning parameters} \label{sec:data_driven_selec_rule}
 Assumptions~\ref{as:prelim_rate} and \ref{as:tuning_param} provide theoretical guidance for selecting the tuning parameters $\psi_{NT}$ and $c_{NT}$. This guidance, however, is based on asymptotic approximations. In applications with finite sample sizes, an infinite number of choices consistent with the theory will lead to different results. This issue is analogous to bandwidth selection in nonparametric density estimation, or to selecting the number of factors or groups in interactive fixed effects panel data models using AIC and BIC criteria as done in \cite{NgBai_2002} and \cite{BM2015}. The latter involves two tuning parameters: an upper bound for the number of factors or groups and a penalization function characterized by its asymptotic behaviour.\footnote{The eigenvalue ratio test of \cite{AhnHorenstein2013} requires only a known upper bound, but it validity is proved under rectangular array asymptotics.} 

A data-driven rule is proposed below, whose  advantage is to eliminate the risk of specifying an incorrect upper bound for the number of groups. This type of misspecification may lead to poor inference, both in finite samples and asymptotically, as demonstrated in Section~\ref{sec:mc_sim}.

The selection rule for the thresholding parameter $c_{NT}$ assumes the availability of a consistent estimator $\widehat{\beta}^1$ for $\beta^0$, with a rate of convergence of at least $r_{NT}^{-1}\equiv\sqrt{\min(N,T)}/\log(\log(T))$. By Proposition~\ref{prop:nucnormest} in Appendix Section~\ref{sec:nuc_norm_cond},  the nuclear-norm regularized estimator $\widehat\beta^1(\psi_{NT})$, with $\psi_{NT}=\log(\log(T))/\sqrt{16\min(N,T)}$, satisfies this requirement under weak conditions.\footnote{The values obtained for $\widehat\beta^1(\psi_{NT})$ across all Monte Carlo simulations are very close to the nuclear norm estimates, whose computation does not require any tuning parameter.} 

The selection rule for the thresholding parameter $c_{NT} $ also requires an estimate of the noise dispersion. Under cross-sectional homoskedasticity in the sense that 
\[
	\frac{1}{T}\sum_{t=1}^Tv_{it}^2\overset{p}{\to}\sigma^2_v, \quad i=1,\ldots, N,
\]
as $\min(N,T)\to\infty$, an estimator of $\sigma^2_v$ is 
\[
\widehat \sigma_v^2\equiv \min_{(i,j)\in\{1,\ldots,N\}^2, i\neq j}\frac{1}{2T}\sum_{t=1}^T(\widehat v_{it}-\widehat v_{jt})^2.
 \]
It is not difficult to show that $\widehat\sigma_v^2\overset{p}{\to} \sigma_v^2$ under the maintained assumptions \citep[see also Section 4.2.1 in][for a related result]{Zeleneev2020}. It turns out, however, that this estimator exhibits a substantial downward finite-sample bias, which converges to zero very slowly in Monte Carlo simulations: prohibitively large values of $T$ are needed to obtain  reasonable performance. An asymptotically valid method to diminish the bias is to consider the max-min estimator
 \[
 \check \sigma_v^2\equiv \max_{i\in\{1,\ldots,N\}}\min_{j \in\{1,\ldots,N\}, j\neq i}\frac{1}{2T}\sum_{t=1}^T(\widehat v_{it}-\widehat v_{jt})^2.
\]
Again, it is not difficult to show that $\check\sigma_v^2\overset{p}{\to} \sigma_v^2$ under cross-sectional homoskedasticity and the maintained assumptions.  This estimator yields important improvements in Monte Carlo simulations with small to moderate sample sizes. The data-driven selection rule for the thresholding parameter $c_{NT} $
is then
\[c_{NT}\equiv\frac{1.35\times\check\sigma_v\times\log(T)}{\max(K,1)\sqrt{\min(N,T)}}. \]
The theoretical validity of this choice follows from $\check \sigma_v=O_p(1)$ under the maintained assumptions, so that $r_{NT}$ and $c_{NT}$ verify Assumptions~\ref{as:prelim_rate} and \ref{as:tuning_param}.\footnote{An alternative approach could be based on limiting overfitting for different value of $c_{NT}$ as is done in \cite{BonhommeLamadonManresa2022}. This requires computing a regularization path and, therefore, is computationally more costly.}

\section{Monte Carlo simulations}\label{sec:mc_sim}
This section reports the results of some Monte Carlo simulations calibrated to the empirical application. The finite sample properties of the TPWD estimator are assessed across two distinct data generating processes (DGP): a pure GFE model without covariates and a full GFE model with a scalar covariate. In each DGP, the variance of the error is set to the one from \cite{BM2015}'s  calibrated simulations. Grouped fixed-effects and group memberships are set so that the minimum distance between groups matches those of \cite{BM2015}'s  calibrated simulations for $G=4$. The signal-to-noise ratios of the group membership variables match approximately across all DGPs. The simulation design is not identical in order to let the sample size grow.\footnote{\cite{BM2015}'s simulations are based on the dataset from the empirical application, where $N=90$ and $T=7$.} The sample sizes and number of groups vary across $(N,T)\in\{90, 180\} \times \{7, 10, 20, 40\}$ and $G\in\{3, 4\}$, respectively. 

To mirror the empirical findings of \cite{BM2015}, where the observed outcome is a scalar, continuous measure of a country's democratic regime status at time $t$, the grouped fixed effects are set as follows:  for all $t\in\{1,\ldots,T\}$,
\[
\alpha_{1t}\equiv 1, \quad \alpha_{2t}\equiv \frac{t-1}{T-1}, \quad \alpha_{3t}\equiv0, \quad \alpha_{4t}\equiv \frac{\ind{t\geq\lfloor T/2\rfloor}(t-\lfloor T/2 \rfloor)}{T-\lfloor T/2\rfloor}.
\]
The first group could be labeled as ``high democracy,'' the second as ``early transition,'' the third as ``low democracy,'' and the fourth as ``late transition.'' In the baseline setting, groups are balanced:  
\[
g_i = 1 + \sum_{g=1}^{G-1}\ind{i > g\lfloor N/G\rfloor}, \quad i=1,\ldots,N.
\]

The performance of the TPWD estimator is compared with benchmark alternatives: the grouped fixed-effects (GFE) estimator from \cite{BM2015}, the spectral and post-spectral estimators from \cite{ChetverikovManresa2021}, the nuclear-norm (NN) and nuclear-norm regularized (NNR) estimators from \cite{moon2019nuclear}, the least squares interactive fixed effects (IFE) estimator from \cite{Bai2009}, and the infeasible pooled OLS regression that uses the true group memberships (Oracle).

Sections~S.3.1--3.5 contain additional results for models with unbalanced groups, more groups, unit-specific effects, higher signal-to-noise ratio, or time-invariant unobserved heterogeneity. In all experiments, results are averaged across $500$ Monte Carlo samples.

\subsection{Pure grouped fixed effects model}\label{sec:mc_sim_pure}
Consider a pure GFE model without covariates:   
$$y_{it}=\alpha_{g_it}+v_{it}, \quad i=1,\ldots,N,\; t=1,\ldots,T,$$
where $v_{it}\overset{iid}{\sim}\mathcal N(0,(1/3)^2)$ across units and time periods. Let $\widehat\pi_g\equiv N^{-1}\sum_{i=1}^N\ind{g_i=g}$. Since groups are balanced, $\widehat\pi_g=1/G$ and the average signal-to-noise ratio of group membership is 
\[
\frac{\sum_{g=1}^G{\rm Var}(\ind{g_i=g})}{\sigma_v^2}=\frac{\frac{1}{G}\sum_{g=1}^G\widehat\pi_g(1-\widehat\pi_g)}{(1/3)^2}
=2\ind{G=3}+\frac{27}{16}\ind{G=4}.
\]
This is to compare with $1.89\ind{G=3}+1.69\ind{G=4}$, the average signal-to-noise ratio in \cite{BM2015}'s calibrated simulations.

The TPWD estimator with $\widehat\beta^1=0$ and $c_{NT}=1.35\check\sigma_v\log(T)/\sqrt{\min(N,T)}$ is compared to versions without covariates of \cite{ChetverikovManresa2021}'s post-spectral estimator with $g\in\{2,3,4,10\}$ user-specified number of groups (Post-Spectral$^{\bar G=g}$) and \cite{BM2015}'s grouped fixed-effects estimator with $g\in\{2,3,4,10\}$ user-specified number of groups, i.e., their ``Algorithm 1'' with 1,000 randomly generated initialization points (GFE$^{\bar G=g}$). 

The performance of all estimators  is assessed in terms of the average estimated number of groups (if applicable), $500^{-1}\sum_{b=1}^{500}\widehat G^{(b)}$, the average root mean square error (RMSE) of the estimated grouped fixed effects, 
\begin{align*}
	{\rm RMSE}(\widehat\alpha)&\equiv\frac{1}{500}\sum_{b=1}^{500}\sqrt{\frac{1}{NT}\sum_{i=1}^N\sum_{t=1}^T(\widehat\alpha_{\widehat g_i^{(b)}t}^{(b)}-\alpha_{g_it}^{(b)})^2},
\end{align*}
and the average clustering accuracy, measured using the average Precision (P) rate, Recall (R) rate, and Rand Index (RI):
 \[
 P(\widehat g)\equiv\frac{1}{500}\sum_{b=1}^{500}p(\widehat g^{(b)}), \quad  R(\widehat g)\equiv\frac{1}{500}\sum_{b=1}^{500}r(\widehat g^{(b)}),  \quad RI(\widehat g)\equiv\frac{1}{500}\sum_{b=1}^{500}ri(\widehat g^{(b)}),
 \]
 where, by denoting the number of false positives (FP), true positives (TP), false negatives (FN), and true negatives (TN) as
 \begin{align*}
FP(\widehat g)&\equiv\sum_{i<j}\ind{\widehat g_i=\widehat g_j} \ind{g_i\neq g_j}, \\
TP(\widehat g)&\equiv\sum_{i<j}\ind{\widehat g_i=\widehat g_j} \ind{g_i=g_j}, \\ 
FN(\widehat g)&\equiv\sum_{i<j}\ind{\widehat g_i\neq\widehat g_j} \ind{g_i=g_j}, \\ 
TN(\widehat g)&\equiv\sum_{i<j}\ind{\widehat g_i\neq\widehat g_j} \ind{g_i\neq g_j},
\end{align*}
the following three measures of clustering accuracy are invariant to cluster relabelling:
\begin{align*}
	& p(\widehat g)\equiv\frac{TP(\widehat g)}{TP(\widehat g)+FP(\widehat g)}, \quad r(\widehat g)\equiv\frac{TP(\widehat g)}{TP(\widehat g)+FN(\widehat g)}, \\
	&\quad \quad ri(\widehat g)\equiv\frac{TP(\widehat g)+TN(\widehat g)}{TP(\widehat g)+FP(\widehat g)+FN(\widehat g)+TN(\widehat g)}.
\end{align*}
The precision rate measures the proportion of correctly matched unit pairs among all matched pairs. The recall rate measures the proportion of correctly matched unit pairs among all pairs of units belonging to the same population cluster. The Rand index measures the proportion of correct decisions (to match or not to match) among all matching decisions taken by the clustering algorithm (i.e., among all possible pairs of units). It summarizes both Type I and Type II errors in the prediction of group memberships.\footnote{Given the null hypothesis $H_{0,i,j}:g_i=g_j$, $\forall (i,j)$, introduced earlier, the definitions of $FP,TP,FN$, and $TN$ should be flipped, e.g., $FP(\widehat g)\equiv\sum_{i<j}\ind{\widehat g_i\neq \widehat g_j} \ind{g_i= g_j}$. Conceptually, however, a merging HAC algorithm initialized at $N$ singleton groups ``starts'' with the null $\widetilde H_{0,i,j}:g_i\neq g_j$, $\forall (i,j)$,  while a divisive algorithm initialized at a unique group of $N$ units ``starts'' with $H_{0,i,j}$, $\forall (i,j)$.  The multiple-testing problem was stated that way following the standard practice of stating the null hypothesis as an equality. }

\begin{table}[h!]
	\centering
		\caption{Finite sample properties in the pure GFE model (grouped fixed effects)}
	\label{tab:mcsim_pureGFE_tab1}
	\begin{adjustbox}{max width={0.99\linewidth},center}
	\begin{threeparttable}
    \setlength{\tabcolsep}{2pt} 
\begin{tabular}{ccccccccccccccccccccccccccc}
\toprule
{} & {} & {} & \multicolumn{2}{c}{TPWD} & & \multicolumn{2}{c}{Post-Spectral$^{\bar G=2}$} & & \multicolumn{2}{c}{Post-Spectral$^{\bar G=3}$} &&  \multicolumn{2}{c}{Post-Spectral$^{\bar G=4}$} & & \multicolumn{2}{c}{Post-Spectral$^{\bar G=10}$} & & \multicolumn{1}{c}{GFE$^{\bar G=2}$} & & \multicolumn{1}{c}{GFE$^{\bar G=3}$} & & \multicolumn{1}{c}{GFE$^{\bar G=4}$} &  & \multicolumn{1}{c}{GFE$^{\bar G=10}$} && \multicolumn{1}{c}{Oracle} \\
\cmidrule{4-5}\cmidrule{7-8}\cmidrule{10-11}\cmidrule{13-14}\cmidrule{16-17}
$G$ & $N$ & $T$ & \multicolumn{1}{c}{$\widehat G$} & \multicolumn{1}{c}{\parbox{2.5cm}{\centering RMSE \\ [-3pt] \small (CPU time)}}  & & \multicolumn{1}{c}{$\widehat G$} &\multicolumn{1}{c}{\parbox{2.5cm}{\centering RMSE \\ [-3pt] \small (CPU time)}}  && \multicolumn{1}{c}{$\widehat G$} &\multicolumn{1}{c}{\parbox{2.5cm}{\centering RMSE \\ [-3pt] \small (CPU time)}}  && \multicolumn{1}{c}{$\widehat G$} & \multicolumn{1}{c}{\parbox{2.5cm}{\centering RMSE \\ [-3pt] \small (CPU time)}}  && \multicolumn{1}{c}{$\widehat G$} & \multicolumn{1}{c}{\parbox{2.5cm}{\centering RMSE \\ [-3pt] \small (CPU time)}}  && \multicolumn{1}{c}{\parbox{2.5cm}{\centering RMSE \\ [-3pt] \small (CPU time)}}  && \multicolumn{1}{c}{\parbox{2.5cm}{\centering RMSE \\ [-3pt] \small (CPU time)}} && \multicolumn{1}{c}{\parbox{2.5cm}{\centering RMSE \\ [-3pt] \small (CPU time)}}  &&\multicolumn{1}{c}{\parbox{2.5cm}{\centering RMSE \\ [-3pt] \small (CPU time)}} && \multicolumn{1}{c}{\parbox{2.5cm}{\centering RMSE \\ [-3pt] \small (CPU time)}}   \\
\midrule 
3&90&7&\cellcolor{mygreen!25} 6.654 &\cellcolor{mygreen!25} 0.150   & & 1.932 & 0.399   & & \cellcolor{mygreen!25}2.838 & \cellcolor{mygreen!25}0.305   & & 3.276 & 0.286   & & — & — & & 0.270   & & \cellcolor{mygreen!25} 0.088   & & 0.129   & & 0.216   && 0.060  \\ 
&             &             &\cellcolor{mygreen!25}      &\cellcolor{mygreen!25} \text{(}0.046\text{)} & &      & \text{(}0.084\text{)} & & \cellcolor{mygreen!25}     & \cellcolor{mygreen!25}\text{(}0.083\text{)} & &      & \text{(}0.083\text{)} & & — & — & & \text{(}0.427\text{)} & & \cellcolor{mygreen!25} \text{(}0.539\text{)} & & \text{(}0.564\text{)} & & \text{(}0.759\text{)} && \text{(}0.001\text{)}  \\ 
&             &10&\cellcolor{mygreen!25} 4.814 &\cellcolor{mygreen!25} 0.107   & & 1.992 & 0.339   & & \cellcolor{mygreen!25}2.828 & \cellcolor{mygreen!25}0.209   & & 3.352 & 0.187   & & 7.058 & 0.180   & & 0.260   & & \cellcolor{mygreen!25} 0.069   & & 0.108   & & 0.191   && 0.060  \\ 
&             &             &\cellcolor{mygreen!25}      &\cellcolor{mygreen!25} \text{(}0.063\text{)} & &      & \text{(}0.092\text{)} & & \cellcolor{mygreen!25}     & \cellcolor{mygreen!25}\text{(}0.092\text{)} & &      & \text{(}0.092\text{)} & &      & \text{(}0.097\text{)} & & \text{(}0.389\text{)} & & \cellcolor{mygreen!25} \text{(}0.492\text{)} & & \text{(}0.526\text{)} & & \text{(}0.718\text{)} && \text{(}0.001\text{)}  \\ 
&             &20&\cellcolor{mygreen!25} 3.310 &\cellcolor{mygreen!25} 0.066   & & 1.998 & 0.290   & & \cellcolor{mygreen!25}2.998 & \cellcolor{mygreen!25}0.088   & & 3.228 & 0.082   & & 5.138 & 0.091   & & 0.246   & & \cellcolor{mygreen!25} 0.061   & & 0.091   & & 0.161   && 0.061  \\ 
&             &             &\cellcolor{mygreen!25}      &\cellcolor{mygreen!25} \text{(}0.124\text{)} & &      & \text{(}0.162\text{)} & & \cellcolor{mygreen!25}     & \cellcolor{mygreen!25}\text{(}0.161\text{)} & &      & \text{(}0.160\text{)} & &      & \text{(}0.171\text{)} & & \text{(}0.352\text{)} & & \cellcolor{mygreen!25} \text{(}0.407\text{)} & & \text{(}0.459\text{)} & & \text{(}0.657\text{)} && \text{(}0.005\text{)}  \\ 
&             &40&\cellcolor{mygreen!25} 3.012 &\cellcolor{mygreen!25} 0.061   & & 2.000 & 0.258   & & \cellcolor{mygreen!25}3.000 & \cellcolor{mygreen!25}0.062   & & 3.416 & 0.066   & & 4.788 & 0.081   & & 0.243   & & \cellcolor{mygreen!25} 0.061   & & 0.084   & & 0.143   && 0.061  \\ 
&             &             &\cellcolor{mygreen!25}      &\cellcolor{mygreen!25} \text{(}0.322\text{)} & &      & \text{(}0.367\text{)} & & \cellcolor{mygreen!25}     & \cellcolor{mygreen!25}\text{(}0.459\text{)} & &      & \text{(}0.490\text{)} & &      & \text{(}0.605\text{)} & & \text{(}0.360\text{)} & & \cellcolor{mygreen!25} \text{(}0.405\text{)} & & \text{(}0.454\text{)} & & \text{(}0.667\text{)} && \text{(}0.039\text{)}  \\ 
\midrule
3&180&7&\cellcolor{mygreen!25} 9.268 &\cellcolor{mygreen!25} 0.147   & & 1.970 & 0.403   & & \cellcolor{mygreen!25}2.866 & \cellcolor{mygreen!25}0.304   & & 3.290 & 0.287   & & — & — & & 0.269   & & \cellcolor{mygreen!25} 0.076   & & 0.114   & & 0.196   && 0.042  \\ 
&             &             &\cellcolor{mygreen!25}      &\cellcolor{mygreen!25} \text{(}0.396\text{)} & &      & \text{(}0.117\text{)} & & \cellcolor{mygreen!25}     & \cellcolor{mygreen!25}\text{(}0.115\text{)} & &      & \text{(}0.116\text{)} & & — & — & & \text{(}0.703\text{)} & & \cellcolor{mygreen!25} \text{(}0.782\text{)} & & \text{(}0.940\text{)} & & \text{(}1.507\text{)} && \text{(}0.001\text{)}  \\ 
&             &10&\cellcolor{mygreen!25} 5.988 &\cellcolor{mygreen!25} 0.099   & & 1.984 & 0.346   & & \cellcolor{mygreen!25}2.898 & \cellcolor{mygreen!25}0.205   & & 3.382 & 0.173   & & 5.786 & 0.212   & & 0.258   & & \cellcolor{mygreen!25} 0.053   & & 0.091   & & 0.169   && 0.042  \\ 
&             &             &\cellcolor{mygreen!25}      &\cellcolor{mygreen!25} \text{(}0.568\text{)} & &      & \text{(}0.129\text{)} & & \cellcolor{mygreen!25}     & \cellcolor{mygreen!25}\text{(}0.128\text{)} & &      & \text{(}0.128\text{)} & &      & \text{(}0.141\text{)} & & \text{(}0.622\text{)} & & \cellcolor{mygreen!25} \text{(}0.747\text{)} & & \text{(}0.916\text{)} & & \text{(}1.441\text{)} && \text{(}0.003\text{)}  \\ 
&             &20&\cellcolor{mygreen!25} 3.674 &\cellcolor{mygreen!25} 0.052   & & 2.000 & 0.291   & & \cellcolor{mygreen!25}3.000 & \cellcolor{mygreen!25}0.072   & & 3.258 & 0.068   & & 4.682 & 0.065   & & 0.243   & & \cellcolor{mygreen!25} 0.044   & & 0.072   & & 0.136   && 0.043  \\ 
&             &             &\cellcolor{mygreen!25}      &\cellcolor{mygreen!25} \text{(}1.764\text{)} & &      & \text{(}0.239\text{)} & & \cellcolor{mygreen!25}     & \cellcolor{mygreen!25}\text{(}0.235\text{)} & &      & \text{(}0.233\text{)} & &      & \text{(}0.267\text{)} & & \text{(}0.558\text{)} & & \cellcolor{mygreen!25} \text{(}0.656\text{)} & & \text{(}0.851\text{)} & & \text{(}1.385\text{)} && \text{(}0.016\text{)}  \\ 
&             &40&\cellcolor{mygreen!25} 3.058 &\cellcolor{mygreen!25} 0.043   & & 1.998 & 0.257   & & \cellcolor{mygreen!25}3.000 & \cellcolor{mygreen!25}0.043   & & 3.254 & 0.045   & & 4.510 & 0.054   & & 0.240   & & \cellcolor{mygreen!25} 0.043   & & 0.064   & & 0.115   && 0.043  \\ 
&             &             &\cellcolor{mygreen!25}      &\cellcolor{mygreen!25} \text{(}3.846\text{)} & &      & \text{(}0.715\text{)} & & \cellcolor{mygreen!25}     & \cellcolor{mygreen!25}\text{(}0.851\text{)} & &      & \text{(}0.846\text{)} & &      & \text{(}1.091\text{)} & & \text{(}0.547\text{)} & & \cellcolor{mygreen!25} \text{(}0.667\text{)} & & \text{(}0.875\text{)} & & \text{(}1.517\text{)} && \text{(}0.130\text{)}  \\ 
\midrule
4&90&7&\cellcolor{mygreen!25} 6.926 &\cellcolor{mygreen!25} 0.164   & & 1.934 & 0.373   & &  2.714 & 0.325   & & \cellcolor{mygreen!25} 3.282 & \cellcolor{mygreen!25} 0.299   & & — & — & & 0.252   & & 0.120   & &  \cellcolor{mygreen!25} 0.144   & & 0.223   && 0.070  \\ 
&             &             &\cellcolor{mygreen!25}      &\cellcolor{mygreen!25} \text{(}0.044\text{)} & &      & \text{(}0.083\text{)} & &      &\text{(}0.082\text{)} & &  \cellcolor{mygreen!25}    &  \cellcolor{mygreen!25}\text{(}0.083\text{)} & & — & — & & \text{(}0.517\text{)} & &\text{(}0.559\text{)} & &  \cellcolor{mygreen!25}  \text{(}0.624\text{)} & & \text{(}0.789\text{)} && \text{(}0.001\text{)}  \\ 
           &             &10&\cellcolor{mygreen!25} 4.910 &\cellcolor{mygreen!25} 0.137   & & 1.988 & 0.344   & & 2.788 &0.237   & & \cellcolor{mygreen!25} 3.374 &  \cellcolor{mygreen!25}0.215   & & 6.738 & 0.219   & & 0.242   & &  0.123   & & \cellcolor{mygreen!25} 0.117   & & 0.199   && 0.070  \\ 
&             &             &\cellcolor{mygreen!25}      &\cellcolor{mygreen!25} \text{(}0.061\text{)} & &      & \text{(}0.091\text{)} & &     & \text{(}0.091\text{)} & &  \cellcolor{mygreen!25}     & \cellcolor{mygreen!25} \text{(}0.091\text{)} & &      & \text{(}0.096\text{)} & & \text{(}0.573\text{)} & &  \text{(}0.558\text{)} & & \cellcolor{mygreen!25} \text{(}0.633\text{)} & & \text{(}0.762\text{)} && \text{(}0.002\text{)}  \\ 
           &             &20&\cellcolor{mygreen!25} 3.866 &\cellcolor{mygreen!25} 0.102   & & 2.000 & 0.321   & & 2.972 &0.145   & & \cellcolor{mygreen!25} 3.250 &  \cellcolor{mygreen!25}0.131   & & 6.300 & 0.141   & & 0.230   & &  0.116   & & \cellcolor{mygreen!25} 0.084   & & 0.163   && 0.070  \\ 
&             &             &\cellcolor{mygreen!25}      &\cellcolor{mygreen!25} \text{(}0.119\text{)} & &      & \text{(}0.157\text{)} & &     & \text{(}0.159\text{)} & &  \cellcolor{mygreen!25}     & \cellcolor{mygreen!25} \text{(}0.159\text{)} & &      & \text{(}0.176\text{)} & & \text{(}0.598\text{)} & &  \text{(}0.542\text{)} & & \cellcolor{mygreen!25} \text{(}0.617\text{)} & & \text{(}0.733\text{)} && \text{(}0.008\text{)}  \\ 
           &             &40&\cellcolor{mygreen!25} 3.986 &\cellcolor{mygreen!25} 0.077   & & 2.000 & 0.308   & & 3.000 &0.118   & & \cellcolor{mygreen!25} 3.472 &  \cellcolor{mygreen!25}0.119   & & 6.174 & 0.135   & & 0.227   & &  0.117   & & \cellcolor{mygreen!25} 0.072   & & 0.140   && 0.070  \\ 
&             &             &\cellcolor{mygreen!25}      &\cellcolor{mygreen!25} \text{(}0.290\text{)} & &      & \text{(}0.367\text{)} & &     & \text{(}0.454\text{)} & &  \cellcolor{mygreen!25}     & \cellcolor{mygreen!25} \text{(}0.485\text{)} & &      & \text{(}0.649\text{)} & & \text{(}0.618\text{)} & &  \text{(}0.576\text{)} & & \cellcolor{mygreen!25} \text{(}0.627\text{)} & & \text{(}0.763\text{)} && \text{(}0.055\text{)}  \\ 
\midrule
4&180&7&\cellcolor{mygreen!25} 8.126 &\cellcolor{mygreen!25} 0.148   & & 1.952 & 0.287   & &  2.646 & 0.277   & & \cellcolor{mygreen!25} 3.250 & \cellcolor{mygreen!25} 0.272   & & — & — & & 0.192   & & 0.116   & &  \cellcolor{mygreen!25} 0.145   & & 0.214   && 0.049  \\ 
&             &             &\cellcolor{mygreen!25}      &\cellcolor{mygreen!25} \text{(}0.391\text{)} & &      & \text{(}0.117\text{)} & &      &\text{(}0.116\text{)} & &  \cellcolor{mygreen!25}    &  \cellcolor{mygreen!25}\text{(}0.117\text{)} & & — & — & & \text{(}0.961\text{)} & &\text{(}1.157\text{)} & &  \cellcolor{mygreen!25}  \text{(}1.422\text{)} & & \text{(}1.733\text{)} && \text{(}0.001\text{)}  \\ 
           &             &10&\cellcolor{mygreen!25} 5.376 &\cellcolor{mygreen!25} 0.120   & & 1.994 & 0.281   & & 2.800 &0.194   & & \cellcolor{mygreen!25} 3.370 &  \cellcolor{mygreen!25}0.174   & & 5.168 & 0.240   & & 0.183   & &  0.113   & & \cellcolor{mygreen!25} 0.122   & & 0.190   && 0.049  \\ 
&             &             &\cellcolor{mygreen!25}      &\cellcolor{mygreen!25} \text{(}0.526\text{)} & &      & \text{(}0.127\text{)} & &     & \text{(}0.127\text{)} & &  \cellcolor{mygreen!25}     & \cellcolor{mygreen!25} \text{(}0.128\text{)} & &      & \text{(}0.142\text{)} & & \text{(}0.880\text{)} & &  \text{(}1.167\text{)} & & \cellcolor{mygreen!25} \text{(}1.462\text{)} & & \text{(}1.719\text{)} && \text{(}0.003\text{)}  \\ 
           &             &20&\cellcolor{mygreen!25} 3.930 &\cellcolor{mygreen!25} 0.083   & & 2.000 & 0.276   & & 2.988 &0.113   & & \cellcolor{mygreen!25} 3.198 &  \cellcolor{mygreen!25}0.104   & & 5.534 & 0.110   & & 0.175   & &  0.099   & & \cellcolor{mygreen!25} 0.078   & & 0.147   && 0.049  \\ 
&             &             &\cellcolor{mygreen!25}      &\cellcolor{mygreen!25} \text{(}1.283\text{)} & &      & \text{(}0.218\text{)} & &     & \text{(}0.222\text{)} & &  \cellcolor{mygreen!25}     & \cellcolor{mygreen!25} \text{(}0.225\text{)} & &      & \text{(}0.271\text{)} & & \text{(}0.768\text{)} & &  \text{(}1.093\text{)} & & \cellcolor{mygreen!25} \text{(}1.545\text{)} & & \text{(}1.721\text{)} && \text{(}0.023\text{)}  \\ 
           &             &40&\cellcolor{mygreen!25} 3.976 &\cellcolor{mygreen!25} 0.058   & & 2.000 & 0.272   & & 3.000 &0.100   & & \cellcolor{mygreen!25} 3.670 &  \cellcolor{mygreen!25}0.102   & & 4.658 & 0.105   & & 0.172   & &  0.099   & & \cellcolor{mygreen!25} 0.053   & & 0.118   && 0.050  \\ 
&             &             &\cellcolor{mygreen!25}      &\cellcolor{mygreen!25} \text{(}3.736\text{)} & &      & \text{(}0.706\text{)} & &     & \text{(}0.828\text{)} & &  \cellcolor{mygreen!25}     & \cellcolor{mygreen!25} \text{(}0.866\text{)} & &      & \text{(}1.079\text{)} & & \text{(}0.804\text{)} & &  \text{(}1.039\text{)} & & \cellcolor{mygreen!25} \text{(}1.554\text{)} & & \text{(}1.855\text{)} && \text{(}0.197\text{)}  \\ 
\bottomrule
\end{tabular}
	\begin{tablenotes}
		\footnotesize \item {\em Notes:} This table reports the estimated number of groups ($\widehat G$), the root mean square error (RMSE), and the execution time in seconds (CPU time) for the triad pairwise-difference (TPWD) estimator computed with $\widehat\beta^1=0$, average linkage, and cut-off $c_{NT}=1.35\check\sigma_v\log(T)/\sqrt{\min(N,T)}$ and for the post-spectral estimator proposed in \cite{ChetverikovManresa2021} with $\tilde\beta^0=\tilde\beta^1=0$, user-specified number of groups $g\in\set{2,3,4,10}$, and smallest $\lambda$ chosen in the grid  $\{1,1.5,2,2.5,\ldots\}$ such that $m(\lambda)\leq g$ (Post-Spectral$^{\bar G=g}$). It also reports RMSE and CPU time for \cite{BM2015}'s grouped fixed-effects estimator (their ``Algorithm 1'') with a user-specified number of groups $g\in\set{2,3,4,10}$ and $500$ random initialization points (GFE$^{\bar G=g}$) and for the infeasible (Oracle) estimator using the ``true'' group memberships. A green-shaded cell corresponds to a well-specified estimator. Results for Post-Spectral$^{\bar G=10}$ are missing for $T=7$ since Post-Spectral$^{\bar G=g}$ is not properly defined if $g>T$ as it requires computing the $g$ largest eigenvectors of a $T\times T$ matrix. Results are averaged across 500 Monte Carlo samples. 
	\end{tablenotes}
	\end{threeparttable}
	\end{adjustbox}
\end{table}

Table~\ref{tab:mcsim_pureGFE_tab1} reports the average estimated number of groups (if applicable), the RMSE of the estimated grouped fixed effects, and the execution time in seconds for each estimator. The performance of TPWD  in terms of RMSE uniformly dominates that of Post-Spectral$^{\bar G=g}$ across all values of $(G,N,T)$ and user-specified $g\in\{2,3,4,10\}$.\footnote{Note that Post-Spectral$^{\bar G=10}$ cannot be computed for $T=7$ because it requires computing the first ten largest eigenvalues of a $7\times7$ matrix.} For small $T\in\{7,10\}$, the RMSE of the well-specified (green-shaded) Post-Spectral$^{\bar G=G}$ is about twice as large as that of TPWD. Except for $T=7$ or $(G,N,T)\in\{$(3,180,10), (4,90,10),(4,180,10)$\}$, the performance of TPWD  in terms of RMSE dominates that of each misspecified GFE estimator. The performance of TPWD  in terms of RMSE is close but dominated by that of the well-specified (green-shaded) GFE$^{\bar G=G}$ across almost all values of $(G,N,T)$. For $G=3$ (resp.~$G=4$), it is within a $0.071$ (resp.~$0.020$) distance. 

For $(G,N,T)=(3,90,7)$, the TPWD estimator is computed in $1/20$ seconds, has an RMSE of $0.150$, and estimates $6.654$ groups on average. The well-specified post-spectral estimator, Post-Spectral$^{\bar G=3}$, has an RMSE of $0.305$ and estimates $2.838$ groups on average, the well-specified GFE estimator, GFE$^{\bar G=3}$, has an RMSE of $0.088$, the misspecified GFE estimators, GFE$^{\bar G=2}$, GFE$^{\bar G=4}$, and GFE$^{\bar G=10}$, have RMSEs of $0.270$, $0.129$, and $0.216$, respectively, and the Oracle estimator has an RMSE of $0.060$.  As $T$ grows, the number of groups estimated by TPWD converges to the ground truth, and its RMSE steadily decreases to reach the same level of performance as the Oracle estimator, with precision 10$^{-2}$ provided $T\geq 20$ (for $G=3$) or $T\geq 40$ (for $G=4$). While GFE estimators exhibit a similar pattern, post-spectral estimators reach this level of precision only for $G=3$ and $T\geq 40$. Unreported simulations suggest that GFE estimators are quite sensitive to the choice of initializers: if those are randomly drawn around zero (e.g., from a normal distribution), convergence breaks down in the simulations.  This is likely because the clustering problem solved by GFE estimators becomes more difficult due to its NP-hardness as $G$ and $T$ grow. In particular, when $G=3$, the well-specified GFE estimator never attains the Oracle performance and does not seem to converge with $T$ (numerical errors dominate statistical errors). When $G=4$, its convergence is slow so that it does not attain the Oracle performance and similar numerical challenges can be expected for larger values of $T$. 

For $(G,N,T)=(3,90,40)$, the TPWD estimator is computed in $1/3$ seconds, has an RMSE of $0.061$, and estimates $3.012$ groups on average. The well-specified post-spectral estimator, Post-Spectral$^{\bar G=3}$, has an RMSE of $0.062$ and estimates $3.000$ groups on average, the well-specified GFE estimator, GFE$^{\bar G=3}$, has an RMSE of $0.061$, and the misspecified GFE estimators, GFE$^{\bar G=2}$, GFE$^{\bar G=4}$, and GFE$^{\bar G=10}$, have RMSEs of $0.243$, $0.084$, and $0.143$, respectively. For $(G,N,T)=(4,90,40)$, the TPWD estimator is computed in $1/3$ seconds, has an RMSE of $0.077$, and estimates $3.986$ groups on average. In comparison, Post-Spectral$^{\bar G=4}$ has an RMSE of $0.119$ and estimates $3.472$ groups on average, and GFE$^{\bar G=4}$, GFE$^{\bar G=2}$, GFE$^{\bar G=3}$, GFE$^{\bar G=10}$, have RMSEs of $0.072$, $0.227$, $0.117$, and $0.140$, respectively. Similar patterns are observed for $N=180$,  where TPWD requires at most 4 seconds to compute.\footnote{Experiments were run in parallel on the computing cluster of ENSAE Paris, but the estimator itself was not parallelized and takes a similar amount of time to compute on any professional laptop.} 

In some settings, the well-specified post-spectral estimator gets the number of groups closer (e.g., if $G=3$), but this is by construction since the number of groups is an input of the algorithm. The TPWD estimator may slightly overestimate or underestimate the number of groups, depending on the choice of the thresholding parameter in finite samples, but this has little impact on the RMSE. Intuitively, similar units are grouped together based on a finer grid, rather than by imposing a fixed value of \( G \).

\begin{table}[H]
	\centering
		\caption{Finite sample properties in the pure GFE model (group memberships)}
	\label{tab:mcsim_pureGFE_tab2}
	\begin{adjustbox}{max width={0.99\linewidth},center} 
	\begin{threeparttable}
\begin{tabular}{lll *{24}{S[table-format=-1.3]}}
\toprule
{} & {} & {} & \multicolumn{3}{c}{TPWD} & & \multicolumn{3}{c}{Post-Spectral$^{\bar G=2}$} & & \multicolumn{3}{c}{Post-Spectral$^{\bar G=3}$} & & \multicolumn{3}{c}{Post-Spectral$^{\bar G=4}$} & & \multicolumn{3}{c}{Post-Spectral$^{\bar G=10}$}  \\
\cmidrule{4-6}\cmidrule{8-10}\cmidrule{12-14}\cmidrule{16-18}\cmidrule{20-22} 
$G$ &$N$ & $T$ &  P & R &  RI && P &   R &  RI && P &   R &  RI && P &   R &  RI && P &   R &  RI \\
\midrule 
3&90&7& \cellcolor{mygreen!25} 0.970 & \cellcolor{mygreen!25} 0.642 & \cellcolor{mygreen!25} 0.877 & & 0.362 & 0.860 & 0.446 & & \cellcolor{mygreen!25} 0.516 & \cellcolor{mygreen!25} 0.744 & \cellcolor{mygreen!25} 0.670 & & 0.557 & 0.716 & 0.711 & & — & — & —   \\ 
& &10& \cellcolor{mygreen!25} 0.987 & \cellcolor{mygreen!25} 0.848 & \cellcolor{mygreen!25} 0.947 & & 0.453 & 0.891 & 0.579 & & \cellcolor{mygreen!25} 0.735 & \cellcolor{mygreen!25} 0.856 & \cellcolor{mygreen!25} 0.832 & & 0.790 & 0.863 & 0.865 & & 0.865 & 0.776 & 0.873  \\ 
& &20& \cellcolor{mygreen!25} 0.999 & \cellcolor{mygreen!25} 0.988 & \cellcolor{mygreen!25} 0.996 & & 0.521 & 0.915 & 0.686 & & \cellcolor{mygreen!25} 0.963 & \cellcolor{mygreen!25} 0.971 & \cellcolor{mygreen!25} 0.977 & & 0.979 & 0.974 & 0.984 & & 0.995 & 0.926 & 0.975  \\ 
& &40& \cellcolor{mygreen!25} 1.000 & \cellcolor{mygreen!25} 1.000 & \cellcolor{mygreen!25} 1.000 & & 0.569 & 0.954 & 0.747 & & \cellcolor{mygreen!25} 0.998 & \cellcolor{mygreen!25} 0.999 & \cellcolor{mygreen!25} 0.999 & & 1.000 & 0.989 & 0.996 & & 1.000 & 0.950 & 0.984  \\ 
\midrule
3&180&7& \cellcolor{mygreen!25} 0.977 & \cellcolor{mygreen!25} 0.538 & \cellcolor{mygreen!25} 0.843 & & 0.355 & 0.854 & 0.435 & & \cellcolor{mygreen!25} 0.514 & \cellcolor{mygreen!25} 0.748 & \cellcolor{mygreen!25} 0.667 & & 0.546 & 0.717 & 0.701 & & — & — & —   \\ 
& &10& \cellcolor{mygreen!25} 0.992 & \cellcolor{mygreen!25} 0.782 & \cellcolor{mygreen!25} 0.926 & & 0.445 & 0.898 & 0.559 & & \cellcolor{mygreen!25} 0.741 & \cellcolor{mygreen!25} 0.855 & \cellcolor{mygreen!25} 0.836 & & 0.816 & 0.872 & 0.881 & & 0.731 & 0.795 & 0.810  \\ 
& &20& \cellcolor{mygreen!25} 0.999 & \cellcolor{mygreen!25} 0.986 & \cellcolor{mygreen!25} 0.995 & & 0.519 & 0.917 & 0.682 & & \cellcolor{mygreen!25} 0.972 & \cellcolor{mygreen!25} 0.976 & \cellcolor{mygreen!25} 0.982 & & 0.982 & 0.978 & 0.987 & & 0.995 & 0.966 & 0.987  \\ 
& &40& \cellcolor{mygreen!25} 1.000 & \cellcolor{mygreen!25} 0.999 & \cellcolor{mygreen!25} 1.000 & & 0.571 & 0.959 & 0.746 & & \cellcolor{mygreen!25} 1.000 & \cellcolor{mygreen!25} 1.000 & \cellcolor{mygreen!25} 1.000 & & 1.000 & 0.997 & 0.999 & & 1.000 & 0.978 & 0.993  \\ 
\midrule
4&90&7& \cellcolor{mygreen!25} 0.667 & \cellcolor{mygreen!25} 0.620 & \cellcolor{mygreen!25} 0.831 & & 0.264 & 0.893 & 0.360 & & 0.336 & 0.756 & 0.552 & & \cellcolor{mygreen!25} 0.376 & \cellcolor{mygreen!25} 0.716 & \cellcolor{mygreen!25} 0.627 & & — & — & —  \\ 
& &10& \cellcolor{mygreen!25} 0.736 & \cellcolor{mygreen!25} 0.780 & \cellcolor{mygreen!25} 0.875 & & 0.322 & 0.903 & 0.493 & & 0.474 & 0.873 & 0.710 & & \cellcolor{mygreen!25} 0.514 & \cellcolor{mygreen!25} 0.868 & \cellcolor{mygreen!25} 0.751 & & 0.542 & 0.754 & 0.766   \\ 
& &20& \cellcolor{mygreen!25} 0.833 & \cellcolor{mygreen!25} 0.928 & \cellcolor{mygreen!25} 0.930 & & 0.355 & 0.899 & 0.574 & & 0.603 & 0.953 & 0.832 & & \cellcolor{mygreen!25} 0.627 & \cellcolor{mygreen!25} 0.967 & \cellcolor{mygreen!25} 0.851 & & 0.648 & 0.893 & 0.856   \\ 
& &40& \cellcolor{mygreen!25} 0.970 & \cellcolor{mygreen!25} 0.980 & \cellcolor{mygreen!25} 0.987 & & 0.378 & 0.931 & 0.612 & & 0.645 & 0.997 & 0.866 & & \cellcolor{mygreen!25} 0.655 & \cellcolor{mygreen!25} 0.972 & \cellcolor{mygreen!25} 0.869 & & 0.661 & 0.899 & 0.863   \\ 
\midrule
4&180&7& \cellcolor{mygreen!25} 0.761 & \cellcolor{mygreen!25} 0.604 & \cellcolor{mygreen!25} 0.739 & & 0.487 & 0.979 & 0.529 & & 0.532 & 0.847 & 0.593 & & \cellcolor{mygreen!25} 0.545 & \cellcolor{mygreen!25} 0.749 & \cellcolor{mygreen!25} 0.608 & & — & — & —  \\ 
& &10& \cellcolor{mygreen!25} 0.804 & \cellcolor{mygreen!25} 0.802 & \cellcolor{mygreen!25} 0.822 & & 0.528 & 0.978 & 0.596 & & 0.649 & 0.953 & 0.740 & & \cellcolor{mygreen!25} 0.676 & \cellcolor{mygreen!25} 0.957 & \cellcolor{mygreen!25} 0.771 & & 0.612 & 0.800 & 0.683   \\ 
& &20& \cellcolor{mygreen!25} 0.882 & \cellcolor{mygreen!25} 0.949 & \cellcolor{mygreen!25} 0.915 & & 0.551 & 0.947 & 0.632 & & 0.729 & 0.987 & 0.831 & & \cellcolor{mygreen!25} 0.736 & \cellcolor{mygreen!25} 0.989 & \cellcolor{mygreen!25} 0.838 & & 0.742 & 0.965 & 0.835   \\ 
& &40& \cellcolor{mygreen!25} 0.977 & \cellcolor{mygreen!25} 0.982 & \cellcolor{mygreen!25} 0.981 & & 0.563 & 0.958 & 0.652 & & 0.740 & 0.999 & 0.844 & & \cellcolor{mygreen!25} 0.737 & \cellcolor{mygreen!25} 0.968 & \cellcolor{mygreen!25} 0.833 & & 0.739 & 0.964 & 0.834   \\ 

\midrule

{} & {} & {} & \multicolumn{3}{c}{TPWD} & & \multicolumn{3}{c}{GFE$^{\bar G=2}$} & & \multicolumn{3}{c}{GFE$^{\bar G=3}$} & & \multicolumn{3}{c}{GFE$^{\bar G=4}$} & & \multicolumn{3}{c}{GFE$^{\bar G=10}$}  \\
\cmidrule{4-6}\cmidrule{8-10}\cmidrule{12-14}\cmidrule{16-18}\cmidrule{20-22} 
{} & {} & {}   &  P & R &  RI && P &   R &  RI && P &   R &  RI && P &   R &  RI && P &   R &  RI \\
\midrule 
3&90&7& \cellcolor{mygreen!25} 0.970 & \cellcolor{mygreen!25} 0.642 & \cellcolor{mygreen!25} 0.877 & & 0.571 & 0.919 & 0.749 & & \cellcolor{mygreen!25} 0.974 & \cellcolor{mygreen!25} 0.975 & \cellcolor{mygreen!25} 0.983 & & 0.973 & 0.807 & 0.930 & & 0.964 & 0.310 & 0.771   \\ 
& &10& \cellcolor{mygreen!25} 0.987 & \cellcolor{mygreen!25} 0.848 & \cellcolor{mygreen!25} 0.947 & & 0.578 & 0.946 & 0.758 & & \cellcolor{mygreen!25} 0.992 & \cellcolor{mygreen!25} 0.992 & \cellcolor{mygreen!25} 0.995 & & 0.991 & 0.827 & 0.941 & & 0.988 & 0.324 & 0.778  \\ 
& &20& \cellcolor{mygreen!25} 0.999 & \cellcolor{mygreen!25} 0.988 & \cellcolor{mygreen!25} 0.996 & & 0.589 & 0.990 & 0.772 & & \cellcolor{mygreen!25} 1.000 & \cellcolor{mygreen!25} 1.000 & \cellcolor{mygreen!25} 1.000 & & 1.000 & 0.835 & 0.946 & & 1.000 & 0.343 & 0.786  \\ 
& &40& \cellcolor{mygreen!25} 1.000 & \cellcolor{mygreen!25} 1.000 & \cellcolor{mygreen!25} 1.000 & & 0.592 & 0.999 & 0.775 & & \cellcolor{mygreen!25} 1.000 & \cellcolor{mygreen!25} 1.000 & \cellcolor{mygreen!25} 1.000 & & 1.000 & 0.838 & 0.947 & & 1.000 & 0.361 & 0.792  \\ 
\midrule
3&180&7& \cellcolor{mygreen!25} 0.977 & \cellcolor{mygreen!25} 0.538 & \cellcolor{mygreen!25} 0.843 & & 0.574 & 0.915 & 0.748 & & \cellcolor{mygreen!25} 0.976 & \cellcolor{mygreen!25} 0.977 & \cellcolor{mygreen!25} 0.984 & & 0.975 & 0.809 & 0.930 & & 0.969 & 0.305 & 0.768   \\ 
& &10& \cellcolor{mygreen!25} 0.992 & \cellcolor{mygreen!25} 0.782 & \cellcolor{mygreen!25} 0.926 & & 0.583 & 0.948 & 0.759 & & \cellcolor{mygreen!25} 0.993 & \cellcolor{mygreen!25} 0.993 & \cellcolor{mygreen!25} 0.995 & & 0.993 & 0.828 & 0.941 & & 0.990 & 0.315 & 0.773  \\ 
& &20& \cellcolor{mygreen!25} 0.999 & \cellcolor{mygreen!25} 0.986 & \cellcolor{mygreen!25} 0.995 & & 0.594 & 0.993 & 0.774 & & \cellcolor{mygreen!25} 1.000 & \cellcolor{mygreen!25} 1.000 & \cellcolor{mygreen!25} 1.000 & & 1.000 & 0.836 & 0.946 & & 1.000 & 0.329 & 0.779  \\ 
& &40& \cellcolor{mygreen!25} 1.000 & \cellcolor{mygreen!25} 0.999 & \cellcolor{mygreen!25} 1.000 & & 0.596 & 0.999 & 0.776 & & \cellcolor{mygreen!25} 1.000 & \cellcolor{mygreen!25} 1.000 & \cellcolor{mygreen!25} 1.000 & & 1.000 & 0.837 & 0.946 & & 1.000 & 0.344 & 0.784  \\ 
\midrule
4&90&7& \cellcolor{mygreen!25} 0.667 & \cellcolor{mygreen!25} 0.620 & \cellcolor{mygreen!25} 0.831 & & 0.384 & 0.959 & 0.618 & & 0.629 & 0.955 & 0.853 & & \cellcolor{mygreen!25} 0.734 & \cellcolor{mygreen!25} 0.753 & \cellcolor{mygreen!25} 0.874 & & 0.745 & 0.323 & 0.809  \\ 
& &10& \cellcolor{mygreen!25} 0.736 & \cellcolor{mygreen!25} 0.780 & \cellcolor{mygreen!25} 0.875 & & 0.385 & 0.971 & 0.618 & & 0.637 & 0.968 & 0.859 & & \cellcolor{mygreen!25} 0.833 & \cellcolor{mygreen!25} 0.844 & \cellcolor{mygreen!25} 0.921 & & 0.829 & 0.365 & 0.828   \\ 
& &20& \cellcolor{mygreen!25} 0.833 & \cellcolor{mygreen!25} 0.928 & \cellcolor{mygreen!25} 0.930 & & 0.386 & 0.994 & 0.616 & & 0.645 & 0.992 & 0.866 & & \cellcolor{mygreen!25} 0.945 & \cellcolor{mygreen!25} 0.948 & \cellcolor{mygreen!25} 0.974 & & 0.924 & 0.424 & 0.852   \\ 
& &40& \cellcolor{mygreen!25} 0.970 & \cellcolor{mygreen!25} 0.980 & \cellcolor{mygreen!25} 0.987 & & 0.386 & 0.999 & 0.616 & & 0.647 & 0.999 & 0.868 & & \cellcolor{mygreen!25} 0.992 & \cellcolor{mygreen!25} 0.992 & \cellcolor{mygreen!25} 0.996 & & 0.982 & 0.474 & 0.871   \\ 
\midrule
4&180&7& \cellcolor{mygreen!25} 0.761 & \cellcolor{mygreen!25} 0.604 & \cellcolor{mygreen!25} 0.739 & & 0.570 & 0.991 & 0.665 & & 0.729 & 0.925 & 0.815 & & \cellcolor{mygreen!25} 0.754 & \cellcolor{mygreen!25} 0.526 & \cellcolor{mygreen!25} 0.714 & & 0.789 & 0.184 & 0.617  \\ 
& &10& \cellcolor{mygreen!25} 0.804 & \cellcolor{mygreen!25} 0.802 & \cellcolor{mygreen!25} 0.822 & & 0.570 & 0.997 & 0.665 & & 0.734 & 0.950 & 0.826 & & \cellcolor{mygreen!25} 0.824 & \cellcolor{mygreen!25} 0.600 & \cellcolor{mygreen!25} 0.767 & & 0.848 & 0.198 & 0.629   \\ 
& &20& \cellcolor{mygreen!25} 0.882 & \cellcolor{mygreen!25} 0.949 & \cellcolor{mygreen!25} 0.915 & & 0.566 & 0.999 & 0.660 & & 0.739 & 0.992 & 0.841 & & \cellcolor{mygreen!25} 0.955 & \cellcolor{mygreen!25} 0.856 & \cellcolor{mygreen!25} 0.919 & & 0.940 & 0.224 & 0.650   \\ 
& &40& \cellcolor{mygreen!25} 0.977 & \cellcolor{mygreen!25} 0.982 & \cellcolor{mygreen!25} 0.981 & & 0.565 & 1.000 & 0.659 & & 0.740 & 1.000 & 0.844 & & \cellcolor{mygreen!25} 0.996 & \cellcolor{mygreen!25} 0.987 & \cellcolor{mygreen!25} 0.993 & & 0.989 & 0.244 & 0.664   \\ 
\bottomrule
\end{tabular}

	\begin{tablenotes}
		\footnotesize \item {\em Notes:} This table reports the Precision (P) rate, Recall (R) rate, and Rand Index (RI) for the triad pairwise-difference (TPWD) estimator with $\widehat\beta^1=0$, average linkage, and cut-off $c_{NT}=1.35\check\sigma_v\log(T)/\sqrt{\min(N,T)}$, for the post-spectral estimator proposed in \cite{ChetverikovManresa2021} with $\tilde\beta^0=\tilde\beta^1=0$, user-specified number of groups $g\in\set{2,3,4,10}$, and smallest $\lambda$ chosen in the grid  $\{1,1.5,2,2.5,\ldots\}$ such that $m(\lambda)\leq g$ (Post-Spectral$^{\bar G=g}$), and for \cite{BM2015}'s grouped fixed-effects estimator (their ``Algorithm 1'') with a user-specified number of groups $g\in\set{2,3,4,10}$ and $500$ random initialization points (GFE$^{\bar G=g}$). A green-shaded cell corresponds to a well-specified estimator. Results for Post-Spectral$^{\bar G=10}$ are missing for $T=7$ since Post-Spectral$^{ \bar G=g}$ is not properly defined if $g>T$ as it requires computing the $g$ largest eigenvectors of a $T\times T$ matrix. Results are averaged across 500 Monte Carlo samples. 
	\end{tablenotes}
	\end{threeparttable}
	\end{adjustbox}
\end{table}

The results of Table~\ref{tab:mcsim_pureGFE_tab1} are best understood by examining the performance of group membership estimates. Table~\ref{tab:mcsim_pureGFE_tab2} displays the average Precision rate, Recall rate, and Rand index of all reported estimators of group memberships.  For all user-specified $g\in\{2,3,4,10\}$, the performance of Post-Spectral$^{\bar G=g}$ in terms of Precision rate and Rand index is dominated by that of TPWD across all values of $(G,N,T)$. The Rand index of TPWD is most often slightly dominated by but within a $0.141$ distance of that of the well-specified GFE estimator across all values of $(G,N,T)$. This is not the case for post-spectral estimators. For $G=4$, post-spectral estimators have a better Recall rate than TPWD, which is partly explained by the fact that they take the number of groups as input and contrain the estimated number of groups to be less or equal to it. This is also the case for GFE$^{\bar G=g}$ with $g\leq 4$. Similarly, misspecified GFE estimators with $g> G$ almost always achieve the best Precision rate, as they allow for finer clustering at the expense of lower recall. TPWD provides more accurate clustering in terms of the Rand Index by achieving a better balance between precision and recall, while simultaneously determining the number of groups endogenously as a by-product. Although it may occasionally identify a slightly larger number of groups (leading to lower recall), the units within these groups tend to share the same true group membership (resulting in higher precision). As shown in Table~\ref{tab:mcsim_pureGFE_tab1}, this trade-off is sufficient to yield a lower RMSE.

\subsection{Full grouped fixed effects model}\label{sec:mc_sim_full}
Consider now a  full GFE model with a scalar covariate:  
$$y_{it}=x_{it}\beta+\alpha_{g_it}+v_{it}, \quad i=1,\ldots,N, \; t=1,\ldots,T,$$
where $\beta=1$, $x_{it}=0.5\alpha_{g_it}+u_{it}$, $u_{it}\overset{iid}{\sim}\mathcal N(0,1/(2\sqrt{3})^2)$, $v_{it}\overset{iid}{\sim}\mathcal N(0,(1/3)^2)$, and $u_{it}$ and $v_{it}$ are mutually independent across units and time periods. In the dataset from the empirical application and \cite{BM2015}'s calibrated simulations, the lagged income covariate has variance 1.152, resulting in a signal-to-noise ratio of 10.368. Here, the variance of the covariate is less than 0.1, which for $(N,T)=(90,7)$ leads to an average signal-to-noise ratio for the slope coefficient of 
$$9\left(\frac{0.25}{T}\sum_{t=1}^T\frac1G\sum_{g=1}^G\widehat\pi_g(1-\widehat\pi_g)\alpha_{gt}^2+\frac{1}{12}\right)=0.9769\mathds1\{G=3\}+0.9218\mathds1\{G=4\},$$ 
i.e., more than ten times smaller. The level of correlation between the covariate and the fixed effect is similar in both papers.

The performance obtained after one (TPWD) or four (Iterated TPWD) iterations of the TPWD estimator which uses $\widehat\beta^1(\psi_{NT})$ -- the NNR estimator described in Section~\ref{sec:prelim_con_est} with $\psi_{NT}=\log(\log(T))/\sqrt{16\min(N,T)}$ -- and the data-driven thresholding rule $c_{NT}=1.35\check\sigma_v\log(T)/\sqrt{\min(N,T)}$  is compared with that of the NNR and  NN estimators proposed in \cite{moon2019nuclear}, the well-specified spectral and post-spectral estimators proposed in \cite{ChetverikovManresa2021}, the GFE estimator (with or without BIC selection of $G$) proposed in \cite{BM2015}, the well-specified IFE estimator proposed in \cite{Bai2009}, initialized at NNR or random draws, and the infeasible pooled OLS regression that uses the true group memberships  (Oracle). I refer to the original papers, table notes, and the code made available online for details regarding the implementation of each estimator. For each estimator, the bias and root mean square error (RMSE) of the estimated slope coefficient are reported:
\begin{align*}
	{\rm Bias}(\widehat\beta)=\frac{1}{500}\sum_{b=1}^{500}\widehat\beta^{(b)}-\beta \quad \text{ and } \quad {\rm RMSE}(\widehat\beta)&=\sqrt{\frac{1}{500}\sum_{b=1}^{500}(\widehat\beta^{(b)}-\beta)^2}.
\end{align*}
When applicable, the coverage rate of a $95\%$-level confidence interval for $\beta$ based on a consistent estimator of the large-$N$, large-$T$ asymptotic variance clustered at the unit level, the RMSE of grouped fixed effect estimates, the estimated number of groups, the Precision and Recall rates, and the Rand index of the estimated clustering  are reported.
\begin{landscape}
\begin{table}[H]
	\centering
		\caption{Finite sample properties in the full GFE model (slope coefficient and grouped fixed effects)}
	\label{tab:mcsim_fullGFE_tab3}
	\begin{adjustbox}{max width={0.8\linewidth},center}
	\begin{threeparttable}
\begin{minipage}{\textwidth}
\begin{tabular}{lll *{31}{S[table-format=-1.3]}}
\toprule
{} & {} & {} & \multicolumn{5}{c}{TPWD} & & \multicolumn{5}{c}{Iterated TPWD} & & \multicolumn{2}{c}{NNR} & &  \multicolumn{2}{c}{NN} & &  \multicolumn{2}{c}{Spectral} &&  \multicolumn{5}{c}{Post-Spectral} & & \multicolumn{4}{c}{GFE}  \\
\cmidrule{4-8}\cmidrule{10-14}\cmidrule{16-17}\cmidrule{19-20}\cmidrule{22-23}\cmidrule{25-29}\cmidrule{31-34} 
$G$ & $N$ & $T$ & \multicolumn{1}{c}{Bias $\widehat\beta$} & \multicolumn{1}{c}{RMSE $\widehat\beta$}  & \multicolumn{1}{c}{$.95$ $\widehat\beta$}   & \multicolumn{1}{c}{RMSE $\widehat\alpha$} & \multicolumn{1}{c}{$\widehat G$} & & \multicolumn{1}{c}{Bias $\widehat\beta$} & \multicolumn{1}{c}{RMSE $\widehat\beta$} & \multicolumn{1}{c}{$.95$ $\widehat\beta$} & \multicolumn{1}{c}{RMSE $\widehat\alpha$} & \multicolumn{1}{c}{$\widehat G$} & & \multicolumn{1}{c}{Bias $\widehat\beta$} & \multicolumn{1}{c}{RMSE $\widehat\beta$} & &  \multicolumn{1}{c}{Bias $\widehat\beta$} & \multicolumn{1}{c}{RMSE $\widehat\beta$} & &  \multicolumn{1}{c}{Bias $\widehat\beta$} & \multicolumn{1}{c}{RMSE $\widehat\beta$} & &  \multicolumn{1}{c}{Bias $\widehat\beta$}& \multicolumn{1}{c}{RMSE $\widehat\beta$} & \multicolumn{1}{c}{$.95$ $\widehat\beta$}  & \multicolumn{1}{c}{RMSE $\widehat\alpha$} & \multicolumn{1}{c}{$\widehat G$} & & \multicolumn{1}{c}{Bias $\widehat\beta$} & \multicolumn{1}{c}{RMSE $\widehat\beta$} & \multicolumn{1}{c}{$.95$ $\widehat\beta$} & \multicolumn{1}{c}{RMSE $\widehat\alpha$}  \\ 
\midrule 
3&90&7& 0.351 & 0.365 & 0.004 & 0.289 & 4.408 & & 0.028 & 0.068 & 0.808 & 0.154 & 6.500 & & 0.732 & 0.739 & & 0.736 & 0.739 & & 0.054 & 0.238 & & 0.447 & 0.489 & 0.034 & 0.351 & 2.928 & & 0.013 & 0.053 & 0.908 & 0.091  \\ 
& &10& 0.207 & 0.233 & 0.142 & 0.213 & 3.232 & & 0.014 & 0.050 & 0.864 & 0.109 & 4.692 & & 0.598 & 0.604 & & 0.601 & 0.603 & & 0.017 & 0.092 & & 0.242 & 0.293 & 0.228 & 0.242 & 2.786 & & 0.005 & 0.042 & 0.914 & 0.070  \\ 
& &20& 0.020 & 0.043 & 0.876 & 0.083 & 3.028 & & 0.001 & 0.028 & 0.932 & 0.067 & 3.322 & & 0.417 & 0.418 & & 0.407 & 0.408 & & 0.005 & 0.042 & & 0.041 & 0.095 & 0.778 & 0.101 & 2.986 & & 0.000 & 0.028 & 0.930 & 0.061  \\ 
& &40& -0.000 & 0.019 & 0.962 & 0.062 & 3.002 & & -0.001 & 0.019 & 0.964 & 0.061 & 3.018 & & 0.312 & 0.318 & & 0.283 & 0.284 & & 0.001 & 0.025 & & -0.001 & 0.019 & 0.964 & 0.061 & 3.000 & & -0.001 & 0.019 & 0.964 & 0.061  \\ 
\midrule
3&180&7& 0.332 & 0.340 & 0.000 & 0.278 & 6.966 & & 0.020 & 0.044 & 0.830 & 0.148 & 9.010 & & 0.729 & 0.734 & & 0.732 & 0.734 & & 0.048 & 0.134 & & 0.428 & 0.456 & 0.000 & 0.342 & 2.744 & & 0.013 & 0.038 & 0.908 & 0.079  \\ 
& &10& 0.176 & 0.188 & 0.022 & 0.194 & 3.844 & & 0.009 & 0.032 & 0.888 & 0.100 & 5.906 & & 0.591 & 0.601 & & 0.598 & 0.599 & & 0.018 & 0.061 & & 0.205 & 0.244 & 0.130 & 0.221 & 2.844 & & 0.004 & 0.028 & 0.928 & 0.055  \\ 
& &20& 0.015 & 0.029 & 0.892 & 0.068 & 3.084 & & -0.000 & 0.020 & 0.946 & 0.051 & 3.664 & & 0.399 & 0.400 & & 0.397 & 0.398 & & 0.005 & 0.028 & & 0.020 & 0.060 & 0.860 & 0.072 & 2.998 & & -0.001 & 0.019 & 0.948 & 0.043  \\ 
& &40& -0.000 & 0.015 & 0.924 & 0.044 & 3.012 & & -0.001 & 0.015 & 0.918 & 0.044 & 3.060 & & 0.289 & 0.301 & & 0.274 & 0.274 & & 0.001 & 0.018 & & -0.001 & 0.015 & 0.920 & 0.043 & 3.000 & & -0.001 & 0.015 & 0.920 & 0.043  \\ 
\midrule
4&90&7& 0.364 & 0.377 & 0.002 & 0.293 & 4.230 & & 0.058 & 0.090 & 0.662 & 0.170 & 6.154 & & 0.726 & 0.732 & & 0.729 & 0.732 & & -0.678 & 10.513 & & 0.323 & 0.379 & 0.180 & 0.297 & 3.544 & & 0.038 & 0.067 & 0.808 & 0.149  \\ 
& &10& 0.258 & 0.275 & 0.034 & 0.239 & 3.078 & & 0.061 & 0.086 & 0.656 & 0.147 & 4.636 & & 0.598 & 0.604 & & 0.601 & 0.603 & & 0.038 & 0.154 & & 0.306 & 0.365 & 0.198 & 0.269 & 3.236 & & 0.026 & 0.050 & 0.858 & 0.120  \\ 
& &20& 0.115 & 0.134 & 0.178 & 0.157 & 2.970 & & 0.042 & 0.063 & 0.646 & 0.113 & 3.652 & & 0.412 & 0.423 & & 0.409 & 0.410 & & 0.013 & 0.049 & & 0.079 & 0.103 & 0.390 & 0.136 & 3.242 & & 0.007 & 0.030 & 0.930 & 0.086  \\ 
& &40& 0.073 & 0.080 & 0.128 & 0.127 & 3.008 & & 0.014 & 0.034 & 0.832 & 0.083 & 3.866 & & 0.318 & 0.319 & & 0.288 & 0.289 & & 0.008 & 0.027 & & 0.060 & 0.064 & 0.176 & 0.121 & 3.462 & & 0.003 & 0.020 & 0.934 & 0.073  \\ 
\midrule
4&180&7& 0.246 & 0.253 & 0.000 & 0.232 & 5.994 & & 0.031 & 0.051 & 0.758 & 0.153 & 8.388 & & 0.657 & 0.658 & & 0.657 & 0.658 & & -0.527 & 12.230 & & 0.285 & 0.311 & 0.118 & 0.260 & 3.508 & & 0.036 & 0.054 & 0.748 & 0.149  \\ 
& &10& 0.148 & 0.159 & 0.034 & 0.176 & 3.478 & & 0.038 & 0.051 & 0.690 & 0.123 & 5.114 & & 0.528 & 0.529 & & 0.528 & 0.529 & & 0.035 & 0.095 & & 0.190 & 0.241 & 0.292 & 0.198 & 3.300 & & 0.030 & 0.044 & 0.768 & 0.126  \\ 
& &20& 0.079 & 0.093 & 0.210 & 0.126 & 2.950 & & 0.027 & 0.040 & 0.666 & 0.089 & 3.820 & & 0.356 & 0.357 & & 0.354 & 0.355 & & 0.010 & 0.032 & & 0.054 & 0.069 & 0.366 & 0.108 & 3.160 & & 0.010 & 0.023 & 0.884 & 0.079  \\ 
& &40& 0.050 & 0.056 & 0.146 & 0.102 & 3.060 & & 0.007 & 0.021 & 0.864 & 0.060 & 3.924 & & 0.259 & 0.272 & & 0.246 & 0.246 & & 0.006 & 0.018 & & 0.048 & 0.050 & 0.100 & 0.102 & 3.634 & & 0.002 & 0.014 & 0.946 & 0.054  \\ 
\bottomrule
\end{tabular}

\end{minipage}

\vspace{0.5cm}

\begin{minipage}{\textwidth}
\begin{tabular}{lll *{26}{S[table-format=-1.3]}} 
\toprule
{} & {} & {} & \multicolumn{5}{c}{GFE with BIC selection} & & \multicolumn{3}{c}{IFE R} && \multicolumn{3}{c}{IFE R-BC} && \multicolumn{3}{c}{IFE NNR} && \multicolumn{3}{c}{IFE NNR-BC}& &  \multicolumn{4}{c}{Oracle}  \\
\cmidrule{4-8}\cmidrule{10-12}\cmidrule{14-16}\cmidrule{18-20}\cmidrule{22-24}\cmidrule{26-29} 
$G$ & $N$ & $T$ & \multicolumn{1}{c}{Bias $\widehat\beta$} & \multicolumn{1}{c}{RMSE $\widehat\beta$} & \multicolumn{1}{c}{$.95$ $\widehat\beta$} & \multicolumn{1}{c}{RMSE $\widehat\alpha$} & \multicolumn{1}{c}{$\widehat G$} & & \multicolumn{1}{c}{Bias $\widehat\beta$} & \multicolumn{1}{c}{RMSE $\widehat\beta$} & \multicolumn{1}{c}{$.95$ $\widehat\beta$} & &  \multicolumn{1}{c}{Bias $\widehat\beta$} & \multicolumn{1}{c}{RMSE $\widehat\beta$} & \multicolumn{1}{c}{$.95$ $\widehat\beta$} & & \multicolumn{1}{c}{Bias $\widehat\beta$} & \multicolumn{1}{c}{RMSE $\widehat\beta$} & \multicolumn{1}{c}{$.95$ $\widehat\beta$}  & & \multicolumn{1}{c}{Bias $\widehat\beta$} & \multicolumn{1}{c}{RMSE $\widehat\beta$} & \multicolumn{1}{c}{$.95$ $\widehat\beta$} & & \multicolumn{1}{c}{Bias $\widehat\beta$} & \multicolumn{1}{c}{RMSE $\widehat\beta$} & \multicolumn{1}{c}{$.95$ $\widehat\beta$} & \multicolumn{1}{c}{RMSE $\widehat\alpha$}  \\ 
\midrule 
3&90&7& 0.015 & 0.056 & 0.882 & 0.108 & 3.404 & & 0.005 & 0.068 & 0.816 & & 0.005 & 0.069 & 0.790 & & 0.005 & 0.067 & 0.818 & & 0.005 & 0.069 & 0.790 & & 0.000 & 0.048 & 0.934 & 0.062  \\ 
& &10& 0.005 & 0.042 & 0.914 & 0.071 & 3.008 & & 0.005 & 0.052 & 0.844 & & 0.006 & 0.053 & 0.832 & & 0.005 & 0.052 & 0.844 & & 0.006 & 0.053 & 0.832 & & 0.001 & 0.041 & 0.932 & 0.062  \\ 
& &20& 0.000 & 0.028 & 0.930 & 0.061 & 3.000 & & 0.003 & 0.032 & 0.904 & & 0.003 & 0.032 & 0.896 & & 0.003 & 0.032 & 0.904 & & 0.003 & 0.032 & 0.896 & & 0.000 & 0.028 & 0.930 & 0.061  \\ 
& &40& -0.001 & 0.019 & 0.964 & 0.061 & 3.000 & & 0.000 & 0.020 & 0.922 & & 0.000 & 0.020 & 0.922 & & 0.000 & 0.020 & 0.922 & & 0.000 & 0.020 & 0.922 & & -0.001 & 0.019 & 0.964 & 0.061  \\ 
\midrule
3&180&7& 0.017 & 0.042 & 0.864 & 0.140 & 5.000 & & 0.002 & 0.044 & 0.862 & & 0.003 & 0.045 & 0.844 & & 0.002 & 0.044 & 0.862 & & 0.003 & 0.045 & 0.844 & & 0.001 & 0.032 & 0.954 & 0.044  \\ 
& &10& 0.005 & 0.029 & 0.912 & 0.087 & 3.914 & & 0.004 & 0.034 & 0.874 & & 0.004 & 0.035 & 0.860 & & 0.004 & 0.034 & 0.874 & & 0.004 & 0.035 & 0.860 & & 0.001 & 0.028 & 0.930 & 0.044  \\ 
& &20& -0.001 & 0.019 & 0.948 & 0.043 & 3.000 & & -0.000 & 0.022 & 0.912 & & -0.000 & 0.022 & 0.912 & & -0.000 & 0.022 & 0.912 & & -0.000 & 0.022 & 0.912 & & -0.001 & 0.019 & 0.946 & 0.043  \\ 
& &40& -0.001 & 0.015 & 0.920 & 0.043 & 3.000 & & -0.000 & 0.016 & 0.904 & & -0.000 & 0.016 & 0.906 & & -0.000 & 0.016 & 0.904 & & -0.000 & 0.016 & 0.906 & & -0.001 & 0.015 & 0.920 & 0.043  \\ 
\midrule
4&90&7& 0.038 & 0.067 & 0.802 & 0.150 & 4.052 & & 0.012 & 0.079 & 0.758 & & 0.013 & 0.081 & 0.754 & & 0.012 & 0.079 & 0.760 & & 0.014 & 0.081 & 0.756 & & 0.000 & 0.046 & 0.948 & 0.071  \\ 
& &10& 0.029 & 0.053 & 0.836 & 0.120 & 3.880 & & 0.020 & 0.058 & 0.796 & & 0.021 & 0.059 & 0.776 & & 0.020 & 0.058 & 0.796 & & 0.021 & 0.059 & 0.776 & & 0.002 & 0.037 & 0.956 & 0.071  \\ 
& &20& 0.018 & 0.042 & 0.804 & 0.093 & 3.752 & & 0.012 & 0.036 & 0.852 & & 0.013 & 0.036 & 0.846 & & 0.012 & 0.036 & 0.852 & & 0.013 & 0.036 & 0.846 & & 0.001 & 0.028 & 0.938 & 0.070  \\ 
& &40& 0.021 & 0.041 & 0.678 & 0.087 & 3.676 & & 0.009 & 0.023 & 0.894 & & 0.010 & 0.024 & 0.890 & & 0.009 & 0.023 & 0.894 & & 0.010 & 0.024 & 0.890 & & 0.002 & 0.020 & 0.936 & 0.071  \\ 
\midrule
4&180&7& 0.033 & 0.052 & 0.748 & 0.167 & 5.000 & & 0.008 & 0.053 & 0.806 & & 0.007 & 0.054 & 0.780 & & 0.008 & 0.053 & 0.806 & & 0.007 & 0.054 & 0.780 & & -0.000 & 0.032 & 0.954 & 0.050  \\ 
& &10& 0.026 & 0.041 & 0.794 & 0.143 & 5.000 & & 0.010 & 0.039 & 0.838 & & 0.012 & 0.039 & 0.822 & & 0.010 & 0.039 & 0.838 & & 0.012 & 0.039 & 0.822 & & 0.000 & 0.027 & 0.948 & 0.050  \\ 
& &20& 0.010 & 0.023 & 0.888 & 0.080 & 4.028 & & 0.006 & 0.023 & 0.892 & & 0.007 & 0.024 & 0.872 & & 0.006 & 0.023 & 0.892 & & 0.007 & 0.024 & 0.872 & & 0.000 & 0.019 & 0.948 & 0.050  \\ 
& &40& 0.002 & 0.014 & 0.946 & 0.054 & 4.000 & & 0.006 & 0.016 & 0.900 & & 0.006 & 0.016 & 0.890 & & 0.006 & 0.016 & 0.900 & & 0.006 & 0.016 & 0.890 & & 0.001 & 0.014 & 0.952 & 0.050  \\ 
\bottomrule
\end{tabular}

\end{minipage}
	\begin{tablenotes}
		\footnotesize \item {\em Notes:} This table reports the bias and root mean square error (RMSE) of each estimator of $\beta$. When relevant, it also reports the coverage rate of a $95\%$ confidence interval based on large-$N$, large-$T$ consistent estimates of analytical standard errors ($.95$ $\widehat\beta$), the RMSE of the grouped fixed-effects estimator, and the estimated number of groups ($\widehat G$). The triad pairwise-difference (TPWD) estimator is computed with $\widehat\beta^1$ set to the nuclear-norm regularized (NNR) estimator with $\psi_{NT}=\log(\log(T))/\sqrt{16\min(N,T)}$, average linkage, and cut-off $c_{NT}=1.35\check\sigma_v\log(T)/\sqrt{\min(N,T)}$. Iterated TPWD is computed by iterating TPWD $4$ times, replacing 3 times the preliminary estimator $\widehat\beta^1$ by the TPWD estimate obtained at the previous iteration. NNR is obtained by concentrating out the optimization with respect to the unobserved effects and solving the convex optimization problem with MATLAB \texttt{fminsearch} routine. The nuclear norm (NN) estimator is obtained by solving a convex optimization problem with MATLAB \texttt{fminsearch} routine. The Spectral estimator is implemented as in \cite{ChetverikovManresa2021}, with user-specified number of groups for the outcome and regressor equations set to the true number of groups. The Post-Spectral estimator is implemented as in \cite{ChetverikovManresa2021}, with user-specified number of groups set to the true number of groups and smallest $\lambda$ chosen in the grid  $\{1,1.5,2,2.5,\ldots\}$ such that $m(\lambda)\leq G$. The grouped fixed-effects (GFE) estimator proposed in \cite{BM2015} is implemented as in \cite{BM2015}'s  Algorithm 1, with a user-specified number of groups set to the true number of groups and $100$ random initialization points following the method described in Section~S.1.1 \cite{BM2015}'s Supplementary Material with $\theta^{(0)}\sim\mathcal N(0,1)$. The GFE with BIC selection of the number of groups is implemented as in Section~S.3.2 of \cite{BM2015}, with $G_{max}=5$. \cite{Bai2009}'s interactive fixed effects (IFE) estimator is implemented using $100$ random initialization points (IFE R) or taking one optimization step starting from NNR (IFE NNR) and applying bias-correction based on large-$N$, large-$T$ approximations (IFE R-BC, IFE NNR-BC). The infeasible (Oracle) estimator is obtained from a pooled OLS regression of the outcome on the covariates controlling for the interactions of time and ``true'' group dummies. Results are averaged across 500 Monte Carlo samples. 
	\end{tablenotes}
	\end{threeparttable}
	\end{adjustbox}
\end{table}
\end{landscape}

Table~\ref{tab:mcsim_fullGFE_tab3} shows the results in terms of bias, RMSE, estimated number of groups, and coverage rates. These findings confirm the theoretical guarantees stated in Proposition~\ref{prop:sup_norm_cons} and Corollary~\ref{cor:AN}. As the sample size increases, both the TPWD and iterated TPWD estimators exhibit consistency, and the coverage rates of the confidence intervals -- constructed using a consistent estimator of the variance of their asymptotic normal distribution -- converge to their nominal levels. Remarkably, for small values of $T$, four iterations significantly reduce bias and improve coverage. As noted in \cite{moon2019nuclear} and \cite{ChetverikovManresa2021}, the convergence of the NNR and NN estimators is indeed slow. Yet, contrary to what was conjectured in \cite{ChetverikovManresa2021}, this does not prevent the iterated TPWD estimator to reach near-Oracle performance even for small values of $T$ for which post-spectral estimators perform quite poorly. Remarkably, the iterated TPWD estimator uniformly dominates the NNR, NN, and well-specified post-spectral estimators across all values of $(G,N,T)$, in terms of all metrics. While its bias is sometimes slightly greater -- and often slightly smaller -- than that of the spectral estimator (which relies on the factor structure assumption for the covariates), the difference remains within a margin of 0.03. However, its RMSE is uniformly smaller across all values of \( (G, N, T) \) except for $G=4$ and $T\in\{20,40\}$. In the following, the focus is exclusively on the iterated TPWD, which will be referred to simply as TPWD.

For $(G,N,T)=(3,90,7)$, TPWD has a bias of $0.028$ and its coverage rate already peaks at $80.8\%$. The bias of the well-specified post-spectral estimator is $0.447$ and its coverage rate is $3.4\%$. The well-specified GFE estimator has a bias of $0.013$ and coverage rate of $90.8\%$. The infeasible oracle estimator has a bias less than $10^{-3}$ and displays a coverage rate of $93.4\%$. The NNR and NN estimator have biases of $0.732$ and $0.736$ respectively. As $T$ increases, the performance of each estimator steadily improves, and estimates of the number of groups converge to the true number of groups. For $(G,N,T)=(3,90,20)$, TPWD has a bias of $0.001$ and coverage rate of $93.2\%$. The bias of the well-specified post-spectral estimator is $0.041$ and its coverage rate is $77.8\%$. Both the well-specified GFE and Oracle estimator have a bias less than $10^{-3}$ and coverage rate of $93\%$. The NNR and NN estimator have biases of $0.417$ and $0.418$ respectively. As $N$ increases, the discrepancy in the performance of the TPWD and post-spectral estimators remains stable. For $(G,N,T)=(3,180,10)$,  TPWD has $88.8\%$ coverage  to contrast with $13\%$ and $92.8\%$ for the post-spectral and GFE estimators respectively. For $(G,N,T)=(3,180,20)$,  TPWD has $94.6\%$ coverage  to contrast with $86\%$ and $94.8\%$, respectively. 

While the coverage rate converges to the prescribed level for all estimators as $N$ and $T$ reach their maximum value if $G=3$, this is not the case for the post-spectral estimator if $G=4$. For $(G,N,T)=(4,90,7)$, TPWD has $66.2\%$ coverage  to contrast with $18\%$ and $80.8\%$ for the post-spectral and GFE estimators respectively. For $(G,N,T)=(4,90,40)$,  TPWD has $91.4\%$ coverage  to contrast with $27\%$ and $93.4\%$, respectively. For $(G,N,T)=(4,180,7)$,  TPWD has $75.8\%$ coverage to contrast with $11.8\%$ and $74.8\%$, respectively. For $(G,N,T)=(4,180,40)$, TPWD has  $86.4\%$  coverage to contrast with $10\%$ and $94.6\%$, respectively. 

In line with Corollary~1 in \cite{Bai2009}, in this setting with i.i.d.~errors and homoskedasticity, the performance of the well-specified IFE estimators is comparable to that of the GFE estimator in terms of bias, even for small values of $T$, but the former displays systematic under-coverage in comparison with the latter. Similarly, the BIC selection approach of \cite{BM2015} performs quite well, as conjectured by the authors. These three approaches, however, are not immune to misspecifying $G$ or the upper bound $G_{\max}$.
 \begin{table}[H]
	\centering
	\caption{Finite sample properties in the full GFE model (group memberships)}
	\label{tab:mcsim_fullGFE_tab4}
	\begin{adjustbox}{max width={0.99\linewidth},center}
	\begin{threeparttable}
\begin{tabular}{lll *{19}{S[table-format=-1.3]}} 
\toprule
{} & {} & {} & \multicolumn{3}{c}{TPWD} & & \multicolumn{3}{c}{Iterated TPWD} & & \multicolumn{3}{c}{Post-Spectral} & & \multicolumn{3}{c}{GFE} & & \multicolumn{3}{c}{GFE with BIC selection} \\
\cmidrule{4-6}\cmidrule{8-10}\cmidrule{12-14}\cmidrule{16-18}\cmidrule{20-22}
$G$ &$N$ & $T$ &  P & R & RI && P & R & RI && P & R & RI && P & R & RI && P & R & RI  \\
\midrule 
3&90&7& 0.620 & 0.636 & 0.745 & & 0.960 & 0.646 & 0.875 & & 0.485 & 0.781 & 0.613 & & 0.972 & 0.973 & 0.982 & & 0.972 & 0.904 & 0.960   \\ 
& &10& 0.758 & 0.807 & 0.845 & & 0.983 & 0.847 & 0.945 & & 0.677 & 0.841 & 0.791 & & 0.992 & 0.992 & 0.995 & & 0.992 & 0.991 & 0.994   \\ 
& &20& 0.977 & 0.978 & 0.985 & & 0.999 & 0.989 & 0.996 & & 0.944 & 0.962 & 0.965 & & 1.000 & 1.000 & 1.000 & & 1.000 & 1.000 & 1.000   \\ 
& &40& 0.999 & 0.999 & 1.000 & & 1.000 & 0.999 & 1.000 & & 1.000 & 1.000 & 1.000 & & 1.000 & 1.000 & 1.000 & & 1.000 & 1.000 & 1.000   \\ 
\midrule
3&180&7& 0.657 & 0.536 & 0.749 & & 0.973 & 0.552 & 0.847 & & 0.488 & 0.777 & 0.624 & & 0.974 & 0.975 & 0.983 & & 0.971 & 0.647 & 0.878   \\ 
& &10& 0.802 & 0.783 & 0.863 & & 0.990 & 0.789 & 0.928 & & 0.719 & 0.842 & 0.821 & & 0.992 & 0.992 & 0.995 & & 0.992 & 0.842 & 0.946   \\ 
& &20& 0.982 & 0.981 & 0.988 & & 1.000 & 0.986 & 0.995 & & 0.970 & 0.976 & 0.981 & & 1.000 & 1.000 & 1.000 & & 1.000 & 1.000 & 1.000   \\ 
& &40& 0.999 & 0.999 & 1.000 & & 1.000 & 0.999 & 1.000 & & 1.000 & 1.000 & 1.000 & & 1.000 & 1.000 & 1.000 & & 1.000 & 1.000 & 1.000   \\ 
\midrule
4&90&7& 0.408 & 0.642 & 0.675 & & 0.644 & 0.656 & 0.825 & & 0.397 & 0.783 & 0.627 & & 0.726 & 0.745 & 0.870 & & 0.727 & 0.741 & 0.869   \\ 
& &10& 0.467 & 0.802 & 0.713 & & 0.695 & 0.792 & 0.859 & & 0.443 & 0.839 & 0.671 & & 0.826 & 0.838 & 0.918 & & 0.810 & 0.858 & 0.914   \\ 
& &20& 0.592 & 0.928 & 0.820 & & 0.774 & 0.931 & 0.907 & & 0.624 & 0.960 & 0.848 & & 0.940 & 0.943 & 0.972 & & 0.872 & 0.960 & 0.948   \\ 
& &40& 0.641 & 0.987 & 0.863 & & 0.931 & 0.981 & 0.972 & & 0.655 & 0.973 & 0.869 & & 0.992 & 0.993 & 0.996 & & 0.882 & 0.996 & 0.955   \\ 
\midrule
4&180&7& 0.621 & 0.663 & 0.667 & & 0.756 & 0.581 & 0.730 & & 0.567 & 0.855 & 0.636 & & 0.751 & 0.520 & 0.711 & & 0.763 & 0.376 & 0.672   \\ 
& &10& 0.656 & 0.889 & 0.738 & & 0.789 & 0.816 & 0.818 & & 0.651 & 0.925 & 0.738 & & 0.816 & 0.591 & 0.761 & & 0.830 & 0.416 & 0.704   \\ 
& &20& 0.699 & 0.976 & 0.796 & & 0.851 & 0.953 & 0.897 & & 0.733 & 0.988 & 0.835 & & 0.952 & 0.853 & 0.917 & & 0.952 & 0.843 & 0.912   \\ 
& &40& 0.748 & 0.994 & 0.845 & & 0.964 & 0.984 & 0.973 & & 0.737 & 0.972 & 0.834 & & 0.995 & 0.987 & 0.992 & & 0.995 & 0.987 & 0.992   \\ 
\bottomrule
\end{tabular}
\begin{tablenotes}
	\footnotesize \item {\em Notes:} This table reports the precision (P) rate, recall (R) rate, and Rand index (RI) for each estimator. The triad pairwise-difference (TPWD) estimator is computed with $\widehat\beta^1$ set to the nuclear-norm regularized (NNR) estimator with $\psi_{NT}=\log(\log(T))/\sqrt{16\min(N,T)}$, average linkage, and cut-off $c_{NT}=1.35\check\sigma_v\log(T)/\sqrt{\min(N,T)}$. Iterated TPWD is computed by iterating TPWD $4$ times, replacing 3 times the preliminary estimator $\widehat\beta^1$ by the TPWD estimate obtained at the previous iteration. The Post-Spectral estimator is implemented as in \cite{ChetverikovManresa2021}, with user-specified number of groups set to the true number of groups and smallest $\lambda$ chosen in the grid  $\{1,1.5,2,2.5,\ldots\}$ such that $m(\lambda)\leq G$. The grouped fixed-effects (GFE) estimator proposed in \cite{BM2015} is implemented as in \cite{BM2015}'s  Algorithm 1, with a user-specified number of groups set to the true number of groups and $100$ random initialization points following the method described in Section~S.1.1 \cite{BM2015}'s Supplementary Material with $\theta^{(0)}\sim\mathcal N(0,1)$. The GFE with BIC selection of the number of groups is implemented as in Section~S.3.2 of \cite{BM2015}, with $G_{max}=5$. Results are averaged across 500 Monte Carlo samples. 

			\end{tablenotes}
		\end{threeparttable}
	\end{adjustbox}
\end{table}

Table~\ref{tab:mcsim_fullGFE_tab4} reports clustering accuracy metrics. The three measures are little affected by the introduction of a single covariate and the slow rate of convergence of the preliminary consistent estimator, even for small values of $T$  (compare with Table~\ref{tab:mcsim_pureGFE_tab2}).

In summary, given that I am not aware of any theoretical result that would guarantee consistency of the BIC selection approach for some of the asymptotic regimes considered in this paper (with possibly $T<<N$), that naive cross-validation for $G$ is not recommended for unsupervised clustering algorithms tasks with sample dependent parameters, and that heuristics such as the ``elbow method'' or ``gap curve'' are generally not theoretically justified \citep[see, e.g., Chapter 14.3.11 in ][]{hastie2009elements}, I would recommend to use the TPWD estimator for inference, especially when $T$ is small, $N$ is large, or $G$ is expected to be large. In particular, TPWD scales remarkably well. In an unreported Monte Carlo simulation with $N=2000$, $T=7$, and $G=4$, which may be typical in microeconometric applications, the TPWD point estimate of $\beta=1$ is $1.015$ and it takes less than 2 minutes to compute. 

\section{Empirical application: income and (waves of) democracy}\label{sec:emp_app}
Understanding the statistical relationship between income and democracy has been a longstanding issue in political science and economics \citep{Lipset1959,Barro1999}. Using panel data for $N=90$ countries observed at $T=7$ points in time over the period 1970--2000, \cite{Acemoglu2008} found that the statistically significant positive association between income on democracy vanishes when country fixed effects are included in the regression. They argued that these results are consistent with countries having embarked on divergent paths of economic and political development at certain points in history, or critical junctures. Some of the examples they mention are the end of feudalism, the industrialization age, or the process of colonization. In this perspective, the fixed effects are meant to capture these highly persistent historical events. \cite{BM2015} proposed to test this assumption by computing an approximation of the GFE estimator using alternative minimization and reporting the results for several numbers of groups. They argued that the true number of groups would be less than $10$, reporting statistically significant income effects -- elasticities of a measure of democracy to lagged income per capita -- between $0.061$ and $0.089$ .

This section provides a reassessment of their results, consistently estimating the number of groups by applying the TPWD estimator to their preferred specification: a regression model of democracy (measured by the Freedom House indicator) on lagged democracy and lagged log-GDP per capita with unrestricted group-specific time patterns of heterogeneity $\alpha_{g_it}$:
\begin{equation*}
	democracy_{it}=\beta_1democracy_{it-1}+\beta_2 logGDPpc_{it-1}+\alpha_{g_it}+v_{it}. 
\end{equation*}
The data is obtained from the balanced subsample of \cite{Acemoglu2008}.\footnote{Available at: \href{https://www.aeaweb.org/articles?id=10.1257/aer.98.3.808}{https://www.aeaweb.org/articles?id=10.1257/aer.98.3.808}.} The preliminary estimator is a nuclear-norm regularized (NNR) estimator with tuning parameter set to the theoretically valid rule $\psi_{NT}=\log(\log(T))/\sqrt{16\min(N,T)}$. The data-driven thresholding rule of the TPWD estimator is as described in Section~\ref{sec:data_driven_selec_rule}. 

\begin{table}[H]
	\caption{Estimation of the effect of lagged democracy and lagged income on democracy}
	\label{tab:app_res}
	\centering	
	\begin{adjustbox}{max width={0.99\linewidth},center}
	\begin{threeparttable}[h]
\begin{tabular}{l *{8}{c}}
\toprule
& NNR & TPWD$^{\rm 1 iter}$ & TPWD$^{\rm 2 iter}$ &  TPWD$^{\rm 3 iter}$ & TPWD$^{\rm 4 iter}$   & GFE$^{\bar G=2}$& GFE$^{\bar G=3}$& GFE$^{\bar G=10}$ \\
Dependent variable: Democracy \\ 
\midrule
Lagged Democracy $(\beta_1)$ & 0.800 & 0.720 & 0.721 & 0.730 & 0.730 & 0.601 & 0.407 & 0.277  \\ 
& & (0.040) & (0.040) & (0.039) & (0.039) & (0.041) & (0.052) & (0.049)  \\ 
Lagged Income $(\beta_2)$ & 0.016 & 0.071 & 0.070 & 0.070 & 0.070 & 0.061 & 0.089 & 0.075  \\ 
& & (0.012) & (0.012) & (0.012) & (0.012) & (0.011) & (0.011) & (0.008)  \\ 
Cumulative Income $\left(\frac{\beta_2}{1-\beta_1}\right)$ & 0.078 & 0.253 & 0.253 & 0.258 & 0.258 & 0.152 & 0.151 & 0.104  \\ 
& & (0.020) & (0.021) & (0.021) & (0.021) & (0.021) & (0.013) & (0.009)  \\ \\ 
Estimated number of groups ($\widehat G$) & — & 3 & 3 & 4 & 4 & — & — & —  \\ 
\bottomrule
\end{tabular}
	\begin{tablenotes}
		\footnotesize \item {\em Notes:} Balanced sample of \cite{Acemoglu2008}. $N=90$ countries and $T=7$ time periods at the five-year frequency between 1970--2000. Democracy is measured as the Freedom House indicator of democracy. This table reports results for the nuclear-norm regularized
		(NNR) estimator with regularization parameter $\psi_{NT}=\log(\log(7))/(4\sqrt{7})\approx 0.063$, the triad pairwise-differencing estimator (TPWD$^{k {\rm iter}}$) obtained after $k\in\{1,2,3,4\}$ iterations, initialized at the NNR estimator, and with regularization parameter $c_{NT,k}=(1.35/2)\check \sigma_{v,k}\log(7)/\sqrt{7}$, and \cite{BM2015}'s
		grouped fixed-effects estimator (their Algorithm 1) with pre-specified number of groups set to $g$ (GFE$^{\bar G=g}$). When available, analytical standard errors based on large $N,T$ approximations and clustered at the country level are shown in parentheses. 
	\end{tablenotes}
	\end{threeparttable}
	\end{adjustbox}
\end{table}

Table~\ref{tab:app_res} displays the NNR estimates, the TPWD$^{k{\rm it}}$ estimates at iteration $k\in\set{1,2,3,4}$ (until convergence), and the GFE$^{\bar G=g}$ estimates with user-specified number of groups $g\in\set{2,3,10}$. After three iterations, the TPWD estimator converges to four estimated groups and delivers a significant income effect of  $0.070$. This point estimate is relatively close to the GFE$^{\bar G=2}$ and GFE$^{\bar G=10}$ estimates of $0.061$ and $0.075$, respectively. The estimated cumulative income effect ($\beta_2/(1- \beta_1)$) is 0.258, also significant, and more than twice the GFE$^{\bar G=10}$ estimate of $0.104$. The preliminary NNR estimator delivers point estimates of $0.016$ and $0.078$, respectively. 

\section{Conclusion}\label{sec:conc}
Grouped fixed effects models are plagued with an underlying combinatorial classification problem, rendering estimation and inference difficult. This paper proposes a novel strategy for the constructive identification of all model parameters, including the number of groups. The method simultaneously solves the model selection, classification, and  estimation problems. The corresponding three-step estimator has polynomial computational cost and is straightforward to implement, requiring only smooth convex optimization and elementary arithmetic operations. It builds on an initial off-the-shelf consistent estimator of the slope coefficients and applies thresholding to suitable pairwise-differencing transformations of the residualized regression equations. Under mild conditions, the proposed estimator is shown to be uniformly consistent for the latent grouping structure and asymptotically normal as both dimensions of the panel grow jointly. Importantly, the number of groups is consistently estimated without prior knowledge of its support, and the time dimension may grow at a much slower rate than the cross-sectional dimension.

Beyond its stronger large-sample properties -- established under relatively weaker assumptions than in the existing literature -- Monte Carlo simulations demonstrate its finite-sample competitiveness, if not superiority, compared to spectral clustering and grouped fixed effects estimators.

Several open questions remain for future research: Could this approach be used to construct a formal test of the grouping assumption? Can similar differencing strategies be extended to nonlinear structural models or potential outcome frameworks? Might agglomerative methods help address the problem of weak factors in latent group structures? And is it possible to develop finite-sample or uniform inference procedures? 
\clearpage
\appendix

\begin{center}
	\Huge  Appendix
\end{center}
\section{Sufficient conditions for consistency of the nuclear-norm regularized estimator}\label{sec:nuc_norm_cond}
Define $\gamma^0\equiv(\mathbf{1}\{g_i^0=g\})_{i=1,\ldots,N;g=1,\ldots,G^0}\in\set{0,1}^{N\times G^0}$, $\alpha^0\equiv(\alpha_{gt}^0)_{t=1,\ldots,T;g=1,\ldots,G^0}\in\mathcal A^{T\times G^0}$, $x_k\equiv\textrm{vec}(X_k)$, and $x\equiv(x_1,\ldots, x_k)'$. 
\begin{hyp}\label{as:nuclear_norm_est}
	~
	\begin{enumerate}[label=(\alph*)]
		\item \label{as:nuclear_norm_est_a} As $N$ and $T$ tend to infinity: $\psi_{NT}\to0$ such that $\sqrt{\min(N,T)}\psi_{NT}\to\infty$. 
		\item \label{as:nuclear_norm_est_b} Let
		$\mathbb C\equiv\set{A\in\R^{N\times T}:\norm{M_{\gamma^0}AM_{\alpha^0}}_1\leq3\norm{A-M_{\gamma^0}AM_{\alpha^0}}_1}$,
		where $M_B\equiv I-B(B'B)^\dagger B'$, $I$ is the identity matrix of appropriate dimensions,
		and $^\dagger$ refers to the Moore--Penrose generalized inverse. There exists $\mu>0$, independent from $N$ and $T$, such that for any $a\in\R^{NT}$ with ${\rm mat}(a)\in\mathbb C$, $a'M_xa\geq\mu a'a$ holds for $N,T$ sufficiently large.
		\item \label{as:nuclear_norm_est_c} $\norm{(v_{it})_{i=1,\ldots,N;t=1,\ldots,T}}_{\infty}=O_p\left(\sqrt{\max(N,T)}\right)$.
		\item \label{as:nuclear_norm_est_d} As $N$ and $T$ tend to infinity: $\frac{1}{NT}\sum_{i=1}^N\sum_{t=1}^Tx_{it}x_{it}'\overset{p}{\to}\Sigma>0$ and $\frac{1}{\sqrt{NT}}\sum_{i=1}^N\sum_{t=1}^Tv_{it}x_{it}=O_p(1)$. 
	\end{enumerate}
\end{hyp}
Assumption~\ref{as:nuclear_norm_est}\ref{as:nuclear_norm_est_b} is a restricted eigenvalue condition common in high-dimensional modelling \citep[e.g.,][]{BRT09}. Sufficient conditions for Assumption~\ref{as:nuclear_norm_est}\ref{as:nuclear_norm_est_c} are given in Supplementary Appendix S.2 of \cite{MoonWeidner2017}.
\begin{prop}[\cite{moon2019nuclear}]\label{prop:nucnormest}
	Let Assumption~\ref{as:nuclear_norm_est} hold. Then, as $N$ and $T$ diverge jointly to infinity, $\norm{\widehat \beta^1(\psi_{NT})-\beta^0}=O_p(\psi_{NT})$.
\end{prop}
{\noindent \bf Proof of Proposition~\ref{prop:nucnormest}.} The result follows by \cite{moon2019nuclear}'s Theorem 2, after noticing the interactive fixed effects structure of Model~\eqref{eq:model_true_dgp} displayed in Footnote~\ref{footnote:factor_analytic}. 

\section{Extensions}\label{sec:extensions}
The three-step procedure underlying the TPWD estimator can be applied to several extensions of Model~\eqref{eq:model}. This section briefly outlines a few of them.

\subsection{Linear models with heterogeneous slopes or asymptotically close groups}
To allow for unit-specific effects and/or unit/time-specific slopes, one may work with time-differenced or demeaned equations and/or use a computationally simple first-step estimator well suited for heterogeneous/time-varying slope \cite[e.g.,][]{chernozhukov2019inference}. Group separation conditions need to be adjusted similarly as discussed in \cite{BM2015} 's Supplementary Material, but the main arguments convey. 

The approach of this paper allows for some degree of asymptotic closeness in group-specific effects. Suppose that, for all $(g,\widetilde g)\in\set{1,\ldots,G^0}^2$ such that $g\neq \widetilde g$,
$$T^{-1}\sum_{t=1}^T(\alpha_{gt}^0-\alpha_{\widetilde gt}^0)^2= \rho_{NT,g,\widetilde g},$$
for some deterministic sequence $\rho_{NT,g,\widetilde g}\to0$ as $\min(N,T)\to\infty$. Adapting the rate conditions in Assumption~\ref{as:tuning_param}, TPWD may still recover the true group memberships with probability approaching one. The asymptotic behaviour of the oracle OLS estimator, however, may be affected by this weak-factors property.

\subsection{Nonlinear multiplicative models for networks}
Consider dyadic observations $\{(y_{ij}, x_{ij}) : (i,j) \in \{1,\ldots,n\}^2, i\neq j\}$ for $n$ agents such that 
\begin{equation}\label{eq:mult_model}
    y_{ij} = \varphi(x_{ij};\beta_0)u_{ij}, \quad i \neq j,
\end{equation}
where $\varphi : \R^{K} \to \R^{+*}$ is a function known up to the finite-dimensional parameter vector $\beta_0 \in \R^K$, and $u_{ij}\in\R^+$ is a latent disturbance. Suppose
\begin{equation}\label{eq:unobserved}
    u_{ij} = \alpha_i\gamma_j\omega_{g_i,h_j}\varepsilon_{ij}, \quad i \neq j,
\end{equation}
where $\alpha_i \in \R^{+*}$ and $\gamma_j\in\R^{+*}$ are permanent sender (exporter)  and receiver (importer) unobserved effects, $g_i \in \{1,\ldots,G_0\}$ is an unobserved exporter-group membership variable, $G_0$ is the number of groups of exporters (considered exogenous and fixed), $h_j \in \{1,\ldots,H_0\}$ is an unobserved importer group membership variable, $H_0$ is the number of groups of importers (considered exogenous and fixed), $\omega_{g,h} \in \R^{+*}$ is a permanent unobserved effect affecting group $g$ exporting to group $h$, and $\varepsilon_{ij}\in\R^{+}$ is an idiosyncratic disturbance such that $\Pr(\eps_{ij}=0)<1$. Here, $\alpha_i, \gamma_j, \omega_{g,h}, g_i,h_j$ are consider fixed, that is I condition on them. 

Model~\eqref{eq:mult_model} extends \cite{jochmans_2017} by allowing for latent grouped interactions on top of standard sender and receiver effects. Such grouped effects might capture nonlinear latent measures of reciprocity between units $i$ and $j$: trade shocks that are shared by unobserved groups of exporters or groups of importers. Suppose interest lies in estimating $\beta_0$, $\alpha\equiv(\alpha_1,\ldots,\alpha_n)$, $\gamma\equiv(\gamma_1,\ldots,\gamma_n)$, $g\equiv(g_1,\ldots,g_n)$, $h\equiv(h_1,\ldots,h_n)$, and $\Omega\equiv(\omega_{g,h})_{(g,h)\in[G_0]\times[H_0]}$ under the conditional mean restriction 
\begin{equation}\label{eq:CM}
    \E[\eps_{ij}|x_{12},\ldots,x_{n(n-1)}]=1, \quad \forall i\neq j.
\end{equation}
Because $g,h,\alpha,\gamma$ and $\Omega$ are unobserved, one needs a normalization  specified as:\footnote{The first equality is a standard choice in the literature \cite{jochmans_2017,dzemski2019}. It is equivalent to $\sum_{i=1}^{n}\nu_i = \sum_{j=1}^n\xi_j=0$ where $\nu_i\equiv\log(\alpha_i)$ and $\xi_j\equiv\log(\gamma_j)$. The second equality is needed because of the introduction of the group effects. These normalization choices are arbitrary and one could alternatively assume $\alpha_1=1$ and $\gamma_1=1$ without affecting the validity of the approach.} \begin{equation}\label{eq:normalization}
\prod_{i=1}^n\alpha_i = \prod_{j=1}^{n} \gamma_j=1.
\end{equation}
Section S.2.1 outlines identification arguments based on triad pairwise-differencing.

\subsection{Rubin--Holland Potential Outcomes}
Consider a binary treatment $D_{it}\in\{0,1\}$ and potential outcomes $Y_{it}(0)$ and $Y_{it}(1)$ if individual $i$ is non-treated or treated at period $t$, respectively. Suppose a researcher wants to learn about the treatment effect $Y_{it}(1)-Y_{it}(0)$, but they only observe $Y_{it}(D_{it})$ which is the fundamental problem of causal inference. Under a version of parallel trends holding within each unobserved group $g_i\in\{1,\ldots,N\}$, one strategy is to use untreated periods to estimate group memberships under possibly heterogeneous treatment effects by applying Step 2 to time-differenced outcomes. Since $D_{it}=0$ for this subsample, heterogeneous treatment effects do not perturbate the estimation of the latent group structure. With a sufficiently large number of pre-treatment outcomes, one can consistently estimate the group memberships and build counterfactuals in post-treatment periods to identify a menu of average treatment effects. The theoretical analysis of this procedure generalizing \cite{XDH_DCM_2021} and a large-scale application is work in progress.

\section{Proofs of the results}
\subsection{Proof of Proposition~\ref{prop:sup_norm_cons}}\label{apdx:prop:sup_norm_cons}
Let $\widehat W\equiv(\ind{\widehat d_{\infty,1}^2(i,j)\leq c_{NT}})_{i=1,\ldots,N;j=1,\ldots,N}$ and  $W^0\equiv (\ind{g_i^0=g_j^0})_{i=1,\ldots,N;j=1,\ldots,N}$. Under Assumption~\ref{as:linkage}, Equations~\eqref{eq:consistency_group_member} and \eqref{eq:consistency_grp_number} are immediate corollaries of Lemma~\ref{lem:sup_norm_mat_form} below.
 \begin{lem}\label{lem:sup_norm_mat_form} Let Assumptions \ref{as:prelim_rate}--\ref{as:sup_norm_cons} hold. Then, as $N$ and $T$ tend to infinity,
\begin{equation}\label{eq:sup_norm_cons_matrix_form}
	\norm{\widehat W-W^0}_{\max} = o_p(1).
\end{equation}
\end{lem}
{\bf Proof of Lemma~\ref{lem:sup_norm_mat_form}.}
Fix $\epsilon>0$ and $K_1>0$. By Assumption~\ref{as:prelim_rate}, there exists $K_2>0$ such that, letting $\mathcal E_{1NT}\equiv\set{\norm{\widehat\beta^1-\beta^0}>K_2r_{NT}}$, $\Pr(\mathcal E_{1NT})<\epsilon$ for $\min(N,T)$ sufficiently large. Define $Z_{1NT}(i,j)\equiv\widehat W_{ij}(1-W_{ij}^0)$, $Z_{2NT}(i,j)\equiv(1-\widehat W_{ij})W_{ij}^0$, and the probability events 
\begin{align*}
	\mathcal E_{2NT}&\equiv\set{\min_{g\in\set{1,\ldots,G^0}}\sum_{i=1}^N\mathbf{1}\{g_i^0=g\}\geq 2} \cap\set{C\leq c_{NT}T^\kappa\leq T^\kappa C}\cap\set{c_{NT}r_{NT}^{-1}\geq K_1}\\
	&\qquad \cap\set{r_{NT}\leq\min(\eta/K_2,1)} \cap\set{c_{NT}<c/72}
\end{align*}
and $\mathcal E_{NT}\equiv\mathcal E_{1NT}^c\cap\mathcal E_{2NT}$, where $\eta$, $c$, and $C$ are defined in Equation~\eqref{eq:eta}, Assumption~\ref{as:sup_norm_cons}\ref{as:sup_norm_cons_c}, and  Assumption~\ref{as:tuning_param}\ref{as:tuning_param_b}, respectively. By the union bound, for $\min(N,T)$ sufficiently large,
\begin{align}
	& \Pr\left(\max_{(i,j)\in\set{1,\ldots,N}^2}\abs{\widehat W_{ij}-W_{ij}^0}>0\right) \notag \\
	& \quad \leq \Pr(\mathcal E_{NT}^c)+ \sum_{(i,j) \in \set{1,\ldots,N}^2} \Pr\left(\abs{\widehat W_{ij}\neq W_{ij}^0},\mathcal E_{NT}\right) \notag \\
	& \quad \leq \Pr(\mathcal E_{1NT})+\Pr(\mathcal E_{2NT}^c)+ \sum_{(i,j) \in \set{1,\ldots,N}^2}\Pr\left(\abs{\widehat W_{ij}\neq W_{ij}^0},\mathcal E_{NT}\right) \notag \\
	& \quad \leq 2\epsilon + \sum_{(i,j) \in \set{1,\ldots,N}^2}\Pr(Z_{1NT}(i,j)=1, \mathcal E_{NT}) + \Pr(Z_{2NT}(i,j)=1, \mathcal E_{NT}), \label{eq:decomposition_I} 
\end{align}
where the last inequality follows from $\lim_{\min(N,T)\to \infty}\Pr\left(\mathcal E_{2NT}^c\right)=0$ by Assumptions~\ref{as:tuning_param}\ref{as:tuning_param_b} and \ref{as:sup_norm_cons}\ref{as:sup_norm_cons_e}. Below, I prove that, for $\ell \in\set{1,2}$, and as $\min(N,T)\to\infty$,
\begin{align}\label{eq:exp_decay_1}
	& \max_{(i,j)\in\set{1,\ldots,N}^2}\Pr(Z_{\ell NT}(i,j)=1,\mathcal E_{NT})=o(NT^{-\delta}) \text{ for all } \delta>0.
\end{align}
Equation~\eqref{eq:sup_norm_cons_matrix_form} then follows by combining \eqref{eq:decomposition_I}--\eqref{eq:exp_decay_1} with Assumption \ref{as:tuning_param}\ref{as:tuning_param_a} and because $\epsilon$ is unrestricted.

\medskip
\noindent{1. First, I show \eqref{eq:exp_decay_1} for $\ell=1$.\footnote{Actually, I show the stronger result that the maximum is $o(T^{-\delta})$.}} Let $(i,j) \in \set{1,\ldots,N}^2$ and $\delta>0$. 
$$Z_{1NT}(i,j) =  \mathbf{1}\Bigg\{\max_{k\in \set{1,\ldots,N}\backslash\set{i,j}} \abs{\frac{1}{T}\sum_{t=1}^T(\widehat v_{it}-\widehat v_{jt})\widehat v_{kt}}\leq c_{NT}\Bigg\}\mathbf{1}\{g_i^0\neq g_j^0\}.$$
If $G^0=1$, then almost surely $g_i^0=g_j^0$ and $Z_{1NT}(i,j)=0$. Since this holds for each pair $(i,j)$, \eqref{eq:exp_decay_1} holds. Next, suppose that $G^0>1$. Then,
\begin{align*}
	& \mathbf{1}\{Z_{1NT}(i,j)=1,\mathcal E_{NT}\} \\
	&  \quad=\sum_{\substack{(g,\widetilde g) \in \set{1,\ldots,G^0}^2 \\ g\neq \widetilde g}}\mathbf{1}\{\mathcal E_{NT}\} \mathbf{1}\{g_i^0=g\}\mathbf{1}\{g_j^0=\widetilde g\}\mathbf{1}\set{\max_{k\in \set{1,\ldots,N}\backslash\set{i,j}}\abs{\frac{1}{T}\sum_{t=1}^T(\widehat v_{it}-\widehat v_{jt})\widehat v_{kt}}\leq c_{NT}}.
\end{align*}
If $\mathbf{1}\{\mathcal E_{NT}\}\mathbf{1}\{g_i^0\neq g_j^0\}=1$, there exists a pair $(k^*(i,j,g_i^0), l^*(i,j,g_j^0))\in\mathscr P_2(\set{1,\ldots,N}\backslash\set{i,j})$ such that $g_{k^*(i,j,g_i^0)}^0=g_i^0$ and $g_{l^*(i,j,g_j^0)}^0=g_j^0$. It follows that
\begin{align*}
	& \mathbf{1}\{Z_{1NT}(i,j)=1,\mathcal E_{NT}\} \\
	&  \quad \leq\mathbf{1}\{\mathcal E_{NT}\} \times \\
	& \quad \quad \sum_{\substack{(g,\widetilde g) \in \set{1,\ldots,G^0}^2 \\ g\neq \widetilde g}}\mathbf{1}\{g_i^0=g\}\mathbf{1}\{g_j^0=\widetilde g\}\mathbf{1}\set{\abs{\frac{1}{T}\sum_{t=1}^T(\widehat v_{it}-\widehat v_{jt})\widehat v_{k^*(i,j,g_i^0)t}}\leq c_{NT}} \times\\
	& \quad \quad \quad \mathbf{1}\set{\abs{\frac{1}{T}\sum_{t=1}^T(\widehat v_{it}-\widehat v_{jt})\widehat v_{l^*(i,j,g_j^0)t}}\leq c_{NT}} \\
	&  \quad \leq\mathbf{1}\{\mathcal E_{NT}\} \times \\
	&  \quad \quad  \sum_{\substack{(g,\widetilde g) \in \set{1,\ldots,G^0}^2 \\ g\neq \widetilde g}}\mathbf{1}\{g_i^0=g\}\mathbf{1}\{g_j^0=\widetilde g\}\mathbf{1}\set{\abs{\frac{1}{T}\sum_{t=1}^T(\widehat v_{it}-\widehat v_{jt})(\widehat v_{k^*(i,j,g_i^0)t}-\widehat v_{l^*(i,j,g_j^0)t})}\leq2c_{NT}},
\end{align*}
where the first inequality uses the definition of the maximum and the second inequality follows from the triangle inequality. Since there is at most one pair $(g,\widetilde g)\in\set{1,\ldots,G^0}^2$ such that $g\neq \widetilde g$ and $\mathbf{1}\{g_i^0=g\}\mathbf{1}\{g_j^0=\widetilde g\}=1$, developing the product and noting that $\mathbf{1}\{\abs{a}\leq b\}\leq\mathbf{1}\{a\leq b\}$ for all $(a,b) \in \R\times \R$ yields
\begin{align}
	& \mathbf{1}\{Z_{1NT}(i,j)=1, \mathcal E_{NT}\} \notag\\
	&\quad \leq\mathbf{1}\{\mathcal E_{NT}\} \times\notag \\
	& \quad \quad \max_{\substack{(g,\widetilde g) \in \set{1,\ldots,G^0}^2 \notag\\ g\neq \widetilde g}} \mathbf{1}\Bigg\{\frac{1}{T}\sum_{t=1}^{T}\left(\alpha_{gt}^0-\alpha_{\widetilde gt}^0\right)^2 +\frac{1}{T}\sum_{t=1}^T\left(\alpha_{gt}^0-\alpha_{\widetilde gt}^0\right)\left(v_{it}-v_{jt}+v_{k^*(i,j,g_i^0)t}-v_{l^*(i,j,g_j^0)t}\right)  \notag\\
	& \quad \quad + \frac{1}{T}\sum_{t=1}^T\left(\alpha_{gt}^0-\alpha_{\widetilde gt}^0\right)\left(\beta^0-\widehat\beta^1\right)'\left(x_{it}-x_{jt}+x_{k^*(i,j,g_i^0)t}-x_{l^*(i,j,g_j^0)t}\right) \notag\\
	& \quad \quad +\frac{1}{T}\sum_{t=1}^T(v_{it}-v_{jt})\left(v_{k^*(i,j,g_i^0)t}-v_{l^*(i,j,g_j^0)t}\right) \notag\\
	& \quad \quad +\frac{1}{T}\sum_{t=1}^T\left(\beta^0-\widehat\beta^1\right)'\left(x_{it}-x_{jt}\right)\left(\beta^0-\widehat\beta^1\right)'\left(x_{k^*(i,j,g_i^0)t}-x_{l^*(i,j,g_j^0)t}\right)\notag \\
	& \quad \quad +\frac{1}{T}\sum_{t=1}^T(v_{it}-v_{jt})\left(\beta^0-\widehat\beta^1\right)'\left(x_{k^*(i,j,g_i^0)t}-x_{l^*(i,j,g_j^0)t}\right) \notag \\ 
	& \quad \quad +\frac{1}{T}\sum_{t=1}^T\left(v_{k^*(i,j,g_i^0)t}-v_{l^*(i,j,g_j^0)t}\right)\left(\beta^0-\widehat\beta^1\right)'\left(x_{it}-x_{jt}\right)\leq 2c_{NT}\Bigg\} \notag\\
	& \quad=\mathbf{1}\{\mathcal E_{NT}\}\times\max_{\substack{(g,\widetilde g) \in \set{1,\ldots,G^0}^2 \\ g\neq \widetilde g}}\mathbf{1}\Big\{A_{T}(i,j,g,\widetilde g)\leq 2c_{NT}\Big\},\label{eq:bound0}
\end{align}
where $A_{T}(i,j,g,\widetilde g)$ is defined implicitly. 
Define
\begin{align*}
	B_{T}(i,j,g,\widetilde g) & \equiv \abs{A_{T}(i,j,g,\widetilde g)- \frac{1}{T}\sum_{t=1}^{T}\left(\alpha_{gt}^0-\alpha_{\widetilde gt}^0\right)^2 \right. \\
		& \left.\quad-\frac{1}{T}\sum_{t=1}^T\left(\alpha_{gt}^0-\alpha_{\widetilde gt}^0\right)\left(v_{it}-v_{jt}+v_{k^*(i,j,g_i^0)t}-v_{l^*(i,j,g_j^0)t}\right) \right.\\
		&\left.\quad-\frac{1}{T}\sum_{t=1}^T(v_{it}-v_{jt})\left(v_{k^*(i,j,g_i^0)t}-v_{l^*(i,j,g_j^0)t}\right)}
\end{align*}
and $\bar a\equiv\sup_{a\in\mathcal A}\abs{a}$ with $\bar a<\infty$ by Assumption~\ref{as:sup_norm_cons}\ref{as:sup_norm_cons_a}. It follows from the triangle inequality, the Cauchy--Schwarz inequality, and subadditivity of $x\mapsto\sqrt{x}$ that
 \begin{align*}
  B_{T}(i,j,g,\widetilde g) & \leq2\norm{\widehat\beta^1-\beta^0}\set{\frac{\bar a}{T}\sum_{t=1}^T\left(\Vert x_{it}\Vert+\Vert x_{jt}\Vert +\Vert x_{k^*(i,j,g_i^0)t}\Vert+\Vert x_{l^*(i,j,g_j^0)t} \Vert \right)\right. \\
  & \left.\quad +\norm{\widehat\beta^1-\beta^0}\left(\frac{1}{T}\sum_{t=1}^T\Vert x_{it}\Vert+\Vert x_{jt}\Vert\right)\left(\frac{1}{T}\sum_{t=1}^T \Vert x_{k^*(i,j,g_i^0)t}\Vert+\Vert x_{l^*(i,j,g_j^0)t}\Vert\right)\right. \\
  &\left.\quad+\left(\sqrt{\frac{1}{T}\sum_{t=1}^Tv_{it}^2}+\sqrt{\frac{1}{T}\sum_{t=1}^Tv_{jt}^2}\right)\left(\frac{1}{T}\sum_{t=1}^T\Vert x_{k^*(i,j,g_i^0)t}\Vert+\Vert x_{l^*(i,j,g_j^0)t}\Vert\right)\right. \\
  &\left.\quad+\left(\sqrt{\frac{1}{T}\sum_{t=1}^Tv_{k^*(i,j,g_i^0)t}^2}+\sqrt{\frac{1}{T}\sum_{t=1}^Tv_{l^*(i,j,g_j^0)t}^2}\right)\left(\frac{1}{T}\sum_{t=1}^T\Vert x_{it}\Vert+\Vert x_{jt}\Vert\right)}.
 \end{align*}
By Assumption \ref{as:sup_norm_cons}\ref{as:sup_norm_cons_b}, there exists $M^*>0$ such that $\E[v_{it}^2]\leq M^*$ for all $i,t$. Let $\widetilde M>\max(M,\max(M^*,1))$, where $M$ is defined in Assumption~\ref{as:sup_norm_cons}\ref{as:sup_norm_cons_f}, and $\eta>0$ such that
\begin{equation}\label{eq:eta}
	\eta\leq\min\left(1,\frac{c}{48\left(\bar a4\widetilde M+4\widetilde M^2+8(\widetilde M)^{3/2}\right)}\right),
\end{equation}
where $c$ is defined in Assumption~\ref{as:sup_norm_cons}\ref{as:sup_norm_cons_c}. By definition of $\mathcal E_{NT}$, $\norm{\widehat\beta^1-\beta^0}\leq\eta$ on $\mathcal E_{NT}$. Then, since $\eta\leq1$,
 \begin{align*}
 \ind{\mathcal E_{NT}} B_{T}(i,j,g,\widetilde g) & \leq2\eta\set{\frac{\bar a}{T}\sum_{t=1}^T\left(\Vert x_{it}\Vert+\Vert x_{jt}\Vert +\Vert x_{k^*(i,j,g_i^0)t}\Vert+\Vert x_{l^*(i,j,g_j^0)t} \Vert \right)\right. \\
  & \left.\quad +\left(\frac{1}{T}\sum_{t=1}^T\Vert x_{it}\Vert+\Vert x_{jt}\Vert\right)\left(\frac{1}{T}\sum_{t=1}^T \Vert x_{k^*(i,j,g_i^0)t}\Vert+\Vert x_{l^*(i,j,g_j^0)t}\Vert\right)\right. \\
  &\left.\quad+\left(\sqrt{\frac{1}{T}\sum_{t=1}^Tv_{it}^2}+\sqrt{\frac{1}{T}\sum_{t=1}^Tv_{jt}^2}\right)\left(\frac{1}{T}\sum_{t=1}^T\Vert x_{k^*(i,j,g_i^0)t}\Vert+\Vert x_{l^*(i,j,g_j^0)t}\Vert\right)\right. \\
  &\left.\quad+\left(\sqrt{\frac{1}{T}\sum_{t=1}^Tv_{k^*(i,j,g_i^0)t}^2}+\sqrt{\frac{1}{T}\sum_{t=1}^Tv_{l^*(i,j,g_j^0)t}^2}\right)\left(\frac{1}{T}\sum_{t=1}^T\Vert x_{it}\Vert+\Vert x_{jt}\Vert\right)} \\
	& \equiv C_T(i,j).
\end{align*}
Plugging this upper bound into \eqref{eq:bound0} yields
\begin{align*}
	& \mathbf{1}\{Z_{1NT}(i,j)=1,\mathcal E_{NT}\} \\
	& \quad \leq \max_{\substack{(g,\widetilde g) \in \set{1,\ldots,G^0}^2 \\ g\neq \widetilde g}}\mathbf{1}\Big\{\frac{1}{T}\sum_{t=1}^{T}\left(\alpha_{gt}^0-\alpha_{\widetilde gt}^0\right)^2  \\
	&  \quad + \frac{1}{T}\sum_{t=1}^T\left(\alpha_{gt}^0-\alpha_{\widetilde gt}^0\right)\left(v_{it}-v_{jt}+v_{k^*(i,j,g_i^0)t}-v_{l^*(i,j,g_j^0)t}\right) \\
	& \quad +\frac{1}{T}\sum_{t=1}^T(v_{it}-v_{jt})\left(v_{k^*(i,j,g_i^0)t}-v_{l^*(i,j,g_j^0)t}\right)\leq 2c_{NT}+C_T(i,j)\Big\}.
\end{align*}
By the union bound and some probability algebra, it follows that
\begin{align}
	& \Pr\left(Z_{1NT}(i,j)=1, \mathcal E_{NT}\right) \notag  \\
	& \quad\leq \sum_{\substack{(g,\widetilde g) \in \set{1,\ldots,G^0}^2 \\ g\neq \widetilde g}} \Pr\left(\frac1T\sum_{t=1}^{T}\left(\alpha_{gt}^0-\alpha_{\widetilde gt}^0\right)v_{it}\leq -\frac{c}{12}+2c_{NT}+2\eta\left(\bar a4\widetilde M+4\widetilde M^2+8(\widetilde M)^{3/2}\right), \mathcal E_{NT}\right) \notag \\
	& \quad\quad+4G^0\left(G^0-1\right)\left[\max_{g\neq\widetilde g}\Pr\left(\frac{1}{T}\sum_{t=1}^{T}\left(\alpha_{gt}^0-\alpha_{\widetilde gt}^0\right)^2 \leq \frac{c}{2}\right)+ \max_{i\in\set{1,\ldots,N}}\Pr\left(\frac1T\sum_{t=1}^T\Vert x_{it}\Vert \geq\widetilde M\right) \notag \right.\\
	& \left.\quad \quad +\max_{i\in\set{1,\ldots,N},g\neq\widetilde g}\Pr\left(\abs{\frac1T\sum_{t=1}^{T}\left(\alpha_{gt}^0-\alpha_{\widetilde gt}^0\right)v_{it}}\geq\frac{c}{12}\right)+ \max_{i\in\set{1,\ldots,N}}\Pr\left(\frac1T\sum_{t=1}^Tv_{it}^2\geq\widetilde M\right) \notag \right.\\
	&\left. \qquad +\max_{(i,j,k)\in\mathscr P_3(\set{1,\ldots,N})}\Pr\left(\abs{\frac1T\sum_{t=1}^{T}(v_{it}-v_{jt})v_{kt}}\geq\frac{c}{12}\right)\right]. \label{eq:bound1}
\end{align}
First, I bound the maximum terms. By Assumption~\ref{as:sup_norm_cons}\ref{as:sup_norm_cons_c}, it holds that 
$$\lim_{\min(N,T)\to \infty}\frac{1}{T}\sum_{t=1}^T\E\left[(\alpha_{gt}^0-\alpha_{\widetilde gt}^0)^2\right]=c_{g,\widetilde g}>c.$$
So for $\min(N,T)$ large enough, 
$$\frac{1}{T}\sum_{t=1}^T\E\left[\left(\alpha_{gt}^0-\alpha_{\widetilde gt}^0\right)^2\right]\geq\frac{2c}{3}.$$
By applying Lemma B.5 in \cite{BM2015} to $z_t=(\alpha_{gt}^0-\alpha_{\widetilde gt}^0)^2 -\E[(\alpha_{gt}^0-\alpha_{\widetilde gt}^0)^2]$,\footnote{Lemma B.5 in \cite{BM2015} is a direct consequence of Theorem 6.2 in \cite{Rio2000}.} which satisfies appropriate mixing and tail conditions by Assumptions~\ref{as:sup_norm_cons}\ref{as:sup_norm_cons_b} and \ref{as:sup_norm_cons}\ref{as:sup_norm_cons_d}, and taking $z=c/6$ yields, as $\min(N,T)\to\infty$,
\begin{equation}\label{eq:bound2}
	\Pr\left(\frac{1}{T}\sum_{t=1}^{T}\left(\alpha_{gt}^0-\alpha_{\widetilde gt}^0\right)^2\leq\frac{c}{2}\right)=o(T^{-\delta}),
\end{equation}
uniformly across groups $g$ and $\widetilde g$. Similarly, applying
Lemma B.5 to $z_t=v_{it}^2-\E[v_{it}^2]$ and taking $z=\widetilde M-M^*$ yields
\begin{equation}\label{eq:bound3}
	\Pr\left(\frac{1}{T}\sum_{t=1}^{T}v_{it}^2\geq\widetilde M\right)=o(T^{-\delta}),
\end{equation}
uniformly across units $i$, where it is used that $\set{v_{it}^2}_t$ is strongly mixing since $\set{v_{it}}$ is strongly mixing by Assumption~\ref{as:sup_norm_cons}\ref{as:sup_norm_cons_d}. By Assumption~\ref{as:sup_norm_cons}\ref{as:sup_norm_cons_d}, the process $\set{(\alpha_{\widetilde gt}^0 - \alpha_{gt}^0)v_{it}}_t$ has zero mean, and is strongly mixing with faster-than-polynomial decay rate. Moreover, for all $i,t$ and $m>0$,
$$\Pr\left(\abs{\left(\alpha_{gt}^0-\alpha_{\widetilde gt}^0\right)v_{it}}>m\right) \leq \Pr\left(\abs{v_{it}}>\frac{m}{2\bar a} \right),$$
so $\set{(\alpha_{gt}^0 - \alpha_{\widetilde gt}^0)v_{it}}_t$ also satisfies the tail condition of Assumption~\ref{as:sup_norm_cons}\ref{as:sup_norm_cons_b}, albeit with a different constant $b'>0$ instead of $b>0$. Applying Lemma B.5 from \cite{BM2015} again with $z_t=(\alpha_{gt}^0-\alpha_{\widetilde gt}^0)v_{it}$ and taking $z=c/12$ yields
\begin{equation}\label{eq:bound3b}
	\Pr\left(\abs{\frac{1}{T}\sum_{t=1}^{T}\left(\alpha_{gt}^0-\alpha_{\widetilde gt}^0\right)v_{it}}\geq\frac{c}{12}\right)=o(T^{-\delta})
\end{equation}
uniformly across $i$, $g$, and $\widetilde g$. An analogous reasoning yields
\begin{equation}\label{eq:bound4}
	\sup_{(i,j,k)\in\mathscr P_3(\set{1,\ldots,N})}\Pr\left(\abs{\frac1T\sum_{t=1}^{T}(v_{it}-v_{jt})v_{kt}}\geq\frac{c}{12}\right)=o(T^{-\delta}).
\end{equation}
Second, because $c_{NT}\leq c/72$ on $\mathcal E_{NT}$, a similar reasoning yields
\begin{align}
	&\Pr\left(\frac1T\sum_{t=1}^{T}\left(\alpha_{gt}^0-\alpha_{\widetilde gt}^0\right)v_{it}\leq -\frac{c}{12}+2c_{NT}+2\eta\left(\bar a4\widetilde M+4\widetilde M^2+8(\widetilde M)^{3/2}\right),  \mathcal E_{NT}\right) \notag\\
	&\quad \leq \Pr\left(\frac1T\sum_{t=1}^{T}\left(\alpha_{gt}^0-\alpha_{\widetilde gt}^0\right)v_{it}\leq -\frac{c}{72}\right) \notag\\
	& \quad=o(T^{-\delta}),\label{eq:bound5}
\end{align}
uniformly across $g,\widetilde g$, where I have used the value of $\eta$ given in \eqref{eq:eta}. Combining \eqref{eq:bound1}--\eqref{eq:bound5} and using Assumption~\ref{as:sup_norm_cons}\ref{as:sup_norm_cons_f} yields
\begin{equation*}
	\sup_{(i,j)\in\set{1,\ldots,N}^2}\Pr\left(Z_{1NT}(i,j)=1,\mathcal E_{NT}\right)=G^0(1-G^0)\times o_p(T^{-\delta}) = o_p(T^{-\delta}),
\end{equation*}
i.e., \eqref{eq:exp_decay_1} for $\ell=1$ holds.

\medskip
\noindent{2. Second, I show \eqref{eq:exp_decay_1} for $\ell=2$.}
\begin{align*}
&\mathbf{1}\{Z_{2NT}(i,j)=1,\mathcal E_{NT}\}\\ 
&\quad=\mathbf{1}\{\mathcal E_{NT}\}\mathbf{1}\set{\max_{k\in\set{1,\ldots,N}\backslash\set{i,j}}\abs{\frac{1}{T}\sum_{t=1}^T(\widehat v_{it}-\widehat v_{jt})\widehat v_{kt}} > c_{NT}}\mathbf{1}\{g_i^0=g_j^0\}\\
&\quad\leq\mathbf{1}\{\mathcal E_{NT}\}\mathbf{1}\Big\{\max_{k\in\set{1,\ldots,N}\backslash\set{i,j}}\abs{\frac{1}{T}\sum_{t=1}^T(v_{it}-v_{jt})v_{kt}+\frac{1}{T}\sum_{t=1}^T(v_{it}-v_{jt})\alpha_{g_k^0t}^0 \right. \notag\\
& \left. \quad \quad +\frac{1}{T}\sum_{t=1}^T\left(\beta^0-\widehat\beta^1\right)'\left(x_{it}-x_{jt}\right)\left(\beta^0-\widehat\beta^1\right)'x_{kt}\notag \right. \\
& \left. \quad \quad +\frac{1}{T}\sum_{t=1}^T(v_{it}-v_{jt})\left(\beta^0-\widehat\beta^1\right)'x_{kt}+\frac{1}{T}\sum_{t=1}^T\alpha_{g_k^0t}^0\left(\beta^0-\widehat\beta^1\right)'(x_{it}-x_{jt})\notag \right. \\ 
	& \left. \quad \quad +\frac{1}{T}\sum_{t=1}^Tv_{kt}\left(\beta^0-\widehat\beta^1\right)'\left(x_{it}-x_{jt}\right)}> c_{NT}\Big\}.\label{eq:bound00}
\end{align*}
By the union bound, the triangle inequality, and the Cauchy--Schwarz inequality, 
\begin{align*}
&\Pr\left(Z_{2NT}(i,j)=1,\mathcal E_{NT}\right)\\
&\quad\leq (N-2)\sup_{(i,j,k)\in\mathscr P_3(\set{1,\ldots,N})}\set{\Pr\left(\abs{\frac{1}{T}\sum_{t=1}^T(v_{it}-v_{jt})v_{kt}}>\frac{c_{NT}}{6},\mathcal E_{NT}\right) \right. \\
&\qquad \left.+ \Pr\left(\abs{\frac{1}{T}\sum_{t=1}^T(v_{it}-v_{jt})\alpha_{g_k^0t}^0}>\frac{c_{NT}}{6},\mathcal E_{NT}\right) \right. \\
&\left. \qquad+\Pr\left(\left(\frac{1}{T}\sum_{t=1}^T\Vert x_{it}\Vert+\Vert x_{jt}\Vert\right)\left(\frac{1}{T}\sum_{t=1}^T \Vert x_{kt}\Vert\right)>\frac{c_{NT}}{6\sqrt{2}K_2^2r_{NT}^2},\mathcal E_{NT}\right) \right. \\
&\qquad\left.+\Pr\left(\left(\sqrt{\frac{1}{T}\sum_{t=1}^Tv_{it}^2}+\sqrt{\frac{1}{T}\sum_{t=1}^Tv_{jt}^2}\right)\left(\frac{1}{T}\sum_{t=1}^T\Vert x_{kt}\Vert \right)>\frac{c_{NT}}{6K_2r_{NT}},\mathcal E_{NT}\right) \right. \\
&\left. \qquad+\Pr\left(\frac{1}{T}\sum_{t=1}^T\Vert x_{it}\Vert+\Vert x_{jt}\Vert>\frac{c_{NT}}{6\sqrt{2}K_2r_{NT}},\mathcal E_{NT}\right) \right. \\
&\left.\qquad +\Pr\left(\sqrt{\frac{1}{T}\sum_{t=1}^Tv_{kt}^2}\left(\frac{1}{T}\sum_{t=1}^T\Vert x_{it}\Vert+\Vert x_{jt}\Vert\right)>\frac{c_{NT}}{6\sqrt{2}K_2r_{NT}},\mathcal E_{NT}\right)}.
\end{align*}
Under the strong mixing and tail conditions from Assumptions~\ref{as:sup_norm_cons}\ref{as:sup_norm_cons_b} and \ref{as:sup_norm_cons}\ref{as:sup_norm_cons_d}, and because $K_1\leq c_{NT}/r_{NT}\leq c_{NT}/r_{NT}^2$ on $\mathcal E_{NT}$, for $K_1$ sufficiently large, the last four probabilities in the above expression can be shown to be $o(T^{-\delta})$ for all $\delta>0$, uniformly across $(i,j,k)$, by similar arguments as in Step 1. For the first probability, a close inspection of the proof of Lemma B.5 in \cite{BM2015} reveals that, by taking $z_t=(v_{it}-v_{jt})v_{kt}$ and $z = c_{NT}/6$, and because $c_{NT}\geq CT^{-\kappa}$ on $\mathcal E_{NT}$, 
\begin{align}
	\Pr\left(\abs{\frac{1}{T}\sum_{t=1}^T(v_{it}-v_{jt})v_{kt}}\geq\frac{c_{NT}}{10},\mathcal E_{NT}\right) & \leq 4\left(1+\frac{T^{1/2-2\kappa}}{C_1} \right)^{-(1/2)T^{1/2}} \notag \\
	& \qquad +C_2T^{\kappa}\exp\left(-C_3\left(T^{(1/2-\kappa)/C_4}\right)\right), 
\end{align}
where $C_1,C_2,C_3$, and $C_4$ are positive constants that do not depend on $i,j,k$. Since $\kappa<1/2$, the upper bound is $o_p(T^{-\delta})$ for all $\delta>0$. The second probability can be shown to be $o_p(T^{-\delta})$ for all $\delta>0$ by following the same reasoning. This shows \eqref{eq:exp_decay_1} for $\ell=2$.
\hfill$\square$

\subsection{Proof of Corollary~\ref{cor:AN}}\label{apdx:cor:AN}
Let $(\widetilde \beta', \widetilde\alpha_{11},\ldots,\widetilde\alpha_{G^0T})'$  denote the infeasible oracle estimator computed from a pooled OLS regression of $y_{it}$ on $x_{it}$ and interactions of true group and time indicators $\mathbf{1}\{g_i^0=1\}, \ldots,\mathbf{1}\{g_i^0=G^0\}$, $\mathbf{1}\{t=1\}, \ldots,\mathbf{1}\{t=T\}$. By the same reasoning as in Section S.A.1. of \cite{BM2015}'s Supplementary Material, as $N$ and $T$ tend to infinity,
\begin{equation}
		\sqrt{NT}(\widetilde \beta-\beta^0)\ConvNor{\Sigma_\beta^{-1}\Omega_\beta \Sigma_\beta^{-1}},
\end{equation}
and, for all $(g,t)\in\set{1,\ldots,G^0}\times\set{1,\ldots,T}$,
\begin{equation}
	\sqrt{N}(\widetilde\alpha_{gt}-\alpha_{gt}^0) \ConvNor{\frac{\omega_{gt}}{\pi_g^2}}.
\end{equation}
Without loss of generality, suppose that the labelling of each predicted group matches the true group labelling. By Proposition \ref{prop:sup_norm_cons}, for all $(g,t)\in\set{1,\ldots,G^0}\times\set{1,\ldots,T}$,
\begin{align*}
	\Pr\left(\set{\widehat\alpha_{gt}\neq\widetilde \alpha_{gt}}\cup\set{\widehat \beta\neq\widetilde\beta}\right) &\leq \Pr\left(\widehat G\neq G^0\right)+\Pr\left(\max_{i\in\set{1,\ldots,N}}\;\abs{\widehat g_i-g_i^0}>0\right) \\
	&= o(1) + o(1) \\
	&= o(1).
\end{align*}
Equation~\eqref{eq:AN_group_fe} then follows from
\begin{align*}
	& \abs{\Pr\left(\sqrt{N}(\widehat\alpha_{gt}-\alpha_{gt}^0) \leq a\right)-\Pr\left(\sqrt{N}(\widetilde\alpha_{gt} - \alpha_{gt}^0)\leq a\right)} \\
	& \quad\leq\abs{\Pr\left(\sqrt{N}(\widehat\alpha_{gt}-\alpha_{gt}^0)\leq a, \sqrt{N}(\widetilde\alpha_{gt}-\alpha_{gt}^0) > a\right)} \\
	& \qquad + \abs{\Pr\left(\sqrt{N}(\widehat\alpha_{gt} - \alpha_{gt}^0)>a, \sqrt{N}(\widetilde\alpha_{gt}-\alpha_{gt}^0) \leq a\right)}\\
	& \quad \leq  \Pr\left(\widehat \alpha_{gt}\neq\widetilde \alpha_{gt}\right) +\Pr\left(\widehat \alpha_{gt}\neq\widetilde \alpha_{gt}\right)\\
	&\quad =o(1)
\end{align*}
for any $a>0$. Equation~\eqref{eq:AN_theta_cov} follows from a similar argument.

\section{Additional table}
\begin{table}[H]
	\centering	
	\caption{Clustering performance across linkage criteria}
	\label{tab:mc_pureGFE_complinkage}
	\begin{adjustbox}{max width={0.99\linewidth},center}
			\begin{threeparttable}
			\begin{tabular}{lll *{22}{S[table-format=-1.3]}}
\toprule
{} & {} & {} & \multicolumn{4}{c}{Average} & & \multicolumn{4}{c}{Average squared} & & \multicolumn{4}{c}{Complete} &&  \multicolumn{4}{c}{Single} \\
\cmidrule{4-7}\cmidrule{9-12}\cmidrule{14-17}\cmidrule{19-22} 
$G$ & $N$ & $T$ & \multicolumn{1}{c}{$\widehat G$} & \multicolumn{1}{c}{P} & \multicolumn{1}{c}{R} & \multicolumn{1}{c}{RI} & & \multicolumn{1}{c}{$\widehat G$} & \multicolumn{1}{c}{P} & \multicolumn{1}{c}{R} & \multicolumn{1}{c}{RI} & & \multicolumn{1}{c}{$\widehat G$} & \multicolumn{1}{c}{P} & \multicolumn{1}{c}{R} & \multicolumn{1}{c}{RI} & & \multicolumn{1}{c}{$\widehat G$} & \multicolumn{1}{c}{P} & \multicolumn{1}{c}{R} & \multicolumn{1}{c}{RI}   \\
\midrule 
3&90&7& 14.833 & 0.982 & 0.317 & 0.776 & & 3.048 & 0.931 & 0.938 & 0.956 & & 14.680 & 0.978 & 0.248 & 0.753 & & 1.106 & 0.339 & 0.998 & 0.348  \\ 
& &10& 7.833 & 0.992 & 0.603 & 0.869 & & 2.990 & 0.968 & 0.974 & 0.980 & & 10.314 & 0.992 & 0.368 & 0.793 & & 1.114 & 0.357 & 1.000 & 0.376  \\ 
& &20& 4.600 & 1.000 & 0.931 & 0.978 & & 2.992 & 0.994 & 0.997 & 0.996 & & 6.878 & 1.000 & 0.542 & 0.851 & & 1.922 & 0.602 & 1.000 & 0.691  \\ 
& &40& 3.367 & 1.000 & 0.991 & 0.997 & & 3.000 & 1.000 & 1.000 & 1.000 & & 5.132 & 1.000 & 0.719 & 0.909 & & 2.968 & 0.987 & 1.000 & 0.993  \\ 
\midrule
3&180&7& 23.533 & 0.988 & 0.258 & 0.754 & & 3.124 & 0.938 & 0.940 & 0.960 & & 25.584 & 0.983 & 0.150 & 0.719 & & 1.076 & 0.333 & 0.999 & 0.335  \\ 
& &10& 9.667 & 0.995 & 0.561 & 0.854 & & 3.018 & 0.977 & 0.978 & 0.985 & & 16.248 & 0.995 & 0.248 & 0.752 & & 1.048 & 0.337 & 1.000 & 0.342  \\ 
& &20& 5.200 & 1.000 & 0.896 & 0.966 & & 3.000 & 0.998 & 0.998 & 0.998 & & 8.752 & 1.000 & 0.452 & 0.819 & & 1.536 & 0.482 & 1.000 & 0.554  \\ 
& &40& 3.667 & 1.000 & 0.989 & 0.996 & & 3.000 & 1.000 & 1.000 & 1.000 & & 6.292 & 1.000 & 0.592 & 0.866 & & 2.934 & 0.974 & 1.000 & 0.985  \\ 
\midrule
4&90&7& 13.067 & 0.698 & 0.400 & 0.810 & & 2.998 & 0.586 & 0.918 & 0.819 & & 15.110 & 0.725 & 0.247 & 0.795 & & 1.090 & 0.247 & 0.999 & 0.256  \\ 
& &10& 8.400 & 0.775 & 0.574 & 0.856 & & 2.700 & 0.542 & 0.953 & 0.772 & & 10.728 & 0.805 & 0.378 & 0.827 & & 1.082 & 0.253 & 1.000 & 0.272  \\ 
& &20& 4.733 & 0.892 & 0.870 & 0.942 & & 2.140 & 0.421 & 0.997 & 0.650 & & 7.326 & 0.911 & 0.594 & 0.888 & & 1.498 & 0.314 & 1.000 & 0.427  \\ 
& &40& 4.233 & 0.968 & 0.966 & 0.984 & & 2.022 & 0.392 & 1.000 & 0.621 & & 5.806 & 0.981 & 0.794 & 0.947 & & 1.988 & 0.384 & 1.000 & 0.611  \\ 
\midrule
4&180&7& 19.000 & 0.788 & 0.296 & 0.652 & & 2.942 & 0.697 & 0.956 & 0.792 & & 22.750 & 0.792 & 0.120 & 0.596 & & 1.092 & 0.447 & 1.000 & 0.450  \\ 
& &10& 10.000 & 0.824 & 0.514 & 0.733 & & 2.194 & 0.593 & 0.994 & 0.689 & & 15.408 & 0.857 & 0.200 & 0.631 & & 1.104 & 0.455 & 1.000 & 0.464  \\ 
& &20& 5.000 & 0.932 & 0.904 & 0.927 & & 2.000 & 0.565 & 1.000 & 0.659 & & 8.514 & 0.940 & 0.436 & 0.738 & & 1.524 & 0.507 & 1.000 & 0.556  \\ 
& &40& 4.300 & 0.980 & 0.973 & 0.979 & & 2.000 & 0.565 & 1.000 & 0.659 & & 6.460 & 0.987 & 0.601 & 0.820 & & 1.976 & 0.562 & 1.000 & 0.653  \\ 
\bottomrule
\end{tabular}

\begin{tablenotes}
			\footnotesize \item {\em Notes:} This table reports the precision (P) rate, recall (R) rate, and Rand index (RI) for the triad pairwise-difference (TPWD) estimator with $\widehat\beta^1=0$, cut-off $c_{NT}=1.35\check\sigma_v\log(T)/\sqrt{\min(N,T)}$, and linkage $\in\{$\texttt{average, average squared, complete, single}$\}$. The \texttt{average squared} linkage applies the \texttt{average} linkage function to the pointwise squared dissimilarity matrix. Results are averaged across 500 Monte Carlo samples. 
			\end{tablenotes}
		\end{threeparttable}
	\end{adjustbox}
\end{table}

\clearpage
\bibliography{biblio}

\end{document}